\documentclass{tlp}

\usepackage{latexsym}

\newtheorem{theorem}{Theorem}[section]
\newtheorem{lemma}[theorem]{Lemma}

\newtheorem{definition}[theorem]{Definition}
\newtheorem{proposition}[theorem]{Proposition}

\newtheorem{property}[theorem]{Property}

\newcommand{\myif}{\ \mbox{\rm if}\ }

\newcommand{\myiff}{\ \mbox{\rm iff}\ }
\renewcommand{\and}{\ \mbox{\rm and}\ }
\newcommand{\myor}{\ \mbox{\rm or}\ }

\newcommand{\otherwise}{\ \mbox{\rm otherwise}}
\newcommand{\subc}{\succ}
\newcommand{\comp}{\circ}
\newcommand{\cleq}{\sqsubseteq}
\newcommand{\clt}{\sqsubset}

\newcommand{\scleq}{\unlhd}
\newcommand{\eqass}{\approx}

\newcommand{\Cc}[1]{{\it Cc}(#1)}

\newcommand{\Nstar}{{\mbox{\bf N}^\star}}
\newcommand{\N}{{\mbox{\bf N}}}

\newcommand{\pag}{\mbox{\tt Pat($\cal R$)}}

\title[Sequence-Based Abstract Interpretation of Prolog]
	{Sequence-Based Abstract Interpretation of Prolog}

\author[B. Le Charlier, S. Rossi and P. Van Hentenryck]
{BAUDOUIN LE CHARLIER\\
Institut d'Informatique, University of Namur,\\
21 rue Grandgagnage, B-5000 Namur, Belgium
\and SABINA ROSSI\\
Dipartimento di Informatica, Universit\`a di Venezia,\\
via Torino 155, 30172 Venezia, Italy
 \and PASCAL VAN HENTENRYCK\\
Department of Computer Science,
Brown University,\\ P.O. Box 1910, Providence
RI 02912, Usa
}

\begin{document}
\maketitle

\begin{abstract}

Abstract interpretation is a general methodology for 
systematic development of program analyses. An abstract interpretation 
framework is centered around a parametrized
non-standard semantics that can be instantia\-ted by various domains to approximate different
program properties.

Many abstract interpretation frameworks and analyses for Prolog
have been proposed, which seek to extract information useful for program optimization.
Although motivated by practical considerations, notably making Prolog competitive with imperative 
languages, such frameworks fail to capture some of the 
control structures of existing implementations
of the language.

 In this paper we propose a novel framework for the abstract interpretation 
of Prolog which  handles the depth-first search rule and the cut operator.
 It relies on the notion of {\em substitution sequence} to model the result of
the execution of a goal.
The framework consists of  (i) a denotational
concrete semantics, (ii) a  safe abstraction of the concrete semantics
defined in terms of a class of post-fixpoints, and (iii) a generic abstract
interpretation algorithm. We show that traditional abstract domains of
substitutions may easily be adapted  to the new framework, and provide
experimental evidence of the effectiveness of our approach.
We also show that previous work on determinacy analysis, that was not
expressible by existing abstract interpretation frameworks, can be seen as an
instance of our framework.

The ideas developed in this paper can be applied to other logic
languages, notably to  constraint logic languages,
and the theoretical approach  should
be of general interest for the analysis of many 
non-deterministic programming languages.

\end{abstract}

\section{Introduction}
\label{introduction}

{\it Abstract interpretation} \cite{Cousot77} is a general 
methodology for systematic development of program analyses.
It has been applied to
various   formalisms and paradigms  including flow-charts and imperative,
functional, logic, and constraint programming. 


Abstract interpretation of Prolog and, more generally, of logic
programming was initiated by Mellish \shortcite{Mellish87} 
and further developed by numerous researchers,
e.g., Bruynooghe \shortcite{Bruynooghe91},
Cousot and Cousot \shortcite{Cousot92},
Jones and S{\o}ndergaard \shortcite{Jones87},
Le Charlier \emph{et al.}
\shortcite{ICLP91AI}, Marriott and S{\o}ndergaard \shortcite{Marriott89b}.
Many different kinds of practical analyses 
and optimizations have been proposed,
a detailed description of which can be found in
\cite{Cousot92,Getzinger.SAS94}.  Briefly, mode
\cite{Cortesi91,D89,Debray88c,So86}, type
\cite{BG89b,Graph.JLP,GZ86,Janssens92,KH85,K83,Kluzniak87,L83,MK84,XW88,YS91},
and aliasing \cite{Codish91,Jacobs89} analyses collect information
about the state of variables during the execution and are useful to
speed up term unification and make memory allocation more efficient
\cite{Hermenegildo92,Warren88}. Sharing analysis
\cite{Corsini91,Cortesi,Kluzniak88,Muthukumar91} is similar to
aliasing except that it refers to the sharing of memory structures to
which program variables are instantiated; it is useful to perform
compile-time garbage collection \cite{Jensen90,Kluzniak88,Mulkers90}
and automatic parallelization \cite{SAS94.Gras,Chang85,Giacobazzi90,Jacobs92}.
Reference chain analysis \cite{Marien89b,Vanroy92} attempts to determine an
upper bound to the length of the pointer chain for a program variable.
Trai\-ling analysis \cite{Taylor89} aims at detecting variables which do
not need to be trailed.  Liveness analysis \cite{Mulkers91a}
determines when memory structures can be reused
 and is useful to perform update-in-place. 

All these analyses  approximate the set of values (i.e., terms or memory
structures) to which program variables can be instantiated at some
given program point. It is thus not surprising   that almost all
frameworks for the abstract interpretation of Prolog, e.g.,
\cite{Barbuti90,Bruynooghe91,Jones87,Marriott93,Marriott89b,Mellish87,Nilsson90a},
are based on abstractions of  sets of substitutions.  Such 
traditional
frameworks ignore  
important control features of the language, like
 the depth-first
search strategy and the cut operator. 
The reason is that  these control features 
are difficult to model accurately,
and yet not strictly necessary for a {\em variable level} analysis. 
However, modeling Prolog control features has two main advantages. First,
it allows one to perform so-called
{\em predicate level} analyses, like determinacy
\cite{Giacobazzi92,Sahlin91,Ueda,Vanroy87,Vanroy92} and local stack
\cite{Marien89,Meier91} analyses.
These analyses are not captured by traditional abstract interpretation 
frameworks; 
they usually rely on some ad hoc te\-chni\-que
and  require special-purpose proofs of correctness,
e.g., \cite{Debray89a,Sahlin91},
  which may be rather involved.
They are useful to perform
optimizations, such as  the
choice point removal and the simplification of environment creation.
Second,
 the analysis of some classes
of programs, like  programs containing 
multi-directional procedures which use
cuts and meta-predicates
to select among different versions, 
may be widely improved. 
This may provide the compiler with more
chances to perform important optimizations such
as dead-code elimination.

Abstract interpretation of Prolog with control
has been investigated by other authors.
In particular, we know of three main different approaches.
The  approach of  Barbuti {\em et al.}   \shortcite{BarbutiControl}
is based on  an abstract semantics for logic programs 
with control which is
parametric with respect to a ``termination theo\-ry''. The latter
 is intended to be provided from outside, for instance
by applying proofs procedures. 
Fil\`e and  Rossi \shortcite{FileRossi}
propose
an operational and non-compositional
abstract interpretation framework for  Prolog with cut
 consisting of a tabled interpreter to visit OLDT abstract trees
decorated with information about sure success or failure of goals.
Finally,  Spoto   \shortcite{Spoto}
define
 an abstract goal-independent denotational 
semantics for Prolog  handling  control rules  and  cut.
Program denotations are adorned with
``observability'' constraints giving information about 
  divergent computations and cut executions.
We know of no experimental results
 validating the effectiveness of  these
 approaches.

In this paper we present a novel abstract interpretation framework for Prolog
which models the depth-first search rule and the cut
operator. It relies on the notion of {\em substitution sequence}
which allows us to collect the solutions to a goal
together with
information such as sure success and failure, the number of solutions,
and/or termination.
The framework that we propose can  be applied to perform predicate
level analyses, such as determinacy, which were not expressible by classical frameworks, and 
can be also used  to improve the accuracy of
existing analyses.
Experiments on a  sample analysis, namely cardinality analysis, will be
discussed.

\subsection{Some Motivating Examples}

In this section we illustrate by means of  
 small examples the 
 functionality of our static analyzer
 and  we discuss
 how it improves on previous abstract interpretation frameworks.
Experimental results on medium-size programs will be reported  later.

The first two examples show that  predicate level properties,
such as determinacy, 
which are out of the scope of traditional 
 abstract interpretation frameworks can be captured by our analyzer.
To the best of our knowledge,
does not exist any specific analysis which 
can infer determinacy of all the programs that are discussed hereafter.
 
Consider first  the procedure {\tt is\_last}:

\begin{tt} \begin{tabbing} 12\=1234\=\kill 
\> is\_last(X,[X]). \\
\> is\_last(X,[\_|T]) :- is\_last(X,T).  \end{tabbing} \end{tt}

When given the input pattern {\tt is\_last(var,ground)}, where {\tt var}
and {\tt ground}
denote the set of all variables and 
the set of all  ground terms respectively, our analy\-sis
returns the  abstract sequence 
$\langle${\tt is\_last(ground,[ground|ground]),0,1,pt}$\rangle$, where
{\tt is\_last(ground,[ground|ground])}  is the pattern characterizing  the
output substitutions,
0 and 1 are, 
respectively,
the minimum and the maximum number of
returned
 output substitutions, and
 {\tt pt} stands for
``possible termination''. 

Consider now the following two versions of the procedure {\tt partition}. 

\begin{tt}
\begin{tabbing} 
12\=1234\=\kill 
\> partition([],P,[],[]). \\
\> partition([S|T],P,[S|Ss],Bs) :- 
      S $\leq$ P, !, 
      partition(T,P,Ss,Bs). \\
\> partition([B|T],P,Ss,[B|Bs]) :- 
    partition(T,P,Ss,Bs). 
\end{tabbing} 
\end{tt}

\begin{tt}
\begin{tabbing} 
12\=1234\=\kill 
\> partition([],P,[],[]). \\
\> partition([S|T],P,[S|Ss],Bs) :- 
  leq(S,P), partition(T,P,Ss,Bs). \\
\> partition([B|T],P,Ss,[B|Bs]) :- 
      gt(B,P), partition(T,P,Ss,Bs). \\
\> leq(K1-V1,K2-V2) :- K1 $\leq$ K2. \\
\> gt(K1-V1,K2-V2) :- K1 $>$ K2.
\end{tabbing} 
\end{tt}

Note that the second version of the procedure
 calls arithmetic predicates through an
auxiliary predicate and is appropriate for a key sort. Given an input
pattern {\tt partition(ground,ground,var,var)}, our analysis returns in both
cases the abstract sequence 
$\langle${\tt partition(ground,ground,ground,ground),0,1,pt}$\rangle$.  
Input/out\-put patterns are used to
determine that the first clause and the two others are mutually
exclusive in both programs, while the cut (in the first version) and
the abstraction of arithmetic predicates (in the second version)
determine the mutual exclusion of the second and the third clause.
Thus we can infer determinacy of both versions of the procedure
{\tt partition}. 

As stated above,  we don't know of any
 static analysis for logic programs
which can infer 
determinacy of all these programs.
For instance, the analysis 
developed by Debray and Warren   \shortcite{Debray89a}
to detect functional computations of a logic program
cannot infer determinacy of the procedure ${\tt is\_last}$; 
the  determinacy analysis proposed by Dawson {\em et al.} 
\shortcite{Ramakrishnan93},
while  it can handle the second version of the procedure
{\tt partition}, it  cannot handle  the first version of it since it does not
deal with the cut; for the same reason,
the  analysis of Giacobazzi and Ricci 
\shortcite{Giacobazzi92} cannot treat the first version of the procedure
{\tt partition};
and the
 cardinali\-ty analysis defined by Sahlin  \shortcite{Sahlin91} cannot handle
any of the examples discussed above since it ignores predicate
arguments.

The next example shows that the use of abstract sequences can improve
on the analysis of variable level properties such as modes.

Consider  the procedure {\tt compress(L,Lc)}, which relates two
lists  {\tt Lc} and {\tt L} 
 such that {\tt Lc} is a compressed version of {\tt L}. 
For instance, the
compressed version of  the list
{\tt [a, b, b, c, c, c]} is {\tt [a, 1, b, 2, c, 3]}. A
library can contain the definition of a single procedure to handle
both compression and decompression as follows. 

\begin{tt}
\begin{tabbing} 
12\=1234\=\kill 
\> compress(A,B) :- 
      var(A), !, decmp(A,B). \\
\> compress(A,B) :- 
      cmp(A,B).
\end{tabbing}
\end{tt}

\begin{tt}
\begin{tabbing} 
12\=1234\=\kill
\> cmp([],[]). \\
\> cmp([C],[C,1]). \\
\> cmp([C1,C2|T],[C1,1,C2,N|Rest])  :- 
       C1 \hskip-0.14cm <> \hskip-0.14cm C2, 
      cmp([C2|T],[C2,N|Rest]). \\
\> cmp([C1,C1|T],[C1,N1|Rest]) :- 
     cmp([C1|T],[C1,N|Rest]), 
       N1 \hskip-0.14cm  := \hskip-0.14cm  N \hskip-0.14cm + \hskip-0.14cm 1.
\end{tabbing}
\end{tt} 

\begin{tt}
\begin{tabbing} 
12\=1234\=\kill 
\> decmp([],[]).  \\
\> decmp([C],[C,1]). \\
\> decmp([C1,C2|T],[C1,1,C2,N|Rest]) \hskip-0.2cm :- \hskip-0.15cm
      decomp([C2|T],[C2,N|Rest]), 
      C1 \hskip-0.14cm <> \hskip-0.14cm C2. \\
\> decmp([C1,C1|T],[C1,N1|Rest])  \hskip-0.2cm {:-} \hskip-0.15cm
      N1 \hskip-0.14cm   > \hskip-0.14cm 1,  \hskip-0.1cm
      N \hskip-0.14cm  := \hskip-0.14cm  N1 \hskip-0.14cm - \hskip-0.14cm 1,\\
  \hskip6.1cm
      decmp([C1|T],[C1,N|Rest]).
\end{tabbing} 
\end{tt}

Given the input patterns {\tt compress(ground,var)} and {\tt compress(var,ground)}, our
analysis returns the abstract sequence $\langle${\tt
compress(ground,ground),0,1,pt}$\rangle$ for both the inputs.
This example
illustrates many of the functionalities of our sy\-stem, including
input/output patterns, abstraction of arithmetic and meta-pre\-di\-ca\-tes,
and the cut, all of which are necessary to obtain the optimal
precision. In addition, it shows that taking the cut into account
improves the analysis of modes.
Indeed,  a mode analysis
ignoring the cut would return the output pattern ${\tt
compress(novar,ground)}$ for the input pattern {\tt compress(var,ground)}, losing
the groundness information. None of the 
abstract interpretation algorithms for logic programs we know of
 can handle this example with an optimal result. 
 Moreover, if
a program only uses the input pattern {\tt compress(var,ground)}, our
analysis detects that the second clause of {\tt compress} is dead code
without any extra processing since no input/output pattern exists for
{\tt comp}. The second clause, the test {\tt var}, and the cut
of the first clause can  then be removed by an optimizer.

Notice that there exist implemented tools for the static
analysis  of Prolog programs, such as PLAI
\cite{Muthukumar92}, which can achieve as accurate  success
and dead-code information 
as our analyzer. However, such tools usually
integrate
  several analyses based on different techniques which
are
not all justified by the
abstract interpretation framework.
The example of the procedure {\tt compress}  shows that
our analy\-zer can
 handle control features of the language
within the abstract interpretation framework
without the need of any extra consideration.

\subsection{Sequence-Based Abstract Interpretation of Prolog}

An abstract interpretation framework \cite{CousotJLC92} is centered around the 
definition of a  non-standard 
 semantics
approximating a concrete semantics of the language.

Most top-down abstract interpretation
frameworks for logic programs, see, for instance,
  \cite{Bruynooghe91,Codognet92a,%
Jones87,TOPLAS,Marriott89,%
Mellish87,Muthukumar92,Nilsson90a,Warren92,Winsborough92},
can be viewed as abstractions of
a concrete structural operational semantics \cite{Plotkin81}.
Such a semantics defines the meaning of a program as a
{\em transition relation} described in terms of transition rules of the form
 $\langle \theta, o \rangle \longmapsto \theta'$, where the latter expresses the fact
 that $\theta'$ is a possible output from the execution of
the construct $o$ (i.e., a procedure, a clause, etc.) called with input
$\theta$.
This structural operational semantics can  easily be
rephrased as a fixpoint semantics
mapping any input pattern  $\langle \theta, o \rangle$ to the set of all corresponding outputs $\theta'$. The fixpoint semantics
can then be lifted to a {\em collecting semantics} that maps 
sets of inputs to sets of outputs and is defined as
 the least fixpoint of a set-based
transformation. 
The {\em non-standard} (or {\em abstract}) {\em semantics} is identical to the
collecting one except that it uses abstract values instead of
sets and abstract operations instead of
operations over sets.
 Finally, an abstract interpretation algorithm can be  derived by
instantiating a generic fixpoint algorithm
 \cite{universal} to the abstract semantics.

The limitations of traditional top-down frameworks for Prolog stem
from the fact that  structural operational semantics 
 are unable to take the depth-first search rule into account.
Control operators such as the cut cannot be modeled
 and are thus simply ignored. To overcome these limitations, we
propose a concrete semantics of Prolog which describes the result of program executions
in terms of substitution sequences. This allows us to 
model the
depth-first search rule and the cut operator. The semantics
 is defined in the denotational setting to deal
with sequences resul\-ting from the execution of infinite computations. 
 Moreover, it is still compositional
allowing us to reuse most of the material of our previous works,
i.e., the abstract domains and the generic algorithm \cite{TOPLAS}.
\\
However, technical proble\-ms arise when applying the 
abstract interpretation approach
  described above. Let us informally explain 
the main ideas behind the definition of our framework.

First,
  we define a {\em concrete semantics} as the
  least fixpoint of a concrete transformation {\it TCB} 
  mapping every so-called concrete behavior $\longmapsto$ to another
  concrete behavior $\stackrel{{\it TCB}}\longmapsto$.  The notion
  of concrete behavior is our
  denotation choice for a Prolog program: it  is a function
  that maps pairs of the form $\langle \theta, p \rangle $
  to a substitution sequence $S$, which
  intuitively represents the sequence of computed answer substitutions
  returned by the query $p(x_1,\dots,x_n)\theta$.  
  The fixpoint construction of the concrete semantics relies on
  a suitable ordering $\sqsubseteq$  defined on sequences.

Second,
  a collecting transformation {\it TCD} is obtained by lifting the concrete
  transformation {\it TCB} to  sets of substitutions and sets of
  sequences.  The transformation
  {\it TCD} is monotonic with respect to set inclusion.  However,
  its least fixpoint does not safely
  approximate the concrete semantics. In fact,  the least set 
  with respect to inclusion, that is the empty set $\{\}$, does not contain the
  least substitution sequence with respect to $\sqsubseteq$, which is a
  special sequence denoted by $<\bot>$. The problem relies on the fact that
  an ordering 
  on sets of sequences that ``combines'' both the ordering
  $\sqsubseteq$ on sequences and the ordering $\subseteq$ on sets is needed.
  This is an instance of the power domain construction problem
  \cite{Schmidt88}, which is difficult in general. 
  We choose a more pragmatic solution which
 consists in restricting to {\em chain-closed}
  sets of sequences, i.e., sets  containing the limit of
  every increasing chain, with respect to $\sqsubseteq$, of their elements.
  We also introduce the notion of
  {\em pre-consistent} collecting behavior which, roughly
  speaking, contains a lower approximation, with respect to
  $\sqsubseteq$, of the concrete semantics (the least fixpoint of
  {\it TCB}). The transformation {\it TCD} maps
  pre-consistent collecting behaviors to other pre-consistent ones.
  Moreover, assuming that sets of sequences
  are {\em chain-closed},  any pre-consistent post-fixpoint, with
  respect to set inclusion, of {\it TCD} safely approximates the
  concrete semantics.  These results imply that a safe collecting
  behavior can be constructed by iterating on {\it TCD} from any
  initial pre-consistent collecting behavior and by applying some
  widening techniques \cite{Cousot92c} in order to reach a
  post-fixpoint.

Third,
  the {\it abstract semantics} is defined exactly  as
  the collecting one  except that it is parametric with respect to the
  abstract domains.
  In fact, we do not
  explicitly distinguish between the collecting and the abstract
  semantics: in our presentation, the collecting transformation {\it
    TCD} is just a particular instance of the (generic) abstract
  transformation {\it TAB}.

  Finally, a generic abstract interpretation algorithm 
  is derived from the abstract semantics. The algorithm  is
  essentially an instantiation of the universal fixpoint algorithm
  described in \cite{universal}.  

\subsection{Plan of the Paper}
The paper is organized as follows.
Section \ref{CS} and  Section \ref{sec:AS} describe, respectively, 
our concrete and abstract semantics for pure Prolog augmented
with  the cut. 
The generic abstract interpretation algorithm is discussed in Section 
\ref{sec:GAIA}.
Section \ref{sec:FASAS} is a revised and extended version 
of \cite{cardinality}. 
It describes an instantiation of our abstract interpretation
framework 
to approximate the
number of solutions to a goal. Experimental results are reported.
In Section \ref{sec:RDASBAI}
we consider related works on determinacy
analysis.
 Section \ref{conclusion} concludes the paper.

\section{Concrete semantics}
\label{CS}

This section  describes a concrete semantics
 for pure Prolog augmented with the cut.
The concrete semantics is the link between the standard semantics 
of the language and
the abstract one.
Our concrete semantics is denotational and is based on the notion of 
substitution sequence.
Correctness of the concrete semantics with respect to  Prolog 
standard semantics,
i.e., OLD-resolution, is discussed.
Most proofs are omitted here; all details can be found in
 \cite{Cut_rep95}.


\subsection{Syntax}
\label{syntax}

The abstract interpretation framework presented in this paper
assumes that programs are normalized according to the abstract
syntax given in 
Fig.~\ref{abstract_syntax}.  The variables occurring in a literal are distinct;
distinct procedures have distinct names; all clauses of a procedure
have exactly the same head; if a clause uses $m$ different program
variables, these variables are $x_1$, \dots, $x_m$.\\

\begin{figure}[h]
\figrule
\begin{it}
\begin{tabular}{lclllcl}
P  &$ \in$&  Programs & & P  & ::= & pr $|$ pr P \\
pr & $ \in$& Procedures & & pr & ::=& c $|$ c pr  \\
c  &$ \in$&  Clauses & & c  & ::=& h  :- g.      \\
h  &$ \in$&  ClauseHeads & & h  & ::=& p($x_1$, \ldots, $x_n$)    \\
g  &$ \in$&  ClauseBodyPrefixes & & g  & ::=& $<>$ $|$ g , l \\
l  &$ \in$&  Literals  & & l  & ::=& p($x_{i_1}$, \ldots, $x_{i_n}$) $|$ b  \\
b  &$ \in$&  Built-ins & & b  & ::=& $x_i$=$x_j$ $|$ $x_{i_1}$=f($x_{i_2}$, \ldots, $x_{i_n}$) $|$ ! \\
p  &$ \in$&  ProcedureNames\\
f  &$ \in$&  Functors\\
$x_i$ &$ \in$&  ProgramVariables (PV)
\end{tabular} 
\end{it}
\caption{Abstract syntax of normalized programs}
\label{abstract_syntax}
\end{figure}


\subsection{Basic Semantic Domains}
\label{main:BSD}

This section presents the basic semantic domains of substitutions.
 Note that we assume a preliminary knowledge of
logic programming; see, for instance \cite{Apt97,Lloyd}.\\

\noindent
{\bf Variables and Terms.}
We assume the existence of two disjoint and infinite sets of variables, denoted by
${\it PV}$ and ${\it SV}$.
Elements of ${\it PV}$ are called {\it program variables} and are denoted by
$x_1$, $x_2$, \dots, $x_i$, \dots. The set ${\it PV}$ is totally ordered;
 $x_i$ is the $i$-th element of ${\it PV}$.
Elements of ${\it SV}$ are called {\em standard variables} and  are denoted by
 letters $y$ and $z$ (possibly subscripted).
Terms are built using standard variables only.
\\

\noindent
{\bf Standard Substitutions.}
Standard substitutions are substitutions in the usual
sense 
which use standard variables only.
The set of standard substitutions is denoted by ${\it SS}$. 
Renamings are standard substitutions that define a permutation
of standard variables.
The domain and the codomain of a standard substitution $\sigma$
are denoted by ${\it dom}(\sigma)$ and ${\it codom}(\sigma)$, respectively.
We denote by ${\it mgu}(t_1,t_2)$ the set of standard substitutions
that are a most general unifier of terms $t_1$ and $t_2$.\\

\noindent
{\bf Program Substitutions.}
 A program substitution  is a  set 
$\{x_{i_1}/t_1,\dots,x_{i_n}/t_n\}$, where $x_{i_1},\ldots,x_{i_n}$ are distinct program
variables and
$t_1$, \dots, $t_n$ are terms.
Program substitutions are not substitutions in the usual sense;
they are best understood as a form of program store
which expresses the  state of the computation at a given
program point. It is meaningless to compose them as usual substitutions
or to use them to express most general unifiers.
The domain of a program substitution 
$\theta=\{x_{i_1}/t_1,\dots,x_{i_n}/t_n\}$, denoted by
${\it dom}(\theta)$, is the set of
program variables $\{x_{i_1},\dots,x_{i_n}\}$.
The codomain of $\theta$, denoted by
${\it codom}(\theta)$, is the set of
standard variables occurring in $t_1,\dots,t_n$.
Program and standard substitutions
cannot be composed. Instead, standard substitutions
are {\em applied} to program substitutions.
The application of a standard substitution $\sigma$ to
a program substitution $\theta=\{x_{i_1}/t_1,\dots,x_{i_n}/t_n\}$
is the program substitution 
$\theta\sigma=\{x_{i_1}/t_1\sigma,\dots,x_{i_n}/t_n\sigma\}$.
The set of program substitutions is denoted by ${\it PS}$.
The application $x_i\theta$ of a program substitution $\theta$
to a program variable $x_i$ is defined only if  $x_i\in{\it dom}(\theta)$;
it denotes the term bound to $x_i$ in $\theta$.
Let $D$ be a finite subset of ${\it PV}$ and
 $\theta$ be a program substitution such that 
$ D \subseteq {\it dom}(\theta)$.
The {\em restriction} of $\theta$ to  $D$, denoted by $\theta_{/D}$,
is the program substitution such that ${\it dom}(\theta_{/D})= D$ and
$x_i(\theta_{/D})=x_i\theta$,
for all $x_i \in D$.
We denote by ${\it PS}_D$ the set of program substitutions with
domain $D$.\\

\noindent
{\bf Canonical Program Substitutions.}
We say that two program substitutions 
 $\theta$ and $\theta'$  are {\it equivalent} if and only if
 there exists a  renaming
$\rho$ such that $\theta\rho=\theta'$.
We assume that,
for each program substitution $\theta$, we are given a canonical
representative, denoted  by $[\![\theta]\!]$,
of the set of all program substitutions that are equivalent to $\theta$.
We denote by ${\it CPS}$
the set of all 
{\em canonical 
program substitutions} $[\![\theta]\!]$.
For any finite set of program variables $D$,
we denote by ${\it CPS}_D$ the set ${\it PS}_D\cap {\it CPS}$.

\subsection{Program Substitution Sequences}
\label{PSS}
Program substitution sequences are intended to model the sequence of
computed answer substitutions returned by a goal, a clause, or a procedure.
\\

\noindent
{\bf Program Substitution Sequences.}
Let us denote by  $\Nstar$  the set of positive natural numbers.
A program substitution sequence 
 is either a {\em finite}
sequence of the form $<\theta_1, \ldots, \theta_n>$ ($n \geq 0$) or 
an {\em incomplete} sequence of the form $<\theta_1, \ldots, \theta_n,\bot>$
($n \geq 0$) or an {\em infinite} sequence of the form
$<\theta_1, \ldots, \theta_i, \ldots>$ ($i\!\in\!\Nstar $), where the 
$\theta_i$ are program substitutions with the same domain.
We use the notation  $<\theta_1, \ldots, \theta_i, \_>$ to represent
a program substitution sequence when it is not known whether it is finite,
 incomplete or infinite.
Let $S$ be a program substitution sequence.
We denote by   ${\it Subst}(S)$  the 
set of program substitutions that are elements of  $S$. 
The domain of 
$S$ is defined when $S\neq<>$ and $S\neq<\bot>$. In this case, 
${\it dom}(S)$ is the domain of
the program substitutions belonging to ${\it Subst}(S)$.
The set of all program substitution sequences is denoted by {\it PSS}.
Let $D$ be a finite set of program variables. We denote by
{\it  PSS}$_D$  the set of all program substitution sequences 
with domain $D$ augmented
with $<>$ and $<\bot>$.
Let $S\in{\it  PSS}_D$ be a sequence $<\theta_1, \ldots, \theta_i, \_>$
and $D'\subseteq D$. 
The {\em restriction} of $S$ to $D'$, denoted by $S_{/{D'}}$,
is the program substitution sequence $<{\theta_1}_{/{D'}}, \ldots, {\theta_i}_{/{D'}}, \_>$.
The number of elements of $S$,
including the special element $\bot$, is denoted by ${\it Ne}(S)$.
The number of elements of $S$ that are substitutions
is denoted by ${\it Ns}(S)$. 
Sequence concatenation is denoted by $::$ and it is used only when its
first argument is a  finite sequence.\\

\noindent
{\bf Canonical Substitution Sequences.}
The canonical mapping $[\![\cdot]\!]$ is lifted to sequences as
follows.
 Let $S$ be a program substitution sequence $<\theta_1, \ldots, \theta_i, \_>$.
We define 
 $[\![S]\!]=
<[\![\theta_1]\!], \ldots, [\![\theta_i]\!], \_>$.
We denote  by {\it CPSS} the set of all
{\em canonical substitution sequences} $[\![S]\!]$
and by {\it  CPSS}$_D$ the set ${\it  PSS}_D \cap {\it CPSS}$,
for any finite subset  $D$ of {\it PV}.
\\

\noindent
{\bf CPO's of Program Substitution Sequences.}
The sets {\it  PSS}, {\it  PSS}$_D$, {\it  CPSS} and {\it  CPSS}$_D$
can be endowed with a structure of {\it pointed cpo} as described below.

\begin{definition}[Relation $\sqsubseteq$ on Program Substitution Sequences]
\label{seq_ordering}
Let $S_1,S_2\!\in\!{\it PSS}$. We define
\begin{center}
\begin{tabular}{llll}
$S_1 \sqsubseteq S_2$ &  iff & either & $S_1=S_2$ \\
 & & or &there exists $S,S'\!\in\!
{\it PSS}$  such that $S$ is finite,\\
& & &$S_1=S::<\bot>$ and  $S_2=S::S'$.
\end{tabular}
\end{center}
\end{definition}
%

The relation $\sqsubseteq$ on program substitution sequences is an ordering
and
 the pairs $\langle {\it PSS},\sqsubseteq\rangle $,
$\langle {\it CPSS},\sqsubseteq\rangle $,
$\langle {\it PSS}_D,\sqsubseteq\rangle $, and
$\langle {\it CPSS}_D,\sqsubseteq\rangle $
are all pointed cpo's.
\\

We denote by
  $(S_i)_{i\in\N}$  an increasing  chain, 
$S_0  \sqsubseteq S_1  \sqsubseteq \ldots \sqsubseteq S_i  \sqsubseteq
\ldots$ in
  {\it PSS}; whereas we denote by  $\{S_i\}_{i\in\N}$ a,
 non necessarily increasing,
sequence  of elements of
  {\it PSS}.
\\

\noindent
{\bf Lazy Concatenation.}
Program substitution sequences are combined 
through the  operation $\Box$ and its extensions $\Box_{k=1}^{n}$ and $\Box_{k=1}^{\infty}$
defined below.

\begin{definition}[Operation $\Box$]
Let $S_1$, $S_2\!\in\!  {\it PSS}$.
\begin{center}
\begin{tabular}{llcll}
$S_1 \Box S_2$ & $=$ & $S_1::S_2$ & if & $S_1$ is finite\\
               & $=$ & $S_1$      & if & $S_1$ is incomplete or infinite.\\
\\
\end{tabular}
\end{center}
\end{definition}

\begin{definition}[Operation $\Box_{k=1}^{n}$]
Let $\{S_k\}_{{k\in\Nstar}}$ be an 
infinite sequence of program substitution sequences (not neces\-sarily a chain).
For any $n \geq 1$, we define:
\begin{center}
\begin{tabular}{lll} 
$\Box_{k=1}^{0}S_k$ & $=$ & $<\;>$  \\
 $\Box_{k=1}^{n}S_k$ & $=$ & $(\Box_{k=1}^{n-1}S_k)\Box S_n.$\\
\\
\end{tabular}
\end{center}
\end{definition}

\begin{definition}[Operation $\Box_{k=1}^{\infty}$]
Let $\{S_k\}_{{k\in\Nstar}}$ be an 
infinite sequence of program substitution sequences.
The infinite sequence $\{S'_i\}_{i\in\N}$ where
$S'_i= (\Box_{k=1}^{i}S_k)\Box<\bot>$ $({i\in\N})$ is 
a chain. So we are allowed to define:
\begin{center}
\begin{tabular}{lllll}
$\Box_{k=1}^{\infty}S_k$ & = & $\sqcup_{i=0}^{\infty} S'_i$
                         & = & 
                               $\sqcup_{i=0}^{\infty}((\Box_{k=1}^{i}S_k)\Box
                                                                      <\bot>)$.\\
\\
\end{tabular}
\end{center}
\end{definition}
%

The operation $\Box$ is associative; hence,  it is meaningful to 
write $S_1 \Box \ldots \Box S_n$ instead of
$\Box_{k=1}^{n} S_k$. Operations $\Box$,
$\Box_{k=1}^{n}$, and $\Box_{k=1}^{\infty}$ are continuous
with respect to the ordering $\sqsubseteq$ on program substitution sequences.\\

%

\noindent
{\bf Program Substitution Sequences with Cut Information.}
Program substitution sequences with cut information 
 are used to model the result of a  clause together with
  information on cut executions. 

Let ${\it CF }$ be the set of cut flags $\{{\it cut},
 {\it nocut}\}$.
A  program substitution sequence with cut information
 is a pair $\langle S, cf\rangle$
where $S\!\in\!{\it PSS}$ and $cf\!\in\!{\it CF }$.

\begin{definition} [Relation $\sqsubseteq$ on  Substitution Sequences
                                with~Cut~Information]
\label{leqpssc} 
Let $\langle S_1,{\it cf}_1 \rangle , 
     \langle S_2,{\it cf}_2 \rangle \!\in\! {\it PSS}\times {\it CF}$.
We define
\begin{center}
\begin{tabular}{llll}
 $\langle S_1,{\it cf}_1 \rangle 
\sqsubseteq \langle S_2,{\it cf}_2 \rangle $ 
      & iff & either &$S_1\sqsubseteq S_2$ and ${\it cf}_1 = {\it cf}_2$ \\ 
      &     & or    & $S_1=<\bot>$ and ${\it cf}_1={\it nocut}$.
\\
\\
\end{tabular}
\end{center}
\end{definition}

The relation $\sqsubseteq$ on program substitution sequences 
with cut information is an or\-de\-ring.
 Moreover,
 the pairs 
 $\langle {\it PSS}\times {\it CF},\sqsubseteq \rangle $, 
$\langle {\it PSS}_D\times {\it CF},\sqsubseteq \rangle $, 
$\langle {\it CPSS}\times {\it CF},\sqsubseteq \rangle $ and 
$\langle {\it CPSS}_D\times {\it CF},\sqsubseteq \rangle $ 
are all  pointed cpo's. 

We extend the definition of the operation $\Box$  to
program substitution sequences with cut information. 
The extension is  continuous
in both the arguments.

\begin{definition}[Operation $\Box$ with Cut Information]
\label{OCI}
Let $\langle S_1,{\it cf}\rangle\!\in\!{\it PSS}\times {\it CF}$ 
                                   and $S_2\!\in\!{\it PSS}$. We define
\begin{center}
\begin{tabular}{lllll}
 $\langle S_1,{\it cf}\rangle \Box S_2$
            & $=$ & $S_1\Box S_2$ & if & {\it cf} = {\it nocut}\\
            &     & $S_1$         & if & {\it cf} = {\it cut}.
\end{tabular}
\end{center}
\end{definition}

\subsection{Concrete Behaviors}
\label{CB}

The notion of concrete behavior
 provides a mathema\-ti\-cal model for the input/output behavior of programs.
To simplify the presentation, we do not parameterize the semantics
with respect to programs. Instead, we assume a given fixed underlying program
{\it P}.

\begin{definition} [Concrete Underlying Domain]
The {\em concrete underlying domain}, 
denoted by {\it CUD},
 is the set of all pairs $\langle \theta,p \rangle$ such that 
$p$ is the name of a procedure {\it pr}
of $P$ and 
$\theta\!\in\!{\it CPS}_{\{x_1,\dots,x_n\}}$,
where $x_1,\dots,x_n$ are the variables
 occurring in the head of every clause of {\it pr}.
\end{definition}

Concrete behaviors are functions but we 
 denote them by the relation symbol $\longmapsto$ in order to
stress the similarities  between the concrete semantics and a 
structural operational semantics for logic programs defined in  \cite{ACTA95}.

\begin{definition} [Concrete Behaviors]
A {\em concrete behavior}
is a total function $\longmapsto: {\it CUD}\longrightarrow {\it CPSS}$
mapping every pair  $\langle \theta,p \rangle\!\in\!{\it CUD}$  to a
canonical program  substitution
sequence $S$ such that,
for every  $\theta'\!\in\!{\it Subst}(S)$, there exists a standard substitution
$\sigma$ such that $\theta'=\theta \sigma$. 
We denote by $\langle \theta,p \rangle\longmapsto S$  the fact that
$\longmapsto$ maps the pair $\langle \theta,p \rangle$ to $S$.
The set of all concrete behaviors is denoted by
{\it CB}. 
\end{definition}

The ordering $\sqsubseteq$ on program substitution sequences is
lifted to concrete behaviors in a standard way \cite{Schmidt88}.

\begin{definition}  [Relation $\sqsubseteq$ on Concrete Behaviors]
Let $\longmapsto_1 , \longmapsto_2\in\!{\it CB}$. We define
\begin{center}
\begin{tabular}{lclll}
$ \longmapsto_1 \sqsubseteq \longmapsto_2$ & iff & 
                   ($\langle \theta,p \rangle\longmapsto_1 S_1$ and
                    $\langle \theta,p \rangle\longmapsto_2 S_2$)
             imply
   $\ S_1 \sqsubseteq S_2$,  \\
            & & for all
                                  $\langle \theta,p \rangle\!\in
                                  \!{\it CUD}.$
\end{tabular}
\end{center}
\end{definition}

The following result is straightforward.

\begin{proposition}
\label{CB-cpo}
$\langle {\it CB}, \sqsubseteq \rangle$  is a pointed cpo, i.e.,
\begin{enumerate}
\itemsep 2pt
\item the relation $\sqsubseteq$  on
          {\it CB} is a partial order;
\item {\it CB} has a minimum element, which is
                         the concrete behavior
                         $\longmapsto_\bot$ such that
                      for all $\langle \theta,p \rangle\!\in\! {\it CUD}$,
                            $\langle \theta,p \rangle\longmapsto_\bot<\bot>$;
\item every chain $(\longmapsto_i)_{i\in\N}$ in {\it CB} 
       has a least upper bound, denoted by
                                $\sqcup_{i=0}^{\infty}\longmapsto_i$;
      $\sqcup_{i=0}^{\infty}\longmapsto_i$ is the concrete behavior
      $\longmapsto$  such that,
       for all $ \langle \theta,p \rangle\!\in \!{\it CUD}$, 
      $\langle \theta,p \rangle\longmapsto \sqcup_{i=0}^{\infty}S_i$,
      where $\langle \theta,p \rangle\longmapsto_i S_i$
                              $(\forall i\!\in\!\N)$.

\end{enumerate}
\end{proposition}

\subsection{Concrete Operations}
\label{CO}

We specify here the concrete operations which are used in the
definition of the concrete semantics.
The choice of these particular operations is motivated by the fact that
they have useful (i.e., practical)
abstract counterparts  (see Sections~\ref{sec:AS}, \ref{sec:GAIA}
and~\ref{sec:FASAS}). 
%
The concrete operations are polymorphic since their
exact signature depends on a clause $c$ or a literal $l$ or both.\\

Let $c$ be a clause,  $D=\{x_1,\ldots,x_n\}$ be the set of all variables 
occurring in the head of $c$, and
$D'=\{x_1,\ldots,x_m\}$ ($n\leq m$) be the set of all variables 
occurring in $c$.\\

\noindent
{\bf Extension at  Clause Entry}\  :\ \  
     {\tt EXTC}$(c,\cdot):{\it CPS}_D
\rightarrow({\it CPSS}_{D'}\times {\it CF}) $\ \\ 
This operation extends a substitution $\theta$ on 
the set of variables in $D$ 
to the set of variables in $D'$.
Let $\theta\!\in\!{\it CPS}_{D}$.
\begin{quote}
\begin{tabular}{llll}
${\tt EXTC}(c,\theta)$ & $ = $ & 
$ \langle  < [\![\theta']\!]>, {\it nocut} \rangle$
                   & \\

\end{tabular}
\end{quote}
\noindent
 where $x_i\theta'=x_i\theta$ $(\forall i:1\leq i \leq n)$ and
  $x_{n+1}\theta'$, \dots , $x_{m}\theta'$ are
               distinct standard    variables    not belonging to ${\it codom}(\theta)$.

\vskip0.3cm

\noindent
{\bf Restriction at Clause Exit}\ :\ \ 
 {\tt RESTRC}$(c,\cdot):({\it CPSS}_{D'}\times {\it CF})
                         \rightarrow({\it CPSS}_D\times {\it CF}) $ \ \\ 
This operation restricts a pair $\langle S,{\it cf}\rangle$, 
representing the result of the execution of $c$ on the set of
variables in $D'$,
to the set of variables in $D$.
Let  $\langle S, {\it cf}\rangle
\!\in\!({\it CPSS}_D'\times {\it CF})$.
\begin{quote}
\begin{tabular}{llll}
${\tt RESTRC}(c,\langle S,{\it cf}\rangle)$ & $=$ & 
                  $  \langle [\![S']\!],{\it cf}\rangle$ &
where $ S'=S_{/{D}} $.
\end{tabular}
\end{quote}
\vskip0.3cm

Let $l$ be a literal occurring in the body of $c$,
$D''=\{x_{i_1},\ldots,x_{i_{r}}\}$ be the set of 
variables occurring in $l$, and
$D'''$ be equal to $\{x_{1},\ldots,x_{r}\}$.

\vskip0.3cm
\noindent
{\bf Restriction before a Call}\ :\ \ 
 {\tt RESTRG}$(l,\cdot):{\it CPS}_{D''}
                         \rightarrow{\it CPS}_{D'''} $\ \\ 
This operation expresses a substitution $\theta$ on the
parameters $x_{i_1},\ldots,x_{i_r}$ of a call $l$ 
in terms of the formal parameters $x_1,\ldots,x_r$ of $l$. Let
  $\theta \!\in\!{\it CPS}_{D''}$.
\begin{quote}
\begin{tabular}{lll}
${\tt RESTRG}(l,\theta)$ & $=$ & 
     $  [\![\{x_1/x_{i_1}\theta,\ldots,x_{r}/x_{i_{r}}\theta\}]\!]  .$
\end{tabular}
\end{quote}

\vskip0.3cm
\noindent
{\bf Extension of the Result of a Call}\ :\ \ 
 {\tt EXTG}$(l,\cdot,\cdot): {\it CPS}_{D'}\times {\it CPSS}_{D'''}
                         \not\rightarrow{\it CPSS}_{D'} $ \ \\ 
\noindent
This operation extends a substitution $\theta$
with a substitution sequence $S$ repre\-senting the result of
executing a call $l$ on $\theta$.
Hence, it is only used in contexts where the substitutions that are elements
of $S$ are (roughly speaking) instances of $\theta$. 
 Let 
 $\theta\in {\it CPS}_{D'}$.
Let $S\in {\it CPSS}_{D'''} $ be of the form
$<\theta'\sigma_1,\ldots,\theta'\sigma_i,\_>$ where
$x_j\theta'=x_{i_j}\theta$
$(1\leq j\leq {r})$
and the
$\sigma_i$ are standard substitutions such that
${\it dom}(\sigma_i)\subseteq {\it codom}(\theta')$.
%
%
%
Let $\{z_1,\ldots,z_s\}={\it codom}(\theta)
                             \setminus{\it codom}(\theta')$.
      Let $y_{i\mbox{\rm \tiny ,}1},\ldots,y_{i\mbox{\rm \tiny ,}s}$ 
      be distinct standard variables
      not belonging to ${\it codom}(\theta)\cup {\it codom}(\sigma_i)$
      $(1\leq i\leq {\it Ns}(S))$.
      Let $\rho_i$ be a renaming of the form $\{z_1/y_{i\mbox{\rm \tiny ,}1},\ldots,
                    z_s/y_{i\mbox{\rm \tiny ,}s},
                    y_{i\mbox{\rm \tiny ,}1}/z_1,\ldots,
                    y_{i\mbox{\rm \tiny ,}s}/z_s\}$.
\begin{quote}
\begin{tabular}{lll}
${\tt EXTG}(l,\theta,S)$ & $=$ & 
 $  [\![<\theta\rho_1\sigma_1,\ldots,\theta\rho_i\sigma_i,\_>]\!]  .$
\end{tabular}
\end{quote}
%

It is easy to see that the value of  ${\tt EXTG}(l,\theta,S)$  does not depend on the
choice of the $y_{i\mbox{\rm \tiny ,}j}$.
Moreover,  it  is not defined when $S$ is not
of the above mentioned form.\\

\noindent
{\bf Unification of Two Variables}\ :\ \ 
   {\tt UNIF-VAR}$: {\it CPS}_{\{x_1,x_2\}} 
                     \rightarrow {\it CPSS}_{\{x_1,x_2\}}$\  \\ 
Let $\theta\!\in\!{\it CPS}_{\{x_1,x_2\}}$.
This operation unifies $x_1\theta$ with $x_2\theta$.
\begin{quote}
\begin{tabular}{llll}
{\tt UNIF-VAR}$(\theta)$ & $ =$  & $<> $ &
                       if $x_1\theta$ and $x_2\theta$ are not unifiable,\\
  &  $ =$ & $[\![<\theta\sigma>]\!]$ & where $\sigma\!\in\! {\it mgu}
                                                   (x_1\theta,x_2\theta)$,
                                       otherwise.
\end{tabular}  
\end{quote}

\vskip0.3cm
\noindent 
{\bf Unification of a Variable and a Functor}\ :\ \ 
   {\tt UNIF-FUNC}$(f,\cdot): {\it CPS}_{D} 
                     \rightarrow {\it CPSS}_{D}$\ \\ 
Given a functor $f$ of arity $n-1$ and a substitution 
$\theta\!\in\!{\it CPS}_{D}$ where $D=\{x_1,\ldots,x_n\}$, the
{\tt UNIF-FUNC} operation unifies $x_1\theta$ with $f(x_2,\ldots,x_n)
\theta$.
\begin{quote}
\begin{tabular}{llll}
{\tt UNIF-FUNC}$(f,\theta)$ & $ =$  & $<> $ &
                       if $x_1\theta$ and $f(x_2,\ldots,x_{n}\theta)$ are not unifiable,\\
  &  $ =$ & $[\![<\theta\sigma>]\!]$ & where $\sigma\!\in\! {\it mgu}
                                                   (x_1\theta,f(x_2,\ldots,
x_{n})\theta)$,
                                       otherwise.
\end{tabular}  
\end{quote}
\vskip0.3cm

All   operations above
are monotonic and continuous.
We assume that
Sets of program substitutions are  endowed with the 
ordering $\sqsubseteq$ such that $\theta\sqsubseteq\theta'$ iff
$\theta=\theta'$. 

\subsection{Concrete Semantic Rules}
\label{CSR}

The concrete semantics of the underlying program $P$ is the least fixpoint
of a conti\-nu\-ous transformation on {\it CB} 
(the set of concrete behaviors).
This transformation is defined in terms of a set of semantic rules
that naturally extend a concrete behavior to a continuous function
defining the input/output behavior of every prefix of the body of a
clause,
every clause, every suffix
 of a procedure and every procedure of $P$.
This function is called {\em extended concrete behavior}
and maps each element of the extended concrete underlying domain
to a substitution sequence, possibly with cut information, as defined below.

\begin{definition} [Extended Concrete Underlying Domain]
The {\em extended concrete underlying domain},  
denoted by {\it ECUD}, consists of
\begin{enumerate}
\itemsep 2pt
\item
 all triples $\langle \theta, g, c \rangle $, where 
  $c$ is a clause of $P$,
  $g$ is a prefix of the body of $c$,
  and $\theta$ is a canonical program substitution over the variables 
  in the head of $c$; 
\item
 all pairs $\langle \theta,  c \rangle $, where 
  $c$ is a clause of $P$
  and $\theta$ is a canonical program substitution over the variables 
  in the head of $c$; 
\item
 all pairs $\langle \theta,  {\it pr} \rangle $, where 
  ${\it pr}$ is a suffix of a procedure of $P$
  and $\theta$ is a canonical program substitution over the variables 
  in the head of the
  clauses of {\it pr}.
\end{enumerate}
\end{definition}

\begin{definition} [Extended Concrete Behaviors]
An {\em extended concrete behavior} is a total function from
{\it ECUD} to the set ${\it CPSS}\cup({\it CPSS}\times{\it CF})$ such that
\begin{enumerate}
\itemsep 2pt
\item
every triple $\langle \theta, g, c \rangle $ from {\it ECUD}
      is mapped to a program substitution sequence with cut information $\langle S, {\it cf}\rangle $ such that
      ${\it dom}(S)$ is the set of all varia\-bles in  $c$;
\item every pair $\langle \theta,  c \rangle $ from {\it ECUD}
      is mapped to a program substitution sequence with cut information $\langle S, {\it cf}\rangle $ such that
      ${\it dom}(S)$ is the set of variables in the head of  $c$;
\item every pair $\langle \theta,  {\it pr} \rangle $ from {\it ECUD}
      is mapped to a program substitution sequence $S$ such that
      ${\it dom}(S)$ is the set of variables in the head of the
      clauses of  {\it pr}.
\end{enumerate}

\end{definition}

The set of extended concrete behaviors is endowed with a structure of
pointed cpo in the obvious way. It is denoted by {\it ECB}; its
elements  are denoted by $\longmapsto$.

Let $\longmapsto$ be a concrete behavior. The concrete semantic rules
depicted in Figu\-re~\ref{CSRF} define an extended concrete behavior
derived from $\longmapsto$. This extended concrete behavior is
denoted by the same symbol $\longmapsto$. This does not lead to
confusion since the inputs of the two functions belong to different
sets. The definition proceeds by induction on the syntactic structure
of $P$.

The concrete semantic rules  model Prolog operational semantics
through the notion of program substitution sequence.
Rule {\bf R1} defines the program substitution sequence with 
cut information at the entry point of a clause.
Rules {\bf R2} and {\bf R3} define the effect of the execution of a cut
at the clause level. Rules {\bf R4},
{\bf R5} and {\bf R6} deal with execution of literals;
procedure calls are solved by using the concrete behavior
$\longmapsto$ as an oracle.
Rule {\bf R7} defines the result of a clause.
Rules {\bf R8} and {\bf R9} define the result of a procedure
by structural induction on its suffixes.
Rule {\bf R8} deals
with the suffix consisting of the last clause only: it simply
forgets the cut information, which is not meaningful at the procedure
level.
Rule {\bf R9} combines the result of a clause with the
(combined) result of the next clauses in the same procedure:
 it deals with the execution of a cut at the procedure level.
The expression ${\Box}^{{\it Ne}(S)}_{k=1} S_k$ used in
 Rules {\bf R4},
{\bf R5} and {\bf R6}  deserves an
explanation: when the sequence $S$ is incomplete, it is assumed that
$S_{_{{\it Ne}(S)}}=<\bot>$. This convention
is necessary to propagate the non-termination of $ g'$
to $g$.


The following results are instrumental for  proving the well-definedness
of the concrete semantics.

\begin{proposition}[Properties of the Concrete Semantic Rules]
  \begin{enumerate}
\itemsep 2pt
  \item Given a concrete behavior, the concrete semantic rules define
        a unique extended concrete behavior, i.e., 
        a unique mapping from
        {\it CB} to {\it ECB}. 
This mapping is continuous.
  \item Rules {\bf R1} to {\bf R6} have a conclusion of the form
        $\langle \theta, g, c \rangle \longmapsto 
         \langle S, {\it cf}\rangle $.
        In all cases, $S$ is of the
        form $<\theta'\sigma_1,\dots,\theta'\sigma_i,\_>$,
        where the $\sigma_i$ are standard substitutions and
        $\langle\theta',{\it nocut}\rangle={\tt EXTC}(c,\theta)$.
  \\  Rules {\bf R7} to {\bf R9} have a conclusion
    of the form  $\langle \theta, \cdot\rangle \longmapsto  S$.
In all cases, $S$ is of the form
	 \mbox{$<\theta\sigma_1,\dots,\theta\sigma_i,\_>$},
	 where the $\sigma_i$ are standard substitutions.
      \end{enumerate}
 \end{proposition}

 \subsection{Concrete Semantics}
 \label{TCS}

 The concrete semantics of the underlying program $P$ is
defined as the least fixpoint of the following concrete transformation.

 \begin{definition}[Concrete Transformation]
 \label{CT}

 \noindent 
The transformation 
${\it TCB} : {\it CB}\rightarrow{\it CB}$ 
is defined
as follows: for all $\longmapsto \in {\it CB}$,
\vskip0.2cm
 \begin{center}
 \begin{tabular}{rcl} 
   & {\it pr} is a procedure of {\it P}                         & \\ 
   & $p$ is the name of  {\it pr}                                &\\
   & $\langle \theta, {\it pr} \rangle
    \longmapsto S$                                            &  \\
 {\bf T1} & \raisebox{.7ex}{\underline{\hspace*{3.7cm}}}      
							 &    \\
  &  $\langle \theta, p \rangle
     \stackrel{{\it TCB}}\longmapsto 
		   S$                                             \\ 
 \end{tabular}
 \end{center}
where $\stackrel{{\it TCB}}\longmapsto $ stands for ${\it TCB}(\longmapsto)$.
Remember that
$\langle \theta, {\it pr} \rangle
    \longmapsto S$ is defined by means of the previous rules
 which use the concrete behavior $\longmapsto$ as an oracle to solve the procedure calls.
 \end{definition}
{\begin{figure}[top]
\figrule
\begin{it}

\begin{center}
\begin{tabular}{rc}
\\
&  g\  ::=\ $<>$                                                 \\
{\bf R1}  & \raisebox{.7ex}{\underline{\hspace*{3.7cm}}}     
    \\
& $\langle \theta, g, c \rangle
   \longmapsto
 {\tt EXTC}(c,\theta)$    \\                                     \\
\end{tabular}
\end{center}
\hspace{-7cm}
\begin{tabular}{cc}
\begin{tabular}{lc}
&  g\  ::=\ g' , !                                             \\
&  $\langle \theta, g',c \rangle
   \longmapsto
   \langle S, {\it cf} \rangle$                               \\
 & $S\in\{<\bot>,<>\}$                                         \\
{\bf R2}  & \raisebox{.7ex}{\underline{\hspace*{3.7cm}}}      
 \\
& $\langle \theta, g, c \rangle
   \longmapsto
   \langle S, {\it cf} \rangle$                                \\
\end{tabular}
&
\hspace{-6cm}
\begin{tabular}{lc}
 & g\  ::=\ g' , !                                             \\
 & $\langle \theta, g',c \rangle
   \longmapsto
   \langle S, {\it cf} \rangle$                               \\
  & $S = <\theta'>::S'$                                         \\
{\bf R3} &\raisebox{.7ex}{\underline{\hspace*{3.7cm}}} 
    \\
& $\langle \theta, g, c \rangle
   \longmapsto
   \langle <\theta'>, {\it cut} \rangle$                       \\
\end{tabular}
\\ \\ \\
\begin{tabular}{lc}
 & g\  ::=\ g' , l                                             \\
  & l\  ::=\ $x_i$=$x_j$                                        \\

  & $\langle \theta, g',c \rangle
   \longmapsto
   \langle S, {\it cf} \rangle$                               \\
  & $S = <\theta_1,\dots,\theta_i,\_>$                          \\

& $\left\{\begin{array}{c}
 \theta'_k = {\tt RESTRG}(l,\theta_k)                        \\ 
S_{k}^{' }=\mbox{\tt UNIF-VAR}(\theta'_k)                      \\      
S_k = {\tt EXTG}(l,\theta_k,S_{k}^{' })                     \\
 (1\leq k\leq {\it Ns}(S))           \end{array}\right\}$       \\ 

{\bf R4} & \raisebox{.7ex}{\underline{\hspace*{3.7cm}}}      
                \\
& $\langle\theta, g, c \rangle 
  \longmapsto  
 \langle {\Box}^{{\it Ne}(S)}_{k=1} S_k,
                          {\it cf} \rangle$                   \\
\end{tabular}
&
\hspace{-6cm}
\begin{tabular}{lc}
 & g\  ::=\ g' , l                                             \\
 & l\  ::=\ 
   $x_{i_1}$=f($x_{i_2}$, \ldots, $x_{i_n}$)                  \\

  & $\langle \theta, g',c \rangle
   \longmapsto
   \langle S, {\it cf} \rangle$                               \\
  & $S = <\theta_1,\dots,\theta_i,\_>$                          \\

& $\left\{\begin{array}{c}
\theta'_k = {\tt RESTRG}(l,\theta_k)                        \\ 
S_{k}^{' }=\mbox{\tt UNIF-FUNC}(f,\theta'_k)                   \\ 
S_k = {\tt EXTG}(l,\theta_k,S_{k}^{' })                     \\ 
  (1\leq k\leq {\it Ns}(S) )   \end{array}\right\}$         \\ 

{\bf R5} & \raisebox{.7ex}{\underline{\hspace*{3.7cm}}}      
          \\
& $\langle\theta, g, c \rangle 
  \longmapsto  
 \langle {\Box}^{{\it Ne}(S)}_{k=1} S_k,
                          {\it cf} \rangle$                   \\
\end{tabular}
\\ \\ \\ 
\begin{tabular}{lc}
 &  g\  ::=\ g' , l                                             \\
  & l\  ::=\ 
    p($x_{i_1}$, \ldots, $x_{i_n}$)                           \\

  & $\langle \theta, g',c \rangle
   \longmapsto
   \langle S, {\it cf} \rangle$                               \\
  & $S = <\theta_1,\dots,\theta_i,\_>$                          \\

 & $\left\{\begin{array}{c}
\theta'_k = {\tt RESTRG}(l,\theta_k)                        \\ 
\langle  \theta'_k ,p\rangle
    \longmapsto  S_{k}^{'}                                   \\ 
S_k = {\tt EXTG}(l,\theta_k,S_{k}^{' })                     \\ 
   (1\leq k\leq {\it Ns}(S))       \end{array}\right\}$  \\ 

{\bf R6} & \raisebox{.7ex}{\underline{\hspace*{3.7cm}}}    
                        \\
& $\langle\theta, g, c \rangle 
  \longmapsto  
 \langle {\Box}^{{\it Ne}(S)}_{k=1} S_k,
                          {\it cf} \rangle$                   \\
\end{tabular}
&
\hspace{-6cm}
\begin{tabular}{lc}
\\
\\
\\
\\
\\
                                                              \\
  & c \  ::=\  h  :- g.                                         \\ 
  & $\langle \theta, g ,c \rangle
   \longmapsto
   \langle S, {\it cf} \rangle$                               \\
{\bf R7} &\raisebox{.7ex}{\underline{\hspace*{3.7cm}}}      
                            \\
  & $\langle \theta, c \rangle
   \longmapsto
{\tt RESTRC}(c, \langle S, {\it cf} \rangle)$                 \\    
\end{tabular}
\\ \\ \\
\begin{tabular}{lc}
                                                              \\
 & pr \  ::=\  c                                               \\ 
  & $\langle \theta, c \rangle
   \longmapsto
   \langle S, {\it cf} \rangle$                               \\
{\bf R8} & \raisebox{.7ex}{\underline{\hspace*{3.7cm}}}      
                                      \\
  & $\langle \theta, {\it pr} \rangle
   \longmapsto S$                                             \\ 
\end{tabular}
&
\hspace{-6cm}
\begin{tabular}{lc}
  & pr \  ::=\  c {\it pr}'                                     \\ 
  & $\langle \theta, c \rangle
   \longmapsto
   \langle S, {\it cf} \rangle$                               \\
  & $\langle \theta, {\it pr}' \rangle
   \longmapsto S'$                                            \\
{\bf R9} & \raisebox{.7ex}{\underline{\hspace*{3.7cm}}}      
                                    \\
  & $\langle \theta, {\it pr} \rangle
   \longmapsto 
   \langle S, {\it cf} \rangle \Box S'$                       \\ 
\end{tabular}\\ \\

\end{tabular}
\end{it}
\caption{Concrete semantic rules}
\label{CSRF}
\end{figure}}

The transformation ${\it TCB}$
 is well-defined  and continuous. 

 \begin{definition}[Concrete Semantics]
 \label{DCS}

 \noindent The concrete semantics of the underlying program $P$ is the least
 concrete beha\-vior $\longmapsto$ such that
 $$ \longmapsto\ =\ \stackrel{{\it TCB}}\longmapsto.$$
 \end{definition}

 \subsection{Correctness of the Concrete Semantics}
 \label{CCS}

 Since OLD-resolution \cite{Lloyd,Tamaki86} is the standard semantics of
 pure Prolog augmented with cut,  our concrete
 semantics and OLD-resolution  have to be proven equivalent.
 The proof is fairly complex because OLD-resolution is not compositional.
 Consequently, the two semantics do not  naturally  match.
 The equivalence proof is given in \cite{Cut_rep95}. In this section, we
 only give the principle of the proof.
 \begin{enumerate}
 \itemsep 2pt
 \item We assume that OLD-resolution uses standard variables to
 rename clauses apart.
 The  initial queries are also assumed to contain
 standard variables only.
 \item The notion of {\em incomplete OLD-tree limited to depth $k$}
 is defined ({\it IOLD}$_k$-tree, for short).
 Intuitively, an {\it IOLD}$_k$-tree is an OLD-tree modified according
 to the following rules:
       \begin{enumerate}
\itemsep 2pt
       \item procedure calls may be unfolded only down to depth $k$;
       \item branches that end at a node whose leftmost literal may not
	    be unfolded are called {\em incomplete};
       \item a  depth-first left-to-right traversal of the tree is performed in
	    order to determine the cuts that are reached by the standard
	    execution and to prune the tree accordingly; see \cite{Lloyd};
       \item the traversal ends when the whole tree has been visited  or
	     when a node that may not be unfolded is reached;
       \item the branches on the right of the left-most incomplete branch
	     are pruned (if such a branch exists).
 \end{enumerate}
 \item Assuming a query of the form $p(t_1,\dots,t_n)$ and denoting
 the concrete beha\-vior
${\it TCB}^k({\longmapsto_{\bot}})$
 by $\longmapsto_k$, it can be shown that the sequence of computed
 answer substitutions $<\sigma_1,\dots,\sigma_i,\_>$ for the
 {\it IOLD}$_k$-tree of $p(t_1,\dots,t_n)$ is such that
 $\langle \theta, p\  \rangle \longmapsto_k
	 [\![ <\theta\sigma_1,\dots,\theta\sigma_i,\_>]\!]$
 where $\theta=\{x_1/t_1,\dots,x_n/t_n\}$.
 \item The equivalence of our concrete semantics and OLD-resolution
    is a simple consequence of the previous result.\\ For every query  
$p(t_1,\dots,t_n)$, $<\sigma_1,\dots,\sigma_i,\_\!>$
is the sequence of computed answer substitutions of $p(t_1,\dots,t_n)$
according to OLD-resolution if and only if
$\langle \theta, p\  \rangle \longmapsto
	 [\![ <\theta\sigma_1,\dots,\theta\sigma_i,\_>]\!]$
where $\theta=\{x_1/t_1,\dots,x_n/t_n\}$ and 
$\longmapsto$ is the concrete behavior of the program
    according to our concrete semantics.
 \end{enumerate}

 In fact, the correctness of our concrete semantics should be
 close to obvious to anyone who knows about both Prolog and denotational
 semantics.
So, the equivalence proof
 is 
 a formal technical exercise, which adds little to our basic 
 understan\-ding of the concrete semantics.

 \subsection{Related Works}
 \label{RW_CS}

 Denotational semantics for Prolog have been proposed before
  \cite{debruin,Debray88b,JonesMy}. Our concrete semantics is not
 intended to improve on these works from the language understanding
 standpoint. Instead, it is merely designed as a basis for an abstract
 interpretation framework; in particular, it uses concrete operations
 that are as close as possible to the operations used by the structural
 operational semantics presented in \cite{ACTA95} upon which 
 our previous frameworks
 are based. This allows us to reuse
 much of the material from our existing abstract domains and generic
 algorithms; see, \cite{SPE,ICLP91AI,TOPLAS,ACTA95}.
 The idea of distinguishing between
 finite, incomplete, and infinite sequences is originally due to
  Baudinet \shortcite{Baudinet}.
 %
 %

\section{Abstract semantics}
\label{sec:AS}

As we have already
 explained in the introduction, our abstract semantics is  not
 defined as a
least fixpoint of an abstract transformation
but instead as a set of post-fixpoints 
 that fulfill a safety requirement, namely
pre-consistency. Moreover, the abstract domains are assumed to 
 represent so-called
chain-closed sets of concrete elements as specified below.


\subsection{Abstract Domains}
\label{sub:AD}

We  state here  
the mathematical assumptions that are required to be satisfied by the  
abstract domains.
Specific abstract domains will be described
in  Section \ref{sec:FASAS}.\\

\noindent
{\bf Abstract Substitutions.}
For   every finite set $D$ of program variables, we denote by 
${\it CS}_D$ the set 
$\wp({\it PS}_D)$.
%
%
\noindent A domain of abstract substitutions is a family of sets ${\it AS}_D$
indexed by the finite sets $D$ of program variables. 
Elements of ${\it AS}_D$ are called {\em abstract substitutions};
they are denoted by $\beta$.
Each set ${\it AS}_D$ is endowed with a partial order $\leq$
and a monotonic  {\em concretization} function 
${\it Cc}:{\it AS}_D\rightarrow {\it CS}_D$
associating to each  abstract substitution $\beta$ the set 
${\it Cc}(\beta)$ of program
substitutions it denotes.\\

\noindent
{\bf Abstract Sequences.}
For every finite set $D$ of program variables, we denote by 
${\it CSS}_D$ the set $\wp({\it PSS}_D)$.
 Abstract sequences denote chain-closed subsets of ${\it CSS}_D$.\\
%
%
%
\noindent A domain of abstract sequences is a family of sets ${\it ASS}_D$
indexed by the finite sets $D$ of program variables. 
Elements of ${\it ASS}_D$ are called {\em abstract sequences};
they are denoted by $B$.
Each set ${\it ASS}_D$ is endowed with a partial order $\leq$
and a monotonic {\em concretization} function 
${\it Cc}:{\it ASS}_D\rightarrow {\it CSS}_D$.
Moreover, the following properties are required to be satisfied: (1)
every ${\it ASS}_D$ contains an abstract sequence $B_{\bot}$
such that $<\bot>\,\in {\it Cc}(B_{\bot})$; 
(2)
for every $B\in{\it ASS}_D$, ${\it Cc}(B)$ is {\em chain-closed}, i.e.,
      for every chain $(S_i)_{i\in {\bf N}}$ of elements of ${\it Cc}(B)$,
      the limit $\sqcup_{i=0}^{\infty}S_i$ also belongs to ${\it Cc}(B)$.
The disjoint union of all the ${\it ASS}_D$ is denoted by ${\it ASS}$.\\

\noindent
{\bf Abstract Sequences with Cut Information.}
Let ${\it CSSC}_D$ denote $\wp({\it PSS}_D\times {\it CF})$.
%
%
\noindent A domain of abstract sequences 
with cut information is a family of sets ${\it ASSC}_D$
indexed by the finite sets $D$ of program variables. 
Elements of ${\it ASSC}_D$ are called 
{\em abstract sequences with cut information};
they are denoted by $C$.
Every set ${\it ASSC}_D$ is endowed with a partial order $\leq$
and a monotonic {\em concretization} function 
${\it Cc}:{\it ASSC}_D\rightarrow {\it CSSC}_D$.
The disjoint union of all the ${\it ASSC}_D$ is denoted by ${\it ASSC}$.\\
%

\noindent
{\bf Abstract Behaviors.}
Abstract behaviors are the abstract counterpart of the concrete
behaviors introduced in Section \ref{CB}. They are endowed with a
weaker mathe\-matical structure as described below.
As in the case of concrete behaviors, a fixed underlying program $P$ is
assumed.

\begin{definition} [Abstract Underlying Domain]
\label{AAUD}
The {\em abstract underlying domain}, 
denoted by {\it AUD},
 is the set of all pairs $\langle \beta,p \rangle$ such that 
$p$ is  a procedure name
in $P$ of arity $n$ and 
$\beta\!\in\!{\it AS}_{\{x_1,\dots,x_n\}}$.
\end{definition}
\begin{definition}[Abstract Behaviors]
\label{def:A}
An {\em abstract behavior}
is a total function ${\it sat}: {\it AUD}\longrightarrow {\it ASS}$
mapping each pair  $\langle \beta,p \rangle\!\in\!{\it AUD}$  to an
abstract sequence $B$ with $B\in{\it ASS}_{\{x_1,\dots,x_n\}} $,
where $n$ is the arity of $p$.
The set of all abstract behaviors is denoted by
{\it AB}. 
The set {\it AB} is endowed with the partial ordering $\leq$ such that,
for all ${\it sat}_1,{\it sat}_2\in{\it AB}$:
$$
\begin{array}{lclr}
{\it sat}_1 \leq {\it sat}_2 
& \;
\mbox{\rm iff} \;
& 
{\it sat}_1\langle \beta,p \rangle \leq
{\it sat}_2\langle \beta,p \rangle, &\;
   \forall \langle \beta,p \rangle \in {\it AUD}.
\end{array}
$$
\end{definition}

It would be reasonable to assume that abstract behaviors
are monotonic functions but this is not necessary for the safety
results. The notation {\it sat} stands for ``set of
abstract tuples''. It is used because the abstract interpretation
algorithm, derived from the abstract semantics, actually computes a set
of tuples of the form $\langle \beta,p ,B\rangle $, i.e., a part of the
table of an abstract behavior.

\subsection{Abstract Operations}
\label{sub:AO}

In this section, we give the specification of the primitive
  abstract operations 
used  by the abstract semantics. 
The specifications are safety assumptions which, roughly speaking,
state that the abstract operations safely simulate the corresponding concrete
ones.
In particular,
operations {\tt EXTC}, {\tt RESTRG}, {\tt RESTRC},
      {\tt UNIF-VAR}, {\tt UNIF-FUNC} are faithful abstract counterparts of
      the corresponding concrete operations. Hence,
      their specification simply states
      that, if some concrete input belongs to the concretization of their  
    (abstract) input, then the corresponding concrete output  belongs to 
      the concretization
       of their (abstract) output. Moreover, overloading the opera\-tion
      names is natural in these cases.
       Operation  {\tt AI-CUT}
      deals with the cut; its specification is also straightforward.
Operations {\tt EXTGS} and {\tt CONC} are related to the concrete
      operations {\tt EXTG} and $\Box$ in a more involved way. We will discuss them  in more detail.
Finally, operations {\tt SUBST} and {\tt SEQ} are simple conversion
      operations to convert  an abstract domain into another.

 Let us  specify the operations, using the notations of 
Section \ref{CO}.
\\
\\
\noindent
{\bf Extension at  Clause Entry}\ :\ \  
     {\tt EXTC}$(c,\cdot):{\it AS}_D
\rightarrow{\it ASSC}_{D'} $\  \\
Let $\beta \in {\it AS}_{D}$ and $\theta \in {\it CPS}_{D}$.
The following property is required to hold.
\begin{quote}$
\begin{array}{lll}
\theta\in{\it Cc}(\beta)
&
\Rightarrow
&
{\tt EXTC}(c,\theta)\in {\it Cc}({\tt EXTC}(c,\beta)).
\end{array}$
\end{quote}
\vskip0.3cm
\noindent
{\bf Restriction at Clause Exit}\ :\ \ 
 {\tt RESTRC}$(c,\cdot):{\it ASSC}_{D'}
                         \rightarrow{\it ASSC}_D $\  \\
Let $C\in {\it ASSC}_{D'}$ and $\langle S, {\it cf}\rangle
\in({\it CPSS}_D'\times {\it CF})$. 
\begin{quote}$
\begin{array}{lll}
\langle S, {\it cf}\rangle \in{\it Cc}(C)
&
\Rightarrow
&
{\tt RESTRC}(c,\langle S, {\it cf}\rangle)\in 
{\it Cc}({\tt RESTRC}(c,C)).
\end{array}$
\end{quote}
\vskip0.3cm
\noindent
{\bf Restriction before a Call}\ :\ \ 
 {\tt RESTRG}$(l,\cdot):{\it AS}_{D''}
                         \rightarrow{\it AS}_{D'''} $\ \\ 
Let $\beta \in{\it AS}_{D''}$  and $\theta \in{\it CPS}_{D''}$. 
\begin{quote}$
\begin{array}{lll}
\theta \in{\it Cc}(\beta)
&
\Rightarrow
&
{\tt RESTRG}(l,\theta)\in 
{\it Cc}({\tt RESTRG}(l,\beta)).
\end{array}$
\end{quote}
\vskip0.3cm
\noindent
{\bf Unification of Two Variables}\ :\ \ 
   {\tt UNIF-VAR}$: {\it AS}_{\{x_1,x_2\}} 
                     \rightarrow {\it ASS}_{\{x_1,x_2\}}$\  \\ 
Let $\beta \in {\it AS}_{\{x_1,x_2\}}$ and
 $\theta\in{\it CPS}_{\{x_1,x_2\}}$. 
\begin{quote}$
\begin{array}{lll}
\theta \in{\it Cc}(\beta)
&
\Rightarrow
&
\mbox{\tt UNIF-VAR}(\theta)\in 
{\it Cc}(\mbox{\tt UNIF-VAR}(\beta)).
\end{array}$
\end{quote}
\vskip0.3cm
\noindent
{\bf Unification of a Variable and a Functor}\ :\ \ 
   {\tt UNIF-FUNC}$(f,\cdot): {\it AS}_{D} 
                     \rightarrow {\it ASS}_{{D}}$\  \\ 
Let $\beta \in {\it AS}_{D}$
and  $\theta \in {\it CPS}_{D}$. Let also $f$ be a functor
of arity $n-1$. 
\begin{quote}$
\begin{array}{lll}
\theta \in{\it Cc}(\beta)
&
\Rightarrow
&
\mbox{\tt UNIF-FUNC}(f,\theta)\in 
{\it Cc}(\mbox{\tt UNIF-FUNC}(f,\beta)).
\end{array}$
\end{quote}
\vskip0.3cm
\noindent
{\bf Abstract Interpretation of the Cut}\ :\ \ 
   {\tt AI-CUT}$: {\it ASSC}_{{D'}}
                         \rightarrow {\it ASSC}_{{D'}}$\  \\ 
Let $C \in {\it ASSC}_{D'}$,
    $\theta \in {\it CPS}_{D'}$,  
    $S \in {\it CPSS}_{D'}$,
    ${\it cf} \in {\it CF}$. 
\begin{quote}$
\begin{array}{rcl}
\langle <>,{\it cf}\rangle \in {\it Cc}(C) & \Rightarrow &
\langle <>,{\it cf}\rangle \in {\it Cc}(\mbox{\tt AI-CUT}(C)), \\ 
\langle <\bot>,{\it cf}\rangle \in {\it Cc}(C) & \Rightarrow &
\langle <\bot>,{\it cf}\rangle \in {\it Cc}(\mbox{\tt AI-CUT}(C)),\\ 
\langle <\theta>::S ,{\it cf}\rangle \in {\it Cc}(C) & \Rightarrow &
\langle <\theta>,{\it cut}\rangle \in {\it Cc}(\mbox{\tt AI-CUT}(C)).
\end{array}$
\end{quote}
\vskip0.3cm
\noindent
{\bf Extension of the Result of a Call}\ :\ \  
{\tt EXTGS}$(l,\cdot,\cdot):{\it ASSC}_{D'}\times {\it ASS}_{D'''}
                         \rightarrow{\it ASSC}_{D'} $\  \\ 
The specification of this operation is more complex because it
abstracts in a single operation the calculation of all sequences
$S_k = {\tt EXTG}(l,\theta_k,S'_{k})$ and of their concatenation
${\Box}^{{\it Ne}(S)}_{k=1} S_k$, performed by the
 rules {\bf R4}, {\bf R5}, {\bf R6}
(see Figure \ref{CSRF}).
At the abstract level, it may be too expensive or even impossible
to simulate the execution of $l$ for all elements of $S$,
as defined in the rules. Therefore,
we abstract $S$ to its substitutions,
losing the ordering. The abstract execution will be the following.
 Assuming that $C$ abstracts the
program substitution sequence with cut information 
$\langle S,{\it cf}\rangle$ before $l$, we compute $\beta={\tt SUBST}(C)$;
then we compute $\beta'={\tt RESTRG}(l,\beta)$ and, subsequently,
we get the abstract sequence $B$ resulting from the abstract exe\-cution of 
 $l$ with
input $\beta'$.
The set ${\it Cc}(B)$ contains
all sequences $S'_k$ of rules {\bf R4}, {\bf R5}, {\bf R6}. Then,
an over approximation of the set of all possible values
${\Box}^{{\it Ne}(S)}_{k=1} S_k$ is computed from the information provided 
by $C$
and $B$.
This is realized by the following operation {\tt EXTGS}.
 Let $C\in {\it ASSC}_{D'}$, $B\in {\it ASS}_{D'''}$,
 $\langle S,{\it cf}\rangle\in ({\it CPSS}_{D'}\times {\it CF})$ and
$S'_1,\dots,S'_{{\it Ns}(S)}\in{\it CPSS}_{D'''}$.
\begin{quote}$
\hskip-0.8cm
\begin{array}{rcl}
\left.
\begin{array}{c}
\langle S,{\it cf}\rangle  \in {\it Cc}(C),\\
S=<\theta_1,\dots,\theta_i,\_>,\\
       \left( \begin{array}{c}
\forall k:1\leq k\leq  {\it Ns}(S):
       S'_k\in{\it Cc}(B)\\
       \mbox{ and } S_k={\tt EXTG}(l,\theta_k,S'_k) 
\end{array}
\right)
\end{array}
\right\}
                &\Rightarrow&
\langle \Box_{k=1}^{{\it Ne}(S)}S_k ,{\it cf}\rangle \in 
                       {\it  Cc}(\mbox{\tt EXTGS}(l,C,B)).
\end{array}$
\end{quote}
%
%
\vskip0.3cm
\noindent
{\bf Abstract Lazy Concatenation}\ :\ \ 
    ${\tt CONC}: ({\it AS}_D \times {\it ASSC}_D  \times {\it ASS}_D)
                 \rightarrow {\it ASS}_D$\  \\
This operation is the abstract counterpart of the concatenation operation
$\Box$. It is however extended with an
additional argument to increase the accuracy.
 Let $B'={\tt CONC}(\beta,C,B)$ where
 $\beta$ describes a set of input substitutions for
a procedure; $C$ describes the set of substitution sequences with cut
information 
obtained by executing a clause of the procedure on $\beta$;  $B$ describes the set of
substitution sequences obtained by executing the
subsequent clauses of the procedure on $\beta$.
Then, $B'$ describes the set of substitution sequences obtained by
concatenating the results according to the concrete 
concatenation operation $\Box$.

Let us discuss a simple example to
 understand the role of $\beta$.
  Assume that
$$
\begin{array}{rcl}
{\it Cc}(C)=\{ \langle <>, {\it nocut} \rangle ,
            \langle <\{x_1/a\}>, {\it nocut} \rangle\} &
\mbox{\rm and }                                         &
{\it Cc}(B)=\{ <>, 
               <\{x_1/b\}> \}.
\end{array}
$$
If the input mode of $x_1$ is unknown, it must be assumed that
all combinations of elements in ${\it Cc}(C)$ and ${\it Cc}(B)$ are
possible. Thus,
$${\it Cc}(B')=\{ <>, <\{x_1/a\}>,
               <\{x_1/b\}>,  <\{x_1/a\},\{x_1/b\}> \}.$$
On the contrary, if the input mode of $x_1$ is known to be ground,
the outputs $\langle <\{x_1/a\}>, {\it nocut} \rangle$ and
$<\{x_1/b\}>$ are incompatible since $x_1$
cannot be bound to both $a$ and $b$ in the input substitution.
In this case, we have
$${\it Cc}(B')=\{ <>, <\{x_1/a\}>,
               <\{x_1/b\}>\}.$$
The first argument $\beta$ of the operation {\tt CONC} 
 provides information on the input values: it may be useful to improve
 the accuracy of the result.
The above discussion motivates the following specification of
operation {\tt CONC}.
Note that the statement 
\mbox{$(\exists\sigma\in{\it SS}: \theta' = \theta\sigma)$} is abbreviated by
$\theta'\leq \theta$ in the specification. Let
$\beta\in{\it AS}_D$,
$C\in{\it ASSC}_D$,
$B\in{\it ASS}_D$,
$\theta\in{\it CPS}_D$,
$\langle S_1,{\it cf}\rangle\in ({\it CPSS}_{D}\times {\it CF})$ and
$S_2\in{\it CPSS}_{D}$. 
\begin{quote}$
\begin{array}{rcl}
      \left.
      \begin{array}{c}
         \theta\in {\it Cc}(\beta),\\
         \langle S_1,{\it cf}\rangle \in {\it Cc}(C),\\
         S_2\in {\it Cc}(B),\\
         \forall \theta'\in {\it Subst}(S_1)\cup {\it Subst}(S_2): 
                                               \theta'\leq \theta
      \end{array}
      \right\}
                &\Rightarrow&
\langle S_1,{\it cf}\rangle \Box S_2 \in {\it Cc}({\tt CONC}(\beta,C,B)).
\end{array}$
\end{quote}

\vskip0.3cm
\noindent
{\bf Operation}\ \ ${\tt SEQ}: {\it ASSC}_{D}
                         \rightarrow{\it ASS}_{D}$\  \\ 
This operation forgets the cut information 
contained in an abstract sequence with cut information $C$. 
It is applied to the
result of the last clause of a procedure before 
combining this result with
the results of the other clauses. \\
Let $C\in {\it ASSC}_{D}$ and
 $\langle S,{\it cf}\rangle\in ({\it CPSS}_{D}\times {\it CF})$. 
\begin{quote}$
\begin{array}{rcl}
\langle S,{\it cf}\rangle \in {\it Cc}(C)       &\Rightarrow &
S\in {\it Cc}({\tt SEQ}(C)).
\end{array}$
\end{quote}
\vskip0.3cm
\noindent
{\bf Operation}\ \ ${\tt SUBST}: {\it ASSC}_{D'}
                         \rightarrow{\it AS}_{D'}$\  \\ 
This operation forgets still more
information.  It extracts the ``abstract substitution part'' of $C$.
It is applied before executing a literal in a clause. See operation
{\tt EXTGS}.
Let $C\in {\it ASSC}_{D'}$ and
 $\langle S,{\it cf}\rangle\in ({\it CPSS}_{D'}\times {\it CF})$. 
\begin{quote}$
\begin{array}{rcl}
         \langle S,{\it cf}\rangle \in {\it Cc}(C)
        &\Rightarrow &
{\it Subst(S)}\subseteq {\it Cc}({\tt SUBST}(C)).
\end{array}$
\end{quote}
\subsection{Abstract Semantics}
\label{sub:ASem}

We are now in position to present the abstract semantics.
Note that we are not concerned with
algorithmic issues here: they are dealt with in Section~\ref{sec:GAIA}.
\\

\noindent
{\bf Extended Abstract Behaviors.}
Extended abstract behaviors are the abstract counterpart of the
concrete extended behaviors defined in Section \ref{CSR}.
%

\begin{definition} [Extended Abstract Underlying Domain]
\label{DEAUD}
The {\em extended abstract underlying domain},  
denoted by {\it EAUD},
consists of 
\begin{enumerate}
\itemsep 2pt
\item all triples $\langle \beta, g, c \rangle $, where 
  $c$ is a clause of $P$,
  $g$ is a prefix of the body of $c$,
  $\beta\in{\it AS}_{D}$, and
  $D$ is the set of variables 
  in  the head of $c$;
\item all pairs $\langle \beta,  c \rangle $, where 
  $c$ is a clause of $P$,
  $\beta \in{\it AS}_{D}$, and
  $D$ is the set of variables 
  in  the head of $c$;
\item all pairs $\langle \beta,  {\it pr} \rangle $, where 
  ${\it pr}$ is a procedure of $P$
  or a suffix of a procedure of $P$,
  $\beta\in{\it AS}_{D}$, and
  $D$ is the set of variables 
  in the head of the
  clauses of {\it pr}.
\end{enumerate}
\end{definition}
\begin{definition} [Extended Abstract Behaviors]
An {\em extended abstract behavior} is a  function from
{\it EAUD} to ${\it ASS}\cup{\it ASSC}$ such~that
\begin{enumerate}
\itemsep 2pt
\item every triple $\langle \beta, g, c \rangle $ from {\it EAUD}
      is mapped to an abstract sequence with cut
      information
      $C\in {\it ASSC}_{D'} $, where $D'$ is the set of all variables in $c$;
\item every pair $\langle \beta,  c \rangle $ from {\it EAUD}
      is mapped to an abstract sequence with cut information 
      $C\in {\it ASSC}_D $, where $D$ is the set of variables 
      in the head of $c$;
\item every pair $\langle \beta,  {\it pr} \rangle $ from {\it EAUD}
      is mapped to an abstract sequence $B\in {\it ASS}_D$,  
       where $D$ is the set of variables
       in the head of the
       clauses of {\it pr}.
\end{enumerate}
\end{definition}

The set of extended abstract behaviors is endowed with a structure of
partial or\-der in the obvious way. It is denoted by {\it EAB} and its
elements are denoted by~{\it esat}.\\
%

\noindent
{\bf Abstract Transformation.}
The abstract semantics is defined in terms of two semantic functions
that are depicted in Figure \ref{fig:TAT}.
The first function $E: {\it AB}\rightarrow {\it EAB}$ maps
abstract behaviors to extended abstract behaviors.
It is the abstract counterpart of the concrete semantic rules of
Figure \ref{CSRF}.
The second function ${\it TAB} :{\it AB}\rightarrow{\it AB}$
transforms an abstract behavior into another abstract behavior.
It is the abstract counterpart of  Rule {\bf T1} in Definition
\ref{CT}.\\

\begin{figure}[t]
\figrule
\begin{tabbing}
123\=4567891\=23123\=456123456789012345\=12347891234123456789123456789 \kill

${\it TAB}({\it sat})\langle \beta,p \rangle 
          = E({\it sat})\langle \beta,{\it pr}\rangle$ \\
\>{\bf where}   \> ${\it pr}$  is the procedure defining $p$,\\ \\

$E({\it sat})\langle \beta,{\it pr} \rangle 
          = ${\tt SEQ}$(C)$ \\
\>{\bf where}  
          \> $C=E({\it sat})\langle \beta,c \rangle $ 
             \>\> if {\it pr $::=$   c}\\ \\

$E({\it sat})\langle \beta,{\it pr} \rangle 
          =  {\tt CONC}(\beta,C,B)$ \\
\>{\bf where} 
          \> $B=E({\it sat})\langle \beta,{\it pr}' \rangle $ \\
\>        \> $C=E({\it sat})\langle \beta,c \rangle $ 
             \> \>if {\it pr $::=$ c,pr}$'$\\ \\

$E({\it sat})\langle \beta,c \rangle 
          =$ {\tt RESTRC}$(c,C)$ \\
\>{\bf where} 
          \> $C = E({\it sat})\langle \beta,g,c\rangle$\\
\>        \> $g$ is the body of $c$\\ \\

$E({\it sat})\langle\beta,<>,c \rangle = {\tt EXTC}(c,\beta)$ \\ \\

$E({\it sat})\langle \beta,( g , ! ) ,c\rangle 
          = \mbox{\tt AI-CUT}(C)$ \\
\>{\bf where} 
          \> $C = E({\it sat})\langle \beta,g,c\rangle$ \\ \\
           
$E({\it sat})\langle \beta, ( g , l ) ,c\rangle 
          = {\tt EXTGS}(l,C,B)$ \\
\>{\bf where} 
          \> $B=$ 
             \> {\tt UNIF-VAR}$(\beta')$ 
                \> if {\it l $::=$ x$_i$=x$_j$} \\
\>        \> \> {\tt UNIF-FUNC}$(f,\beta')$ 
                \> if  {\it l $::=$ x$_i$=f$(\ldots)$} \\
\>        \> \> ${\it sat}\langle \beta',p \rangle$ 
                \> if  {\it l $::=$ p$(\ldots)$} \\
\>        \> $\beta' = {\tt RESTRG}(l,\beta'')$\\
\>        \> $\beta''= {\tt SUBST}(C)$\\
\>        \> $C= E({\it sat})\langle \beta,g,c \rangle$.
\end{tabbing}
\caption{The  abstract transformation}
\label{fig:TAT}
\end{figure}

\noindent
{\bf Abstract Semantics.}
The abstract semantics is defined as the set of all abstract
   behaviors that 
are both post-fixpoints of the abstract transformation
{\it TAB} and pre-consistent. 
The corresponding definitions are given first; then
  the rationale underlying 
the definitions  is   discussed.

\begin{definition}[Post-Fixpoints of {\it TAB}]
An abstract behavior ${\it sat}\in{\it AB}$ is called
a {\em post-fixpoint} of {\it TAB} if and only if
${\it TAB}({\it sat})\leq{\it sat} $, i.e., if and only if
$$
\begin{array}{llll}
{\it TAB}({\it sat})\langle \beta,p \rangle
&\leq
&{\it sat}\langle \beta,p \rangle,
&\;\;\forall \langle \beta,p \rangle \in {\it AUD}.\\
\end{array}
$$
\end{definition}
\begin{definition}[Pre-Consistent Abstract Behaviors]
\label{def:PCAB}
\noindent 
Let $\longmapsto$ be the concrete semantics of the underlying
program, according to Defi\-nition \ref{DCS}.
An abstract behavior ${\it sat}\in{\it AB}$ is said to be
 {\em pre-consistent} with respect to $\longmapsto$
if and only if
there exists a concrete behavior $\longmapsto'$
such that 
$$\longmapsto'\ \sqsubseteq \longmapsto$$
and such that,
for all $\langle \beta,p \rangle \in {\it AUD}$ and
$\langle \theta,p \rangle \in {\it CUD}$,
$$
\begin{array}{lll}
  \left.
  \begin{array}{c}
   \theta \in {\it Cc}(\beta),\\
   \langle \theta,p \rangle \longmapsto' S
  \end{array}
   \right\}
&\Rightarrow
&  S \in {\it Cc}({\it sat}\langle \beta,p \rangle).
\\
\end{array}
$$
\end{definition}

In the next section, we show that any pre-consistent
post-fixpoint {\it sat} of {\it TAB} is a safe approximation of
the concrete semantics, i.e., 
it is such that for all $\langle \beta,p \rangle \in {\it AUD}$ and
$\langle \theta,p \rangle \in {\it CUD}$,
$$
\begin{array}{lll}
  \left.
  \begin{array}{c}
   \theta \in {\it Cc}(\beta),\\
   \langle \theta,p \rangle \longmapsto S
  \end{array}
   \right\}
&\Rightarrow
&  S \in {\it Cc}({\it sat}\langle \beta,p \rangle).\\
\\
\end{array}
$$

The abstract semantics is defined as the set of all pre-consistent post-fixpoints.
Indeed, under the current hypotheses on the abstract domains, there is no straightforward way to choose a 
``best'' abstract behavior among all pre-consistent post-fixpoints.
Thus, we consider the problem of computing a reasonably accurate
post-fixpoint as a pragmatic issue to be solved at the algorithmic level.
In fact, the abstract interpretation algorithm 
presented in Section \ref{sec:GAIA} is an improvement of the following construction:
define the abstract behavior ${\it sat}_\bot$ by
$$
\begin{array}{llll}
{\it sat}_\bot \langle \beta, p \rangle 
& = 
& B_\bot,
& \; \;\forall \langle \beta, p \rangle \in {\it AUD}.
\end{array}
$$
Assume that the domain of abstract sequences is endowed with
an 
upper-bound operation ${\tt UB}: {\it ASS}_D \times {\it ASS}_D \rightarrow {\it ASS}_D$
(not necessarily a {\em least} upper bound).
For every ${\it sat}_1,{\it sat}_2\in {\it AB}$, we define
${\tt UB}({\it sat}_1,{\it sat}_2)$ by
$$
\begin{array}{llll}
{\tt UB}({\it sat}_1,{\it sat}_2)\langle \beta, p \rangle
& = 
& {\tt UB}({\it sat}_1\langle \beta, p \rangle,
           {\it sat}_2\langle \beta, p \rangle),
& \;\;\forall \langle \beta, p \rangle \in {\it AUD}.
\end{array}
$$
Let j be an arbitrarily chosen natural number.
An infinite sequence of pre-consistent abstract behaviors
${\it sat}_0,\ldots,{\it sat}_i,\ldots$ is defined as follows:
$$
\begin{array}{llll}
{\it sat}_0 & = & {\it sat}_\bot, \\
{\it sat}_{i+1}& = & {\it TAB}({\it sat}_i)
                   & (0\leq i <j ),\\
{\it sat}_{i+1}& = & {\tt UB}({\it sat}_i,{\it TAB}({\it sat}_i))
                   & ( j\leq i).  \\
\end{array}
$$
The abstract behaviors ${\it sat}_i$ are all pre-consistent because
${\it sat}_\bot$ is pre-consistent by construction,
every  application of {\it TAB} maintains pre-consistency (as proven
in the next section),
and each application of {\tt UB} produces an abstract behavior
whose concretization contains the concretizations of the arguments.
Moreover, assuming that every  partial order ${\it ASS}_D$ is finite
or satisfies the finite ascending chain proper\-ty, the
sequence ${\it sat}_0,\ldots,{\it sat}_i,\ldots$ has a least
upper bound which is the desired pre-consistent post-fixpoint.
In  case the ${\it ASS}_D$ contains chains with infinitely many distinct
elements, {\tt UB} must be a widening operator \cite{Cousot92c}.
\\
The sequence from ${\it sat}_0$ to ${\it sat}_j$
is not ascending in general. In fact, ${\it sat}_\bot$ is not the minimum of
{\it AB} and {\it TAB} is not necessarily monotonic nor extensive
(i.e., ${\it sat}\leq {\it TAB}({\it sat})$ does not always hold).
From step $0$ to $j$, the  computation of the ${\it sat}_i$ 
simulates as closely as possible the computation
of the least fixpoint of the concrete transformation.
From step $j$ to convergence, all iterates are ``lumped'' together. 
All concrete behaviors 
$\longmapsto_j, \longmapsto_{j+1}, \ldots$ 
of the Kleene sequence of the concrete semantics,
are thus included in the concretization of the final post-fixpoint {\it sat}.
So, {\it sat} describes properties that are true not only for the
concrete $\longmapsto$
semantics but also for its approximations 
$\longmapsto_j, \longmapsto_{j+1}, \ldots$.
The choice of $j$ is a compromise: a low value  ensures a faster
convergence while a high value provides a better accuracy.
The abstract interpretation algorithm presented in  Section
\ref{sec:GAIA} does not iterate globally over {\it TAB}. It locally
iterates over $E$ for every needed input 
pattern $\langle \beta, p \rangle$
and uses different values of $j$ for different input patterns.
Depending on the particular abstract domain, the value can be guessed
more or less cleverly. This is the role of the 
special widening operator of  Definition~\ref{def:EW}.
A sample widening operator is described in Section \ref{sub:GAO},
showing how the value of $j$ can be guessed in the case of a practical
abstract domain.

\subsection{Safety of the Abstract Semantics}
\label{sub:SAS}

We prove here the safety of our abstract semantics.
First, we formally define the notion of safe approximation. Then, we show
that the abstract transformation is safe in the sense that, whenever
{\it sat} safely approximates $\longmapsto$, 
${\it TAB}({\it sat})$ safely approximates
$\stackrel{{\it TCB}}\longmapsto$ (Theorem \ref{th:SAT}).
From this basic result, we deduce that
{\it TAB} transforms pre-consistent abstract
behaviors
into other pre-consistent abstract behaviors (Theorem \ref{th:p-c}),
and that, when {\it sat} is a post-fixpoint
of the abstract transformation which safely approximates 
a concrete behavior $\longmapsto$, it also safely approximates 
the concrete behavior  $\stackrel{{\it TCB}}\longmapsto$
(Theorem \ref{th:cppost}).
Theorem \ref{th:cpab} states that abstract beha\-viors are,
roughly speaking, chain-closed with respect to concrete behaviors.
Finally, Theorem \ref{th:SAS} states our main result, i.e.,
every pre-consistent post-fixpoint of the abstract transformation
safely approximates the concrete semantics.

\begin{definition}[Safe Approximation]
\label{def:SA}
\noindent Let $\longmapsto \in{\it CB}$ and
    ${\it sat}\in {\it AB}$.
The abstract behavior {\it sat} {\em safely approximates} the concrete behavior
$\longmapsto$ if and only if, for all
$\langle \theta, p \rangle \in {\it CUD}$ and
    $\langle \beta,  p \rangle \in {\it AUD}$, the following
implication holds:
$$
\begin{array}{lll}
  \left.
  \begin{array}{c}
   \theta\in{\it Cc}(\beta),\\
   \langle \theta, p \rangle\longmapsto S
  \end{array}
   \right\}
   &\Rightarrow&
S\in{\it Cc}({\it sat}\langle \beta ,p \rangle).
\end{array}
$$

 Similarly, let $\longmapsto \in{\it ECB}$ and
    ${\it esat}\in {\it EAB}$.
The extended abstract behavior {\it esat} {\em safely approximates}
$\longmapsto$ if and only if, for all
$\langle \theta, {\it pr} \rangle, 
\langle \theta, c \rangle,
\langle \theta, g, c \rangle
\in {\it ECUD}$ and
    $\langle \beta, {\it pr} \rangle, 
\langle \beta, c \rangle,
\langle \beta, g, c \rangle
 \in {\it EAUD}$, the following
implications hold:
$$
\begin{array}{c}
\begin{array}{lll}
  \left.
  \begin{array}{c}
   \theta\in{\it Cc}(\beta),\\
   \langle \theta,  {\it pr} \rangle\longmapsto S
  \end{array}
   \right\}

&\Rightarrow&
S\in{\it Cc}({\it esat}\langle \beta , {\it pr} \rangle),
\end{array}\\ \\
\begin{array}{lll}
  \left.
  \begin{array}{c}
   \theta\in{\it Cc}(\beta),\\
   \langle \theta, c \rangle\longmapsto \langle S, {\it cf} \rangle
  \end{array}
   \right\}

&\Rightarrow&
\langle S, {\it cf} \rangle\in{\it Cc}({\it esat}\langle \beta ,c \rangle),
\end{array}\\ \\
\begin{array}{lll}
  \left.
  \begin{array}{c}
   \theta\in{\it Cc}(\beta),\\
   \langle \theta, g, c \rangle\longmapsto \langle S, {\it cf} \rangle
  \end{array}
   \right\}

&\Rightarrow&
 \langle S, {\it cf} \rangle \in{\it Cc}({\it esat}\langle \beta , g, c \rangle).
\end{array}\\
\\
\end{array}
$$

\end{definition}

\begin{theorem}[Safety of the Abstract Transformation]
\label{th:SAT}
\noindent 
Let $\longmapsto\in{\it CB}$ and
    ${\it sat}\in {\it AB}$. If {\it sat} safely approximates
$\longmapsto$, then ${\it TAB}({\it sat})$ safely approximates
$\stackrel{{\it TCB}}\longmapsto$.
\end{theorem}

We first establish the following result.
 Remember that if $\longmapsto\in{\it CB}$, its extension
in {\it ECB} is also denoted by $\longmapsto$ (see Section \ref{CSR}).

\begin{lemma}[Safety of $E$]
\label{le:SE}

\noindent
Let $\longmapsto\in{\it CB}$ and
    ${\it sat}\in {\it AB}$. If {\it sat} safely approximates
$\longmapsto$, then $E({\it sat})$ safely approximates $\longmapsto$
(the extension of $\longmapsto$ in {\it ECB}).
\end{lemma}

\begin{proof*}[Proof of Lemma \ref{le:SE}]
\label{pr:SE}

We prove the lemma by structural  induction on the syntax of the
underlying program. It uses the concrete semantic rules of
Figure~\ref{CSRF},  the definition of $E$ in Figure~\ref{fig:TAT},
and  the specifications of the abstract operations given in
Section \ref{sub:AO}. The proof is straightforward due to the close
correspondence of the concrete and the abstract semantics.
We only detail the reasoning for the base case and for the case of 
a goal $(g , l)$ where $l$ is an atom of the form
$p(x_{i_1},\dots,x_{i_n})$.
The other cases are similar.

\vskip0.2cm
\noindent
{\bf Base case}.
 Let $\langle \theta, <>, c \rangle
\!\in\! {\it ECUD}$ and
    $\langle \beta, <>, c \rangle
 \!\in\! {\it EAUD}$. Assume that  $\theta\!\in\!{\it Cc}(\beta)$ and
\mbox{
$\langle \theta, <>, c \rangle\longmapsto \langle S, {\it cf}\rangle$}.
It must be proven that 
$$\langle S, {\it cf}\rangle\in 
{\it Cc}(E({\it sat})\langle\beta,<>,c \rangle). $$
This  relation holds because of the three following facts:
\begin{center}
  \begin{tabular}{rllll}
   $\langle S, {\it cf}\rangle$ & $=$ &
   ${\tt EXTC}(c,\theta)$ 
    & &(by {\bf R2}),\\
   ${\tt EXTC}(c,\theta)$&$\in$&$ {\it Cc}({\tt EXTC}(c,\beta))$
    & &(by specification of {\tt EXTC}),\\

   $E({\it sat})\langle\beta,<>,c \rangle$& $=$ &${\tt EXTC}(c,\beta)$
    & &(by definition of $E$).
  \end{tabular}
\end{center}
\vskip0.2cm
\noindent
{\bf Induction step}.
Let $\langle \theta,  ( g , l ), c \rangle
    \in {\it ECUD}$ and
$\langle \beta,  ( g , l ), c \rangle
    \in {\it EAUD}$,
where $l$ is an atom of the form $p(x_{i_1},\dots,x_{i_n})$. 
Assume  that $\theta\in{\it Cc}(\beta)$ and
\mbox{
$\langle \theta,  ( g , l), 
              c \rangle\longmapsto \langle S, {\it cf}\rangle$}.
It must be proven  that 
$$\langle S, {\it cf}\rangle\in 
{\it Cc}(C), \mbox{
where }
 C = E({\it sat})\langle\beta, ( g , l),c \rangle.$$
By Rule {\bf R6},
there exist program substitutions and program sequences such that
$$
\begin{array}{llll}
  \langle \theta, g,c \rangle
   \longmapsto
   \langle S', {\it cf} \rangle      &                           && (C1)\\
S' = <\theta_1,\dots,\theta_i,\_>    &                           && (C2) \\
\theta'_k = {\tt RESTRG}(l,\theta_k) & (1\leq k\leq {\it Ns}(S)) && (C3) \\ 
\langle  \theta'_k ,p\rangle
    \longmapsto  S_{k}^{'}           & (1\leq k\leq {\it Ns}(S)) && (C4)\\ 
S_k = {\tt EXTG}(l,\theta_k,S_{k}^{' })
                                     & (1\leq k\leq {\it Ns}(S)) && (C5)\\ 
S={\Box}^{{\it Ne}(S)}_{k=1} S_k     &                           && (C6)\\
\end{array}
$$
Moreover, by  definition of
$E({\it sat})$, there exist abstract values such that
$$
\begin{array}{rclll}
C        & = &{\tt EXTGS}(l,C',B)                               && (A1)\\
B        & = &{\it sat}\langle \beta',p \rangle                 && (A2)\\
\beta'   & = &{\tt RESTRG}(l,\beta'')                           && (A3)\\
\beta''  & = &{\tt SUBST}(C')                                   && (A4)\\
C'       & = &E({\it sat})\langle \beta,g ,c \rangle            && (A5)
\end{array}
$$
The following assertions hold. 
By $A5$, $C1$, and the induction hypothesis, 
$$
\begin{array}{rclll}
 \langle S', {\it cf} \rangle 
        & \in & {\it Cc}(C') && (B1).
\end{array}
$$
By $A4$, $B1$, $C2$, and the specification of {\tt SUBST},
$$
\begin{array}{rcllll}
\theta_k & \in &  {\it Cc}(\beta'') & (1\leq k\leq {\it Ns}(S))  
                                                                && (B2).
\end{array}
$$
By $A3$, $B2$, $C3$, and the specification of {\tt RESTRG},
$$
\begin{array}{rcllll}
\theta'_k & \in &  {\it Cc}(\beta') & (1\leq k\leq {\it Ns}(S))  
                                                                && (B3).
\end{array}
$$
By $A2$, $B3$, $C4$, and the hypothesis that {\it sat} safely approximates
$\longmapsto$,
$$
\begin{array}{rcllll}
S'_{k}& \in &  {\it Cc}(B) & (1\leq k\leq {\it Ns}(S))  
                                                                && (B4).
\end{array}
$$
Finally, by $A1$, $B1$, $B4$,  $C2$, $C5$, $C6$, and the specification
of {\tt EXTGS},
$$
\begin{array}{rcllll}
\langle S, {\it cf} \rangle & \in &  {\it Cc}(C). 
\end{array}
\mathproofbox $$
\end{proof*}
\begin{proof}[Proof of Theorem \ref{th:SAT}] 
\label{pr:SAT}

The result follows from the definition of {\it TAB} in 
Figure~\ref{fig:TAT}, 
 the definition of {\it TCB} in Section \ref{CT}, and
 Lemma \ref{le:SE}.
\end{proof}

%
%
%
%
%

The next theorem states that the transformation {\it TAB}
maintains pre-consistency.

\begin{theorem}\label{th:p-c}
 Let ${\it sat}\in {\it AB}$. If {\it sat} is
pre-consistent, then ${\it TAB}({\it sat})$
is also pre-consistent.
\end{theorem}
\begin{proof}
Let $\longmapsto$ be the concrete semantics of the underlying program.
 Since {\it sat} is
pre-consistent, there exists a concrete behavior $\longmapsto'$ such
that
\begin{enumerate}

\itemsep 2pt

\item $\longmapsto'\ \sqsubseteq\ \longmapsto$, and

\item {\it sat} safely approximates $\longmapsto'$.
\end{enumerate}
The first condition implies that 
$$\stackrel{{\it TCB}}{\longmapsto'}\ \sqsubseteq\ \longmapsto,$$
since {\it TCB} is monotonic and 
$\stackrel{{\it TCB}}{\longmapsto}\ =\ \longmapsto$.
\\The second condition and Theorem \ref{th:SAT} imply that
\begin{center}
${\it TAB}({\it sat})$\  safely approximates\  
                              $\stackrel{{\it TCB}}{\longmapsto'}.$

\end{center}
The result follows from the two implied statements and Definition 
\ref{def:PCAB}.
\end{proof}

 The next two theorems state closure properties of
abstract behaviors, which are used to prove the 
safety of the abstract semantics.

\begin{theorem}\label{th:cppost}

Let {\it sat} be a post-fixpoint of {\it TAB}.
Let $\longmapsto \in{\it CB}$. If {\it sat} safely approxi\-mates
$\longmapsto$, then {\it sat} also safely approximates
$\stackrel{{\it TCB}}\longmapsto$.
\end{theorem}
\begin{proof*}
Assume that {\it sat} safely approximate
$\longmapsto$.
Let
$\langle \theta, p \rangle \in {\it CUD}$ and
    $\langle \beta,  p \rangle \in{\it AUD}$.
It must be proven that$$
\begin{array}{lll}
  \left.
  \begin{array}{c}
   \theta\in{\it Cc}(\beta),\\
   \langle \theta, p \rangle \stackrel{{\it TCB}}\longmapsto S
  \end{array}
   \right\}
   &\Rightarrow&
S\in{\it Cc}({\it sat}\langle \beta ,p \rangle).\end{array}
$$
Assume that the left part of the implication holds.
Theorem \ref{th:SAT} implies that
$$S\in{\it Cc}({\it TAB}({\it sat})\langle \beta ,p \rangle).$$
Since {\it sat} is a post-fixpoint and {\it Cc} is monotonic,
$$\begin{array}{rcl}
{\it Cc}({\it TAB}({\it sat})\langle \beta ,p \rangle)
\subseteq 
{\it Cc}({\it sat}\langle \beta ,p \rangle),
\end{array}$$
and then
$$S\in{\it Cc}({\it sat}\langle \beta ,p \rangle).
\mathproofbox $$
\end{proof*}

\begin{theorem}\label{th:cpab}
Let ${(\longmapsto_i)}_{i\in{\bf N}}$ be a chain of concrete behaviors.
Let ${\it sat}\in {\it AB}$. If {\it sat} safely appro\-xi\-mates
$\longmapsto_i$, for all $i\in{\bf N}$, then 
{\it sat} safely approximates
$(\sqcup_{i=0}^{\infty}\longmapsto_i).$
\end{theorem}
\begin{proof*}
Let us abbreviate $(\sqcup_{i=0}^{\infty}\longmapsto_i)$ by
$\longmapsto$. 
It is sufficient to prove that, for any
$\langle \beta, p \rangle \in {\it AUD}$ and 
any $\langle \theta, p \rangle \in {\it CUD}$,
$$
\begin{array}{lll}
  \left.
  \begin{array}{c}
   \theta\in{\it Cc}(\beta),\\
   \langle \theta, p \rangle\longmapsto S
  \end{array}
   \right\}
   &\Rightarrow&
S\in{\it Cc}({\it sat}\langle \beta ,p \rangle).
\end{array}
$$
Fix  $\langle \beta, p \rangle $,
$\langle \theta, p \rangle $, and $S$ satisfying the left part of the
implication. By Theorem \ref{CB-cpo},
$$
\begin{array}{llllll}
S = \sqcup_{i=0}^{\infty} S_i &&
\mbox{\rm where}              &&
\langle \theta, p \rangle\longmapsto_i S_i
&
\forall i\in{\bf N}.
 \end{array}
$$
Since {\it sat} safely approximates every $\longmapsto_i$,
\begin{center}
$S_i\in{\it Cc}({\it sat}\langle \beta ,p \rangle)$ for all 
$i\in{\bf N}$.
\end{center}
 Finally, 
since ${\it Cc}({\it sat}\langle \beta ,p \rangle)$ is chained-closed,
$$S\in{\it Cc}({\it sat}\langle \beta ,p \rangle).
\mathproofbox $$
\end{proof*}

The last theorem states our main result. 

\begin{theorem}[Safety of the Abstract Semantics]
\label{th:SAS}
\noindent 
Let {\it sat} be a pre-consistent post-fixpoint of {\it  TAB}.
Then {\it sat} safely approximates $\longmapsto$ where $\longmapsto$
is the concrete semantics of the underlying program.
\end{theorem}

We first establish the following statement.

\begin{lemma}\label{le:SAS}
Let {\it sat} be a pre-consistent post-fixpoint of {\it  TAB}.
There exists a chain of concrete behaviors
${(\longmapsto_i)}_{i\in{\bf N}}$ such that {\it sat} safely approximates
$\longmapsto_i$, for all $i\in{\bf N}$ and
$(\sqcup_{i=0}^{\infty}\longmapsto_i)\ =\ \longmapsto$ where $\longmapsto$
is the concrete semantics of the underlying program.
\end{lemma}
\begin{proof*}[Proof of Lemma \ref{le:SAS}]
The proof is in three steps. First we construct a sequence
${\{\longmapsto'_i\}}_{i\in{\bf N}}$ of lower-approximations of
$\longmapsto$ which is not necessarily a chain; then we
modify it to get a chain ${(\longmapsto_i)}_{i\in{\bf N}}$;
finally, we show that $(\sqcup_{i=0}^{\infty}\longmapsto_i)\ =\
\longmapsto$. The proof uses the following property of
program substitution sequences, whose proof is left to the reader.
If $S_1$, $S_2$ and $S$ are program substitution sequences such that
$S_1\sqsubseteq S$ and $S_2\sqsubseteq S$, then $S_1$ and $S_2$ have
a least upper-bound, which is either $S_1$ or $S_2$. The least
upper-bound is
denoted by $S_1\sqcup S_2$ in the proof.
\begin{enumerate}
\itemsep 2pt
\item Since {\it sat} is pre-consistent, there exists a concrete
      behavior $\longmapsto'$ such that 
      {\it sat} safely approximate $\longmapsto'$ and
      $\longmapsto'\ \sqsubseteq\ \longmapsto$. The sequence
      ${\{\longmapsto'_i\}}_{i\in{\bf N}}$ is defined by
      $$\begin{array}{llll}
        \longmapsto'_0\ =\ \longmapsto' & \mbox{ and } &
        \longmapsto'_{i+1}\ =\ \stackrel{{\it TCB}}{\longmapsto'_i}
                      & ({i\in{\bf N}}). 
      \end{array}$$
      Since $\longmapsto'\ \sqsubseteq\ \longmapsto$,
      {\it TCB} is monotonic and $\longmapsto$ is a fixpoint of {\it TCB},
      it follows that
      $$\begin{array}{ll}
        \longmapsto'_{i}\ \sqsubseteq\ \longmapsto
                      & (\forall{i\in{\bf N}}). 
      \end{array}$$
      Moreover, by Theorem \ref{th:cppost}, {\it sat}
      safely approximates every $\longmapsto'_{i}$.
\item ${(\longmapsto_i)}_{i\in{\bf N}}$ is now constructed by
      induction over $i$. The correctness of the construction process
      requires to prove that, after each induction step, the relation 
      $\longmapsto_{i}\ \sqsubseteq\ \longmapsto$ holds. We first define
      $$\begin{array}{ll}
        \longmapsto_0\ =\ \longmapsto'_0. 
      \end{array}$$
      Let $i\in {\bf N}$. Assume, by induction, that 
      $\longmapsto_0\ \sqsubseteq\ 
                      \ldots\ \sqsubseteq\ \longmapsto_i
                       \ \sqsubseteq\ \longmapsto$.
      For every $\langle \theta, p\rangle \in {\it CUD}$, we define
      $$\begin{array}{lll}
        \langle \theta, p\rangle
        \longmapsto_{i+1} (S_1\sqcup S_2) & \mbox{\rm where } &
        \left\{
        \begin{array}{l}
        \langle \theta, p\rangle
        \longmapsto_{i} S_1,\\
        \langle \theta, p\rangle
        \longmapsto'_{i+1} S_2 .\end{array}\right.
      \end{array}$$
       Since
       $\longmapsto'_{i+1}\ \sqsubseteq\ \longmapsto$ and
       $\longmapsto_i \ \sqsubseteq\ \longmapsto$, we have that
       $\longmapsto_{i+1}$ is well-defined and
       $\longmapsto_{i+1}\ \sqsubseteq\ \longmapsto$.
Moreover, since {\it sat} safely approximates
       $\longmapsto_i$ (by induction)
and $\longmapsto'_{i+1}$, and  $S_1\sqcup S_2$ is
       equal either to $S_1$ or $S_2$, in the definition of 
       $\longmapsto_{i+1}$, we have that {\it sat} safely approximates
       every $\longmapsto_{i+1}$.
\item  The Kleene sequence of the concrete semantics is a chain
       ${(\longmapsto''_i)}_{i\in{\bf N}}$ defined as follows:
       $$\begin{array}{llll}
        \longmapsto''_0\ =\ \longmapsto_\bot & \mbox{ and } &
        \longmapsto''_{i+1}\ =\ \stackrel{{\it TCB}}{\longmapsto''_i}
                      & ({i\in{\bf N}}). 
      \end{array}$$
       Since $\longmapsto_\bot\ \sqsubseteq\ \longmapsto' $ and
       {\it TCB} is monotonic, 
 it follows, by induction, that
        $$ \begin{array}{ll}
        \longmapsto''_i\ \sqsubseteq\ \longmapsto'_i
                       \ \sqsubseteq\ \longmapsto_i
                       \ \sqsubseteq\ \longmapsto 
                      & (\forall{i\in{\bf N}}). 
      \end{array}$$
      Therefore, by definition of the least upper bound and since
      the least fixpoint is the limit of the Kleene sequence,
      $$ \begin{array}{l}
        \longmapsto\ =\ (\sqcup_{i=0}^\infty \longmapsto''_i)
                       \ \sqsubseteq( \sqcup_{i=0}^\infty \longmapsto_i)
                       \ \sqsubseteq\ \longmapsto. 
      \end{array}$$
      Thus,
       $$ \begin{array}{l}
        \longmapsto\ =\ 
                       (\sqcup_{i=0}^\infty \longmapsto_i).
      \end{array} \mathproofbox$$
\end{enumerate}
\end{proof*}

\begin{proof}[Proof of Theorem \ref{th:SAS}]

\noindent The result is an immediate consequence of Theorem \ref{th:cpab}
and Lemma \ref{le:SAS}
\end{proof}

\subsection{Related Works}
\label{sub:D}
In this section we first discuss
the mathematical approach underlying our abstract semantics and relate it with
the higher-order abstract interpretation frameworks
advocated by  Cousot and  Cousot  \shortcite{CousotICCL94}. 
Then, we compare our approach with
the abstract  semantics  for Prolog with control proposed by  Barbuti 
{\em et al.}  \shortcite{BarbutiControl},
by  Fil\`e and Rossi  \shortcite{FileRossi}, 
and by Spoto  \shortcite{Spoto}.
\\

\noindent
{\bf Cousot and Cousot's Higher-order Abstract Interpretation Frameworks.}
%
%
%
As mentioned in the introduction,
the traditional approach to abstract interpretation can not be applied
to approximate  the concrete semantics
of Section~\ref{CS}.
Indeed, we can define a set-based collecting transformation
by lifting the concrete semantics to
sets of program substitution sequences.
However, the
 least fixpoint of the collecting transformation
does not safely approximate the concrete semantics.
The problem can be solved by restricting to sets of
$\wp({\it CPS}_D)$ and $\wp({\it CPSS}_D)$ 
 that enjoy  some closure properties
ensuring safeness of  the least fixpoint.
 This solution is simi\-lar to the choice of
a power-domain structure in denotational semantics
\cite{Schmidt88,Stoy77}: the needed constructions can  in fact be
viewed as power-domains.
However
 there is no best way
to choose the closure pro\-per\-ties.
Different closure properties are 
adequate for different sorts of information.
It is therefore advocated by  Cousot and Cousot
 in \shortcite{CousotICCL94} that, for higher-order
languages, different collec\-ting semantics should be defined
for the same language 
depending on the kind of properties to be inferred.
In our case, at least two dual collecting semantics could be defined. Both of
them use sets of program  substitution sequences that are
chain-closed.
\begin{enumerate}
\itemsep 0pt
\item The first semantics considers {\em downwards-closed} sets
 of program substitution
  sequences, i.e., such that
  for any $S,S'\in{\it CPSS}_D$,
  $$
  \begin{array}{lll}
  \left.
  \begin{array}{c}
   S\in\Sigma,\\
   S' \sqsubseteq S
  \end{array}
   \right\}
  &\Rightarrow
  & S'\in \Sigma.
   \end{array}
   $$
   This domain is ordered by inclusion and its minimum is $\{<\bot>\}$.
   It is adequate to infer non-termination and upper bounds to the
   length of sequences. In particular, it is adequate for determinacy
   analysis. However,
   it is unable to infer  termination since $<\bot>$
   belongs to any set of sequences.
\item   The second semantics considers {\em upwards-closed} sets
of program substitution sequences, i.e.,
   such that
  for any $S,S'\in{\it CPSS}_D$,
  $$
  \begin{array}{lll}
  \left.
  \begin{array}{c}
   S\in\Sigma, \\
   S\sqsubseteq S'
  \end{array}
   \right\}
  &\Rightarrow
  & S'\in \Sigma.
   \end{array}
   $$
   This domain is ordered by 
   $\Sigma\leq\Sigma'\ \Leftrightarrow\ \Sigma'\subseteq\Sigma$ and 
   its minimum is ${\it CPSS}_D$.
   It is able to infer termination and lower bounds to the
   length of sequences. It is less adequate 
   than the previous one to infer precise information about the
   substitutions in the sequences because its least fixpoint
    corresponds to  a greatest fixpoint in a traditional framework
   ignoring the sequence structure.
\end{enumerate}

\noindent   In both cases, the least fixpoint is well-defined because
   the collecting versions of the operations are monotonic, since
   they have to ensure the closure properties. Moreover, the least fixpoint
   of the collecting semantics safely approximates the concrete
   semantics because all iterates are pre-consistent and the sets are
   chain-closed.
   Nevertheless, our formalization has some advantages.
   \begin{enumerate}
\itemsep 0pt
   \item It can be more efficient:  a single analysis is able to infer 
        all the information that can be inferred by the two 
        collecting semantics. 
    \item It can be more accurate: there are pre-consistent post-fixpoints
        that are more precise than the intersection of the two
        collecting semantics. 
   \end{enumerate}

\vskip0.3cm
\noindent
{\bf Barbuti et al.'s Abstract Semantics}.
The abstract semantics proposed by
 Barbuti {\em et al.}   \shortcite{BarbutiControl} aims at modeling 
 control aspects of logic programs
 such as search strate\-gy and selection rule.
Their semantics is  parametric with respect to
a ``termination theory''. The meaning of a program is  obtained by
composing the meaning of its  ``logic component'' together with a 
corresponding ``termination theory'' (the ``control component'').
The latter can be provided either by applying
techniques of abstract interpretation or by applying proof procedures.
In all cases, control information is deduced from outside
in the form of a separated termination analy\-sis.
This is the main difference with
our framework, where control information, i.e. information relative to
termination or non-termination, 
is   modeled  
within the semantic domains through the notion of substitution sequence.
\\

\noindent
{\bf Fil\`e and Rossi's Abstract Interpretation Framework}.
The  framework proposed by 
 Fil\`e and  Rossi   \shortcite{FileRossi} 
consists of a tabled interpreter
which  explores OLDT abstract trees decorated with
 control information about
sure success or failure of the goals.
Such information is used by the cut operation to prune
the OLDT-tree whenever a cut is reached.
Sure success is modeled in our framework by abstract sequences 
representing
 only non-empty sequences.
The abstract semantics defined by Fil\`e and Rossi is 
 operational
and non-compositional while ours is compositional and based on the
fixpoint approach. Moreover, 
 the abstract execution of a goal  $(g,!)$
is  different.
Whenever is known that $g$ surely succeeds, their framework stops after
generating the first "sure" solution, while ours computes the entire
abstract sequence for $g$  and then cuts it to maintain at most
one solution. Our approach may thus imply some redundant work.
However, if $g$  is used in several contexts, their framework
should recognize this situation and expand the  OLDT-tree
further.\\

\noindent
{\bf Spoto's Denotational Abstract Semantics}.
The  related work closest to ours is
the denotational abstract semantics proposed by 
 Spoto \shortcite{Spoto}.
He defines a 
 goal-independent and compositional abstract 
semantics of Prolog modeling the depth-first search rule and 
the cut. His semantics associates to any Prolog program
a sequence of pairs consisting of a ``kernel''
constraint
and its ``observability'' part. Intuitively,  kernel constraints
denote  computed answers, while  observability constraints give
 information about   
  divergent computations and cut executions.
The main difference with our approach is that his semantics is 
goal-independent
while ours is not. 
This is due to the fact that our abstract 
semantics is {\em functional}, i.e., it 
associates to each program P a function (an abstract behavior)
mapping every pair $\langle \beta,p\rangle$ to an abstract sequence $B$.
However, this choice is unrelated to our  concrete semantics: we could
as well abstract the concrete semantics  by a {\em relational}
abstract semantics \cite{CousotJLC92},
 making it possible to express dependencies between
input substitutions and the corresponding output substitution sequences. 
This is the approach of \cite{RP-98-002}
where we express dependencies between the size of
input terms and the number of corresponding output substitutions.
We will go back  to this issue at 
the end of Section \ref{sub:GRADC}.



\section{Generic abstract interpretation algorithm}
\label{sec:GAIA}

A generic abstract interpretation algorithm 
is an algorithm that is
parametric with respect to the  abstract domains. It can
be instantiated by various domains 
to obtain different data-flow analyses.
Several such algorithms have been proposed for Prolog
\cite{Bruynooghe91,SPE,ICLP91AI,WSA93,TOPLAS,ACTA95,Mellish87,Muthukumar92},
but they do not handle the control features of the language such that
 Prolog search rule and cut.

The algorithm presented here is essentially an instantiation of the
universal fixpoint algorithm described in \cite{universal}
to the abstract semantics of Section \ref{sec:AS}.
In particular, 
it is  quite similar to the algorithm presented  in \cite{ICLP91AI,TOPLAS}: 
in fact,
 the abstract semantics  of Section
\ref{sec:AS}
can be
viewed as a proper generalization of the abstract semantics
described in those papers,
where the sequences 
of computed answer substitutions are no longer abstracted to sets of
substitutions.\\
The universal algorithm in \cite{universal} is top-down, i.e.,
it computes a subset of the fixpoint (in the form of a set of tuples)
contai\-ning the output value corresponding to a
distinguished input together with
 all the tuples needed
to compute it.
Top-down algorithms are naturally used to perform data-flow analyses,
where one is interested in collecting the abstract information
corresponding to a class of initial queries described by the
distinguished input. It is more efficient in general to
compute a part of the fixpoint only and this allows one to use
infinite abstract domains, which are more expressive
\cite{Cousot92c}. 
Although the instantiation of \cite{universal} to our
abstract semantics is as mechanical as
in our previous works (a slightly more general
widening operator is needed however),
the correctness of the algorithm  involves some new theoretical 
issues:  the pre-consistency of the post-fixpoint has now to be proven.
Nevertheless, since the novel algorithm is in practice very similar to
the algorithm presented in \cite{TOPLAS}, we only discuss here the extended
widening operator which  ensures a good compromise between
efficiency and accuracy. A detailed description of the algorithm and
its correctness proof can be found
in \cite{sequence}.

\subsection{Extended Widening}
\label{sub:EW}
The extended widening operation used by the novel algorithm is defined as follows.

\begin{definition}[Extended Widening]
\label{def:EW}
\noindent
An {\em extended widening} 
on abstract sequences is a (polymorphic%
\footnote{It is parametrized over $D$.}) 
operation 
$\nabla:{\it ASS}_D \times{\it ASS}_D \rightarrow {\it ASS}_D$ that
enjoys the following properties.
Let $\{B_i\}_{i\in{\bf N}}$ be a sequence of elements of ${\it ASS}_D$.
Consider the sequence $\{B'_i\}_{i\in{\bf N}}$ defined by
\vskip-0.1cm
$$
\begin{array}{llll}
B'_0 & = & B_0,                               \\
{B'}_{i+1} & = & B_{i+1}\nabla{B'}_{i} & (i\in{\bf N}).
\end{array}
$$
The following conditions hold:
\begin{enumerate}
\itemsep 1pt
\item $B'_i \geq B_i$ $(i\in{\bf N})$;
\item the sequence $\{B'_i\}_{i\in{\bf N}}$ is stationary, i.e., 
there exists  $j\ge 0$ such that ${B'}_i={B'}_j$
for all $i$ such that $j\leq i$.
\end{enumerate}
\end{definition}

An extended widening is slightly more general than a widening
\cite{Cousot92c} because the sequence $\{B'_i\}_{i\in{\bf N}}$
is {\em not} required to be a chain.

Let us now explain how the extended widening is used
by the algorithm. 
Given an input
pair $\langle\beta,p \rangle$, the 
algorithm iterates on the computation of
${\it TAB}({\it sat})\langle\beta,p \rangle$ until
convergence, and concurrently updates {\it sat}, as follows
(recursive calls -- which also modify {\it sat} -- are ignored
in the discussion):
\begin{enumerate}
\itemsep 0pt
\item $B'_0$ $=$ $B_{\bot}$ is stored in the initial ${\it sat}$ as
the output for $\langle\beta,p\rangle$;

\item $B_i$ results from the $i$-$th$  execution of 
${\it TAB}({\it sat})\langle\beta,p \rangle$;

\item ${B'}_{i}=B_{i}\nabla{B'}_{i-1}$ is stored in the current ${\it
sat}$ after the $i$-$th$  execution of 
${\it TAB}({\it sat})\langle\beta,p \rangle$;

\item the loop is exited when $B_{i+1}\leq B'_i$.
\end{enumerate}

\noindent
The loop terminates because 
there must be some $i$ such that $B'_{i+1}=B'_i$
(otherwise Condition 2 of Definition \ref{def:EW} would
be violated), and, hence, $B_{i+1}\leq B'_i$ since
$B'_{i+1}\geq B_{i+1}$ by Condition 1.
The loop can be resumed later on
because some values in {\it sat} have been updated
(Step 1 is omitted in these subsequent executions);
all re-executions of the loop terminate for
the same reasons as the first one; moreover, 
the loop can only be resumed finitely many times because
no element in {\it sat} can be improved infinitely many often,
since
there is a $j$ such that $B'_i=B'_j$  
for all $i$ greater or equal to $j$. 
Note that a local post-fixpoint
is attained  each time the loop is exited.
Thus a global post-fixpoint is obtained when all loops
are terminated for all values in {\it sat}.
The formal characterization of Definition \ref{def:EW} elegantly
captures the idea that the algorithm 
sticks as closely as possible to the abstract semantics
during the first iterations, and starts lumping
 the results together only when enough accuracy is obtained,
in order to ensure convergence. The advantage
of this characterization is that no particular value of $j$ is
fixed. So we can think of ``intelligent'' 
extended widenings that observe how the successive iterates
behave and that enforce convergence exactly at the right time.
The extended widening  used
in our experimental evaluation is based on 
this intuitive idea (see Section \ref{sub:GAO}).

\section{Cardinality analysis}
\label{sec:FASAS}

The abstract interpretation framework for Prolog presented in previous sections
has been instantiated by a domain of
 abstract  sequences
to perform so-called cardinality analysis; see \cite{cardinality}.
 Cardinality analysis
approximates the number of solutions to a goal and is useful for  many
purposes such as indexing, cut insertion and elimination
\cite{D89,Sahlin91}, dead code elimination, and memory management
and scheduling in parallel systems \cite{Bueno91b,Hermenegildo86}.
The analysis  subsumes traditional determinacy analysis
 such as those of \cite{Ramakrishnan93,D89,Giacobazzi92,Sahlin91}.

This section is organized as follows.
First we describe how  a generic abstract domain 
for cardinality analysis, which is parametric with respect to
 {\em any} domain of
abstract substitutions, can be built. Then,
we instantiate this generic domain to the domain of abstract substitutions 
{\tt Pattern}
\cite{TOPLAS}.
Finally, we discuss  experimental evaluations of the
analysis from both accuracy and efficiency standpoints.

\subsection{Generic Abstract Domains for Cardinality Analysis}
\label{sub:GADCA}

In this section,  generic  domains of  abstract sequences 
and abstract sequences with cut information
are  built.
The domains are generic with respect to the information on 
the substitutions in the sequences, but they provide
specific information about the sequence structure.
The latter  consists of 
lower and upper bounds to the number of substitutions in the
sequences
and  information about the nature
(i.e.,  finite,
incomplete or infinite) of the sequences.
This information  allows us to
perform non-termination analysis
 and a limited form of termination analysis. 
 Predi\-cate level analyses, like 
 determinacy 
and functionality
\cite{Debray89a}, which  were previously 
considered falling outside the scope of
abstract interpretation, can be performed.
\\

\noindent
{\bf Abstract Substitutions.}
The substitution part of our generic  domain of abstract sequences
is assumed to be an element of an arbitrary domain of abstract substitutions 
${\it AS}_D$. The only requirement on ${\it AS}_D$ is that it  contains a minimum element
$\beta_\emptyset$ such that ${\it Cc}(\beta_\emptyset)=\emptyset$. 
An abstract domain can always be
enhanced with such an element.\\

\noindent
{\bf Abstract  Sequences.}
The generic domain of abstract  sequences manipulates
termination information whose domain is defined below.

\begin{definition}[Termination Information]
\label{def:TI}
\noindent
A termination information $t$ is an element of the set 
 ${\it TI}=\{ {\it st}, \;{\it snt}, \;{\it pt}\}$
 endowed with the ordering $\leq$ defined by
$$
\begin{array}{llll}
t_1\leq t_2 &  \Leftrightarrow & \mbox{\rm either } t_1 = t_2\ \mbox{\rm or}\ t_2 = {\it pt}
             & \;\;\;\forall t_1,t_2\in{\it TI}.
\end{array}
$$
The symbol {\it st} stands for
``sure termination'' and it characterizes finite sequences;
{\it snt} stands for
``sure non termination'' and characterizes incomplete and infinite
sequences; {\it pt} stands for ``possible
termination'' and corresponds to  absence of information.
\end{definition}

\noindent

The domain of abstract substitution sequences is  defined as follows.

\begin{definition}[Abstract Sequences]
\label{def:ASS_D}
\noindent
Let $D$ be a finite set of program variables. We denote by
${\it ASS}_D$ the set of all 4-tuples 
 $\langle \beta, m, M, t \rangle$ such that  
       $\beta \in {\it AS}_D$,
       $m \in {\bf N}$, 
       $M \in {\bf N}\cup \{\infty\}$, 
       and $t \in {\it  TI}$.
\end{definition}

  Informally, $\beta$ describes all substitutions in the
sequences,
$m$ and $M$ are
lower and upper bounds on the number of substitutions in the
sequences,  and
$t$ is an  information on termination.  

The ordering on abstract sequences is defined as follows.

\begin{definition}[Ordering on Abstract Sequences]
\label{def:OAS}
\noindent
Let $B_1,B_2 \in {\it ASS}_D$.
$$
\begin{array}{llll}
B_1 \leq B_2 &\mbox{\rm iff } & \beta_1 \leq \beta_2 \mbox{\rm \ and\ } 
                                 m_1 \geq m_2 \mbox{\rm \ and\ }
                                 M_1 \leq M_2 \mbox{\rm \ and\ } 
                                 t_1 \leq t_2. 
\end{array}
$$
\end{definition}

The set of program substitution sequences described by an abstract sequence 
$B$ is formally defined as follows.

\begin{definition}[Concretization for Abstract Sequences]
\label{def:CASS_D}

\noindent
Let $B\!\!=\!\!\langle \beta, m,M, t \rangle\!\in\! {\it ASS}_D$. 
We define
\begin{center}
\begin{tabbing}
123\=456\=789\=123456\=12345\kill
${\it Cc}(B)={\it Sseq}_1(\beta)\cap{\it
Sseq}_2(m,M) \cap{\it Sseq}_3(t)$ \\
 { where} \\
\>\> \> 
${\it Sseq}_1(\beta) = \{S: S\in {\it PSS}_D\ \mbox{\rm and}\ 
                     {\it Subst}(S)\subseteq{\it Cc}(\beta)\},$\\
        \>\> \>
${\it Sseq}_2(m,M) = \{S: S\in {\it PSS}\ \mbox{\rm and}\ 
                    m \leq {\it Ns}(S) \leq M\},$\\
        \>\> \>
${\it Sseq}_3({\it snt}) = \{S: S\in {\it PSS}\ \mbox{\rm and}\ 
                    S$ is incomplete or infinite\},\\
        \>\> \>
${\it Sseq}_3({\it st}) = \{S: S\in {\it PSS}\ \mbox{\rm and}\ 
                    S$ is finite\},\\
        \>\>\> 
${\it Sseq}_3({\it pt}) = {\it PSS}.$ \end{tabbing} 
\end{center}
\end{definition}

Monotonicity of the concretization function is a simple
consequence of the definition.

We denote by $B_\bot$  the special abstract sequence
$\langle\beta_\emptyset,0,0,{\it snt}\rangle$ which is such that
 ${\it Cc}(B_\bot)=\{<\bot>\}$ as required in Section \ref{sub:AD}.
It is easy to prove that for all abstract sequences 
$B\in{\it ASS}_D$, the set ${\it Cc}(B)$ is chain-closed; see \cite{sequence}.

\vskip0.3cm

\noindent
{\bf Abstract Sequences with Cut Information.}
Abstract sequences with cut information are obtained by enhancing
abstract sequences  with  information about 
execution of cuts.

Let us first define the abstract domain for cut information.

\begin{definition}[Abstract Cut Information]
\label{def:ACI}
\noindent
An abstract cut information {\it acf} is an element of the set
${\it ACF}=\{{\it cut},{\it  nocut}, {\it weakcut}\}$.
\end{definition}

\begin{definition}[Abstract Sequences with Cut Information]
\label{def:ASSC}
\noindent
Let $D$ be a finite set of program variables.
We denote by $ {\it ASSC}_D $  the set of pairs 
$\langle B, {\it acf} \rangle$ where 
$B\in{\it ASS}_D $ and ${\it acf} \in {\it ACF}$.
\end{definition}

Informally, {\it cut} indicates that a cut has been executed in all
sequences, {\it nocut} that no cut has been executed in any sequence,
and {\it weakcut} that a cut has been executed for all sequences
producing at least one solution. More formally, the concretization
of an abstract sequence with cut information is defined as follows.

\begin{definition}[Concretization for Abstract Sequences
  with Cut Information]
\label{def:CASCI}
\noindent
Let $ B\in{\it ASS}_D$. We define\\
\\
$\begin{array}{llllll}
 {\it  Cc}(\langle B,{\it cut}\rangle) & =
& \{ \langle S, cut\rangle : S\in {\it Cc}(B)\}, \\
 {\it Cc}(\langle B,{\it nocut}\rangle) &=&
\{  \langle S, nocut \rangle: S\in {\it Cc}(B)\}, \\
 {\it Cc}(\langle B,{\it weakcut}\rangle) &=&
\{  \langle S, cut \rangle : S\in {\it Cc}(B)\}
                \cup  \\
& &           \{  \langle S, nocut \rangle :\  
                          S\in {\it Cc}(B)
                     \mbox{\rm\ and\ }S\in  \{<>,<\bot>\}\}.
\end{array} $ 
\end{definition}


\subsection{Abstract Operations}
\label{sub:GAO}
Our next task is to provide definitions of all abstract operations
specified in Section~\ref{sub:AO}.
For space reasons, we describe here a subset of the
operations, i.e.,  extended widening,
 unification,   operation  treating cut,
and  concatenation.  
The other operations are described in the appendix.
The reader is referred to \cite{sequence} for the correctness proofs.

The operations on abstract substitutions which are used in the
definition of the
operations on abstract sequences 
 will be recalled when needed.

\vskip0.3cm

\noindent
{\bf Extended Widening:} $\nabla: {\it ASS}_D\times {\it ASS}_D\rightarrow {\it ASS}_D$\\
We require that the abstract domain ${\it AS}_D$ is equipped with a 
widening operation
$\nabla': {\it AS}_D\times {\it AS}_D\rightarrow {\it AS}_D$.
It can be an extended widening, a normal widening, or,
if ${\it AS}_D$ is finite or
enjoys the finite ascending chain property,
any upper bound operation.
 The widening
on sequences is obtained by taking the least upper bound of
the termination components, the minimum of the lower bounds
and setting the upper bound to infinity. \\ 
Assume that $B_{\it old} = \langle \beta_{\it old}, m_{\it old},
M_{\it old}, t_{\it old} \rangle$ and $B_{\it new} = \langle \beta_{\it new},
 m_{\it new},
M_{\it new}, t_{\it new} \rangle$.\\
The
operation 
 $\nabla: {\it ASS}_D\times {\it ASS}_D\rightarrow {\it ASS}_D$
 is defined
as follows.

\begin{small}
$$
\begin{array}{lllll}
B_{\it new}\nabla B_{\it old} & = & \langle \beta_{\it new}\nabla'\beta_{\it old}, m_{\it new}, M_{\it new}, t_{\it new}\rangle
  &\myif &\! \beta_{\it new}\not\leq\beta_{\it old}
        \\
           & = & \langle \beta_{\it old}, m_{\it new}, M_{\it new}, {\it pt}\rangle
  &\myif &\! \beta_{\it new}\leq\beta_{\it old}
                   \and t_{\it new}\not\leq   t_{\it old}    \\
           & = & \langle \beta_{\it old}, 
                         \min(m_{\it new},m_{\it old}),
                                     \infty, t_{\it old}\rangle
  &\myif &\! \beta_{\it new}\leq\beta_{\it old}
                   \and t_{\it new}\leq   t_{\it old}  \and \\
& & & &
                  
                        (m_{\it new} \!< \!m_{\it old}
                                \myor
                         M_{\it new} \!> \!M_{\it old})    \\
           & = & B_{\it old}
  &\myif &\! B_{\it new}\leq   B_{\it old}.     \\
\end{array}
$$
\end{small}

The first case makes sure that the algorithm iterates until the
abstract substitution  part stabilizes. When it is stable,
the widening is applied on sequences.
\\

\noindent
{\it Example}.
Consider the following program:\\
\\
\(
\begin{array}{lll}
{\tt repeat}.\\
{\tt repeat} & \mbox{\tt :-}& {\tt repeat}.
\end{array}
\)\\

The concrete semantics of this program maps the input
$\langle \epsilon, {\tt repeat}\rangle$, where $\epsilon$ is the empty
substitution, to the infinite sequence $<\epsilon,\ldots,\epsilon,\ldots>$.

On this example, because the program has no variables,
our domain of abstract substitutions only contains   two values, say
$\beta_{\emptyset}$ and $\beta_{\top}$, such that\\
\\
\(\begin{array}{llll}
{\it Cc}(\beta_{\emptyset})& =& \emptyset\\
 {\it Cc}(\beta_{\top})& =&\{\epsilon\}.
\end{array}\)\\

Let $B_{\bot}=\langle \beta_{\emptyset},0,0,{\it snt}\rangle$.
 Starting from $B_{\bot}$, the algorithm computes the abstract sequences\\
\\
\(\begin{array}{llllllllllllll}
B_0& =& B_{\bot}& \hskip0.8cm B'_0& = & B_{\bot}\\
B_1& =& \langle \beta_{\top},1,1,{\it snt}\rangle&
\hskip0.8cm 
B'_1& =& B_1 \nabla B'_{0}=\langle \beta_{\top},1,1,{\it snt}\rangle \\
B_2& =& \langle \beta_{\top},2,2,{\it snt}\rangle
&\hskip0.8cm
 B'_2& =& B_2 \nabla B'_{1}=\langle \beta_{\top},1,\infty,{\it snt}\rangle \\
B_3& =& \langle \beta_{\top},2,\infty,{\it snt}\rangle
\end{array}
\)
\\

Notice that the widening on sequences is applied when the 
abstract substitution part stabilizes, i.e., after the computation of the
abstract sequence $B_2$. The next iterate $B_3$ satisfies the property that
$B_3\leq B'_2$. Hence, according to the discussion in Section 
\ref{sub:EW},  the execution terminates
returning the final value
\begin{center}
$B'_2=\langle \beta_{\top}, 1,\infty,{\it snt}\rangle.$
\end{center}

Observe that $B'_2$
    safely approximates the concrete
infinite sequence
$<\epsilon,\ldots,\epsilon,\ldots>$. Moreover, it
 expresses the fact that 
the execution of {\tt repeat} {\em surely} succeeds at least once
and  {\em surely} does not terminate\footnote{This example also shows that
our framework can express non-failure properties such as the ones
described in \cite{BC99,DLH97}.}.

\vskip0.3cm
\noindent
{\bf Unification of Two Variables:}\ \ 
   {\tt UNIF-VAR}$: {\it AS}_{\{x_1,x_2\}} 
                     \rightarrow {\it ASS}_{\{x_1,x_2\}}$\ \\
\noindent
Given an abstract substitution $\beta$ with domain
$\{x_1,x_2\}$, this operation returns an abstract  sequence 
which represents a set of substitution sequences
 of length 0 or~1 (depending upon the
success or failure of the unification). The terms bound to $x_1$ and
$x_2$ are unified in all these sequences.
The operation {\tt UNIF-VAR} on abstract sequences uses an upgraded  
version of the operation {\tt UNIF-VAR}
on abstract substitutions defined in \cite{ICLP91AI,TOPLAS}. 
The latter,  in addition
to the resulting abstract substitution, 
 produces now two flags indicating whether the unification always
succeeds, always fails, or can both succeed and fail. 
 The additional information 
is expressed by the boolean values {\it ss} and {\it sf} as specified below.\\

\noindent
{\em Operation}  $\;${\tt UNIF-VAR}$: {\it AS}_{\{x_1,x_2\}} 
                     \rightarrow ({\it AS}_{\{x_1,x_2\}}
                     \times{\it Bool}\times{\it Bool})$
 
\noindent
Let $\beta\in{\it AS}_{\{x_1,x_2\}}$ and 
    $\langle\beta',{\it ss},{\it  sf}\rangle=
     \mbox{\tt UNIF-VAR}(\beta)$. The following  conditions
hold:
\begin{enumerate}
\itemsep 3pt
\item $
    \begin{array}{llll}
      \forall\theta\in{\it Cc}(\beta):
       \forall\sigma\in{\it SS}:
        (\sigma\in{\it mgu}(x_1\theta,x_2\theta)&\Rightarrow&
     [\![\theta\sigma]\!]\in{\it Cc}(\beta'));
    \end{array}$
\item 
  $\begin{array}{lll}
   {\it ss}={\it true}&\Rightarrow& 
                         (\forall\theta\in{\it Cc}(\beta):\ 
                          x_1\theta\mbox{\rm\ and\ }
                          x_2\theta\mbox{\rm\ \ are~unifiable);}
   \end{array}$
\item 
  $\begin{array}{lll}
   {\it sf}={\it true}&\Rightarrow& 
                         (\forall\theta\in{\it Cc}(\beta):\ 
                          x_1\theta\mbox{\rm\ and\ }
                          x_2\theta\mbox{\rm\ \ are~not~unifiable})
                          .
   \end{array}$
\end{enumerate}

\vskip0.2cm

 Based on the upgraded operation {\tt UNIF-VAR}
 for abstract substitutions, we provide an implementation of the operation
{\tt UNIF-VAR} for abstract sequences, 
which is correct with respect to the corresponding
specification given in
Section \ref{sub:AO}.

The operation {\tt UNIF-VAR}$: {\it AS}_{\{x_1,x_2\}} 
                     \rightarrow {\it ASS}_{\{x_1,x_2\}}$ on abstract sequences
is  defined as follows. 
 Let $\beta\in {\it AS}_{\{x_1,x_2\}} $ and 
$\langle\beta'',{\it ss},{\it  sf}\rangle=
     \mbox{\tt UNIF-VAR}(\beta)$. We have that {\tt UNIF-VAR}$(\beta)=B'$ where
 $B'$ is the abstract sequence $\langle \beta',m',M',t'\rangle$
such that
$$
\begin{array}{lll}
\beta'   & = & \beta''                  \\
m'      & = & \myif~{\it ss}~\mbox{\rm then}~1~\mbox{\rm else}~0\\
M'      & = & \myif~{\it sf}~\mbox{\rm then}~0~\mbox{\rm else}~1\\
t'      & = & {\it st}.
\end{array}
$$

\noindent
{\bf Abstract Interpretation of the Cut:}\ \ 
   {\tt AI-CUT}$: {\it ASSC}_{{D'}}
                        \rightarrow {\it ASSC}_{{D'}}$\ 

\noindent
Let $C=\langle\langle \beta,m,M,t\rangle,{\it acf}\rangle$.
{\tt AI-CUT}$(C)=\langle\langle \beta',m,'M',t'\rangle,{\it acf'}\rangle$
where
%
\begin{quote}
 $ \begin{array}{llll}
\beta' & = & \beta                                         \\ 
                                                             
m'     & = & \min(1,m)                                      \\ 
                                                             
M'     & = & \min(1,M)                                      \\ 
                                                             
t'     & = & {\it st}  &\myif m\geq 1 \myor {\it t}={\it st}  \\
       & = & {\it snt} &\myif M=0 \and {\it t}={\it snt}   \\
       & = & {\it pt}  &\otherwise                          \\
                                                              
{\it acf}'
       & = & {\it cut}  &\myif m\geq 1 \myor {\it acf}={\it cut}\\
       & = & {\it nocut} &\myif M=0 \and {\it acf}={\it nocut}\\
       & = & {\it weakcut} &\otherwise.
  \end{array}$
\end{quote}

\noindent
{\it Example}. Consider the program
\\
\\
\(\begin{array}{lllll}
{\tt p(X)} & \mbox{\tt :-} & {\tt q(X),\;!}.\\
{\tt q(X)} & \mbox{\tt :-} & {\tt X=a}.\\
{\tt q(X)} & \mbox{\tt :-} & {\tt X=b}.
\end{array}
\)\\
\\

For the sake of simplicity we use a 
simple domain of abstract substitutions
which can be seen as the mode component of the
{\tt Pattern} domain \cite{RP-98-002,TOPLAS}.
 The example is intended to illustrate the abstract execution 
of the operation
{\tt AI-CUT}. Hence, we do not enter here into the details of
the other  operations, but the reader is referred
to the appendix for
their definition.

The abstract execution of  the procedure ${\tt p}$ called with
its argument being a variable is as follows.
Let $$\beta={\tt X}\mapsto {\tt var}$$ be the initial abstract substitution.
Let  $c$ be  the clause of the program defining ${\tt p}$.
\\
First, the abstract sequence with cut information $C$ is computed by
$$C={\tt EXTC}(c,\beta)=\langle\langle {\tt X}\mapsto {\tt var}, 1,1,
{\it st}\rangle , {\it nocut}\rangle.$$
Then, the procedure {\tt q} that occurs
in the body of $c$ is executed with 
$\beta={\tt SUBST}(C)$ returning the
abstract sequence 
$$B=\langle {\tt X}\mapsto {\tt ground}, 2, 2, {\it st}\rangle.$$
Hence, the abstract sequence
with cut information $C'$ is computed as follows
$$C'={\tt EXTGS}({\tt q(X)}, C,B)=\langle \langle {\tt X}\mapsto {\tt ground}, 2, 2, {\it st}\rangle,
{\it nocut}\rangle.$$ Now, the operation
{\tt AI-CUT}$(C')$ is applied. Following the definition above, one obtains
 $$\mbox{{\tt AI-CUT}}(C')=\langle \langle {\tt X}\mapsto {\tt ground}, 1, 1, {\it st}\rangle,
{\it cut}\rangle$$  
expressing~the fact that a cut in the body of $c$ is {\em surely} executed.
The final result~is
$$B'={\tt SEQ}(C')=\langle {\tt X}\mapsto {\tt ground}, 1, 1, {\it st}\rangle$$
 stating that the execution of ${\tt p} $ called with
its argument being  a variable surely terminates and
succeeds exactly once.

Consider now the abstract execution of  the procedure ${\tt p}$ called with
a ground argument.
Let $$\beta={\tt X}\mapsto {\tt ground}$$ be the initial abstract substitution.
In this case, the abstract sequence with cut information $C$ is first
computed by
$$C={\tt EXTC}(c,\beta)=\langle\langle {\tt X}\mapsto {\tt ground}, 1,1,
{\it st}\rangle , {\it nocut}\rangle.$$
Then, the procedure {\tt q}  is executed with 
$\beta={\tt SUBST}(C)$ returning
$$B=\langle {\tt X}\mapsto {\tt ground}, 0, 1, {\it st}\rangle.$$
The abstract sequence 
with cut information $C'$ is computed as follows
$$C'={\tt EXTGS}({\tt q(X)}, C,B)=
\langle \langle {\tt X}\mapsto {\tt ground}, 0, 1, {\it st}\rangle,
{\it nocut}\rangle.$$ The operation
{\tt AI-CUT}$(C')$ returns 
 $$\mbox{\tt AI-CUT}(C')=\langle \langle {\tt X}\mapsto {\tt ground}, 0, 1, {\it
st}\rangle,
{\it weakcut}\rangle$$ expressing the fact
 that, in this case, the computation either fails without executing the cut
or succeeds  once after executing the cut.
The final result is
$$B'={\tt SEQ}(C')=\langle {\tt X}\mapsto {\tt ground}, 0, 1, {\it
st}\rangle$$ stating that the execution of ${\tt p}$
called with a ground argument
succeeds at most once and surely terminates.

The {\tt Pattern} domain 
used in our experiments
is more elaborated than
the simple domain of abstract substitutions
used in this example.
However, it does not provide more precision in these
cases.
A more sophisticated domain where an 
abstract sequence is represented 
as $\langle<\beta_1,\ldots,\beta_n>,m,M,t\rangle$
with $<\beta_1,\ldots,\beta_n>$ being an explicit sequence of
 abstract substitutions
 could return in the first case
a more precise result. Indeed, one could obtain
$B=\langle \{{\tt X}\mapsto {\tt a}\},\{{\tt X}\mapsto {\tt b}\}
, 2, 2, {\it st}\rangle$ and then
$B'=\langle \{{\tt X}\mapsto {\tt a}\}, 1, 1, {\it st}\rangle$.
However, such a domain could not improve the result in the second
case since the fact that the output substitution can be
 either ${\tt X}\mapsto {\tt a}$ or
${\tt X}\mapsto {\tt b}$ would be represented by
${\tt X}\mapsto {\tt ground}$ as we have done above.

\vskip0.3cm
\noindent
{\bf Abstract Lazy Concatenation.}
The implementation of the operation {\tt CONC}  is complicated here,
in order  to get accurate results when the domain ${AS}_D$ is instantiated to the domain {\tt Pattern}.
The implementation works on {\em
  enhanced} sets of abstract sequences which allow us to keep 
individual structural information about the results of every clause in order
to detect mutual exclusion of the clauses.

Let us  motivate the lifting of
abstract sequences to enhanced abstract sequences.
Lifting an abstract domain to its power set,
see, for instance, \cite{Cousot79a,FileILPS94}, is sometimes
useful when the original abstract domain is not expressive enough to
gain a given level of
accuracy.
Replacing an abstract domain by its power set is computationally
expensive
however; see \cite{WSA93_Granularity}.
Sometimes, the accuracy is lost only inside a few operations; 
thus, a good compromise can be to lift the domain only locally,
when these operations are executed, and to go back to the simple
domain afterwards. This is exactly what we are going to do for the
operation {\tt CONC}. The lifted version 
of the abstract domain 
that we are about to define
is useful when the abstract domain  is able to express
definite, but not disjunctive,
 structural information about
 terms.
In such a domain, for instance, 
the principal functor of the term bound to a program
variable
can be either
definitely known or not known at all; it is not possible to express
that it belongs to a given finite set.
The domain {\tt Pattern} used in our experiments is an abstract domain
of this kind. Disjunctive structural information is however essential
to implement the operation {\tt CONC} accurately: it allows us to
detect
mutually exclusive abstract sequences, i.e., abstract sequences that
should not be ``abstractly concatenated'' since they correspond to
different concrete inputs.
In order to keep disjunctive structural information, our
implementation of {\tt CONC} works on a finite set of abstract sequences.
This set is ``normalized'' in some way, in order to simplify the
case analysis in the implementation.
Basically, we differentiate between ``surely empty'' abstract sequences,
approximating only sequences of the form  $<>$ or $<\bot>$,
 and ``surely non empty'' abstract
sequences, approximating 
 only sequences of the form $<\theta>::S$. This is useful
because sequences such as $<>$ or $<\bot>$ are possible outputs
for any input, while sequences of the form $<\theta>::S$ are only
possible for some inputs. Therefore we only have to check
incompatibility of ``surely non empty'' abstract
sequences.
This discussion motivates  
the following definitions of semi-simple abstract sequences and simple
abstract sequences.

\begin{definition}[Semi-Simple Abstract Sequences]
\label{def:SNAS}
\noindent 
Let $B\in{\it ASS}_D$. We say that $B$ is a {\em semi-simple} abstract sequence
if 
\begin{enumerate}
\itemsep 0pt
\item either, $\beta=\beta_\emptyset$ and $m=M=0$
\item or, $\beta\neq\beta_\emptyset$ and $1\leq m \leq M$.
\end{enumerate} 
\end{definition}

\begin{definition}[Simple Abstract Sequences]
\label{def:NAS}
\noindent 
Let $B\in{\it ASS}_D$. We say that $B$ is a {\em simple} abstract sequence
if it is semi-simple and $t\in\{{\it snt},{\it st}\}$.
\end{definition}

Semi-simple abstract sequences formalize our idea of distinguishing
between ``surely empty'' and ``surely non empty'' abstract sequences. 
Note that,
assuming that $\beta_\emptyset$ is the only abstract substitution such that
${\it Cc}(\beta_\emptyset)=\emptyset$, we have that ${\it Cc}(B)\neq\emptyset$ for
any semi-simple abstract sequence $B$. 

\begin{definition}[Enhanced Abstract Sequences]
\label{def:EnhAS}
\noindent
Let $D$ be a finite set of program variables.  We denote by
${\it ASS}^{\it enh}_D$  the set of all sets of the form
$\{B_1,\dots,B_n\}$, where $n\geq 0$ and    
$B_1,\dots,B_n$ are semi-simple abstract sequences from ${\it ASS}_D$.
Elements of ${\it ASS}^{\it enh}_D$ are called {\em enhanced abstract
  sequences}; they are denoted by {\it SB} in the following.
The concretization function 
${\it Cc}:{\it ASS}^{\it enh}_D \rightarrow {\it CSS}_D$ is defined by
${\it Cc}({\it SB}) =\bigcup_{B\in{\it SB}}{\it Cc}(B).$
\end{definition}

The operation $ {\tt SPLIT1}$ transforms an
arbitrary abstract sequence into an equivalent enhanced abstract
sequence. 

\vskip0.3cm
\noindent
{\em Operation} $ \;{\tt SPLIT1}: {\it ASS}_D \rightarrow {\it ASS}^{\it enh}_D$\\
\noindent
This operation is required to satisfy the property that
for every $B\in{\it ASS}_D$,
$ {\it Cc}({\tt SPLIT1}(B))={\it Cc}(B)$.
Let $B=\langle \beta,m,M,t\rangle$. We define ${\it SB}'={\tt SPLIT1}(B)$
as
 ${\it SB}'={\it SB}_1\cup{\it SB}_2$ where
\vskip0.2cm
\begin{quote}
  $\begin{array}{llll}
   {\it SB}_1 & = & \{\langle \beta_\emptyset,0,0,t\rangle\} &
                    \myif m=0                             \\
              & = & \emptyset & \otherwise                  \\
   {\it SB}_2 & = & \{\langle \beta,\max(1,m),M,t\rangle\} &
                    \myif \beta\neq\beta_\emptyset
                    \and \max(1,m)\leq M                   \\
              & = & \emptyset & \otherwise .                 \\
  \end{array}$
\end{quote}

\vskip0.2cm

The operation {\tt MERGE} is the converse of {\tt SPLIT1}: it
%
transforms an enhanced
abstract sequence into a plain abstract sequence. 
Most of the time, this operation loses
part of the information expressed by the enhanced abstract
substitution sequence; but it does not lose any information
when the enhanced
abstract sequence results from a single application of {\tt SPLIT1}.

\vskip0.3cm
\noindent
{\em Operation} $ \;{\tt MERGE}: {\it ASS}^{\it enh}_D \rightarrow {\it ASS}_D$\\
The operation 
$ {\tt MERGE}$
 satisfies the following properties:

\begin{enumerate}
\itemsep 2pt
\item For every ${\it SB}\in{\it ASS}^{\it enh}_D$,
$ {\it Cc}({\it SB})\subseteq{\it Cc}({\tt MERGE}({\it SB}))$
\item For every $B\in{\it ASS}_D$,
$ {\it Cc}({\tt MERGE}({\tt SPLIT1}(B)))={\it Cc}(B)$.
\end{enumerate}

The definition of {\tt MERGE} requires choosing a particular abstract
sequence $B_\emptyset$ such that ${\it Cc}(B_\emptyset)=\emptyset$.
We decide that $B_\emptyset=\langle\beta_\emptyset,1,0,{\it st}\rangle$.
This choice is arbitrary since there is no best (least)
representation of the empty set of abstract sequences in this domain.
Moreover, it uses the binary operation 
$ {\tt UNION}: ({\it AS}_D \times {\it AS}_D) \rightarrow {\it AS}_D$,
 which is inherited from
our previous framework. The latter is
 extended  to finite sequences
of abstract substitutions as follows:

\begin{quote}
  \begin{tabular}{lllll}
   ${\tt UNION}(<>)$ & $=$ & $\beta_\emptyset$ &       \\
   ${\tt UNION}(<\beta>)$ & $=$ & $\beta,\;$ 
 for every   
                                        $\beta\in {\it AS}_D$  \\
   ${\tt UNION}(<\beta_1,\dots,\beta_n>)$ & $=$ & 
   ${\tt UNION}(\beta_1,{\tt UNION}(<\beta_2,\dots,\beta_n>))$, \\
& &  for all   
    $\beta_1,\dots,\beta_n\in{\it AS}_D$ $(n\geq 2)$.  \\
  \end{tabular}   
\end{quote}

The operation  {\tt MERGE} can now be defined.
Let $\sqcup$ denote the least upper bound on {\it TI}.
Let  ${\it SB}\in{\it ASS}^{\it enh}_D$ such that 
${\it SB}=\{B_1,\dots,B_n\}$ and 
$B_i=\langle \beta_i,m_i,M_i,t_i\rangle$ $(1\leq i\leq n)$.
The abstract sequence $B'={\tt MERGE}({\it SB})$ is such that

\begin{quote}
  $\begin{array}{llll}
     B' & = & B_\emptyset  & \myif n=0  \\

        & = & B_1 & \myif  n=1 \\

        & = & \langle {\tt UNION}(<\beta_1,\dots,\beta_n>) ,
                      \min(m_1,\dots,m_n),\\
        &   &       \ \max(M_1,\dots,M_n),
                      t_1\sqcup\dots \sqcup t_n\rangle & \myif n\geq 2.
  \end{array}$
\end{quote}


The notion of simple abstract sequence with cut information
 is also useful to simplify the case analysis in the
implementation of {\tt CONC}.

\begin{definition}[Simple Abstract Sequences with Cut Information]

\noindent
Let $B\in{ASS}_D$ and $ {\it acf}\in{\it ACF}$. The abstract sequence
with cut information $\langle B,{\it acf} \rangle$ is said to be {\em simple}
if  $B$ is simple and ${\it acf}\in{\it CF}$.
\end{definition}

The operation {\tt SPLIT2}
converts an arbitrary
abstract sequence with cut information into an equivalent set of 
 simple abstract sequences with cut information. 

\vskip0.3cm
\noindent
{\em Operation} $ {\tt SPLIT2}: {\it ASSC}_D \rightarrow \wp({\it ASSC}_D)$\\
The operation {\tt SPLIT2}  satisfies the following
properties.
For every $C\in{\it ASSC}_D$,
\begin{enumerate}
\itemsep 2pt
\item $\bigcup_{C'\in{\mbox{\footnotesize \tt SPLIT2}} (C)}
               {\it Cc}(C') \,=\,{\it Cc}(C)$;
\item all abstract sequences with cut information in ${\tt SPLIT2}(C)$
  are simple.
\end{enumerate}

\noindent
Its definition is simple.
 We first
apply the operation {\tt SPLIT1} to the abstract sequence part of $C$.
Then we split the cut information. 
Finally we split the termination information. Formally,
${\tt SPLIT2}(C)$ is defined as follows.

\begin{enumerate}
\itemsep 0pt
\item Let $C=\langle B, {\it acf}\rangle\in{\it ASSC}_D$. We define
  \begin{quote}
    $\begin{array}{lll}
        {\tt SPLIT2}(C) & = & \bigcup_{B'\in
              \mbox{\footnotesize \tt SPLIT1}(B)}{\tt
        SPLIT2}(\langle B',{\it acf}\rangle).
    \end{array}$
         \end{quote}
\item Let $B=\langle \beta,m,M,t\rangle\in{\it ASS}_D$. Assume that $B$ is semi-simple. We define
  \begin{quote}
    $\begin{array}{llll}
          {\tt SPLIT2}(\langle B,{\it weakcut}\rangle) & = &
                {\tt SPLIT2}(\langle B,{\it nocut}\rangle)\cup
                {\tt SPLIT2}(\langle B,{\it cut}\rangle) & \myif m=0\\
                                              & = &
             {\tt SPLIT2}(\langle B,{\it cut}\rangle) & \myif m\geq 1.  
  \end{array}$
 \end{quote}

\noindent (Remember that, by Definition \ref{def:SNAS}, we
also have $\beta=\beta_\emptyset$ and $M=0$, in the first case, and
$\beta\neq\beta_\emptyset$ and $m\leq M$, in the second case.)

\item Let $B=\langle \beta,m,M,t\rangle\in{\it ASS}_D$ and ${\it cf}\in{\it CF}$. 
      Assume that $B$ is semi-simple. We define
      \begin{quote}
           $\begin{array}{llll}
                {\tt SPLIT2}(\langle B,{\it cf}\rangle) & = &
                                          \{\langle B,{\it cf}\rangle\}
                       & \myif t\in\{{\it snt},{\it st}\};\\
                                        & = &
            \{\langle \langle \beta,m,M,{\it snt}\rangle,{\it cf}\rangle,
               \langle \langle \beta,m,M,{\it st}\rangle,{\it cf}\rangle\}
                       & \myif t={\it pt}.
             \end{array}$
       \end{quote}
\end{enumerate}

Before presenting the implementation of {\tt CONC},
 we still need to specify the operation {\tt EXCLUSIVE}, which 
is aimed at detecting incompatible outputs.
An implementation of this operation for the domain {\tt Pattern} is
given in Section \ref{sub:IP}.

\vskip0.3cm
\noindent
{\em Operation} ${\;\tt EXCLUSIVE}:({\it AS}_D\times {\it AS}_D\times {\it AS}_D)
\rightarrow {\it Bool}$\\
The operation {\tt EXCLUSIVE}  satisfies the following
property.
For all $\beta,\beta_1,\beta_2\in{\it AS}_D$,
\begin{center}
$\begin{array}{lll}
{\tt EXCLUSIVE}(\beta,\beta_1,\beta_2) & \Rightarrow &
\neg(\exists \theta\in{\it Cc}(\beta)
             ,\theta_1\in{\it Cc}(\beta_1)
             ,\theta_2\in{\it Cc}(\beta_2),
              \sigma_1,\sigma_2\in{\it SS}:\\\
& &  \theta\sigma_1=\theta_1
                                      \and                                              \theta\sigma_2=\theta_2).
  \end{array}$
\end{center}

 We are now ready to describe the operation {\tt CONC}.

\vskip0.3cm
\noindent
{\em Operation} $\;{\tt CONC}: ({\it AS}_D \times {\it ASSC}_D  
                         \times {\it ASS}^{\it enh}_D)
                 \rightarrow {\it ASS}^{\it enh}_D$.\\
\noindent
Let $\beta\in {\it AS}_D$, $C_1\in {\it ASSC}_D  $ and
${\it SB}_2\in {\it ASS}^{\it enh}_D$.
${\it SB}'={\tt CONC}(\beta,C_1,{\it SB}_2)$ is defined as follows.
We assume that $B_i=\langle \beta_i,m_i,M_i,t_i\rangle$.

\begin{enumerate}
\itemsep 0pt
\item 
Let us assume first that  $C_1=\langle B_1,{\it acf}_1\rangle$ is simple and ${\it SB}_2=\{B_2\}
  $.
  \begin{enumerate}
\itemsep 0pt
  \item Suppose that ${\it acf}_1={\it cut}$ or $t_1={\it snt}$. In
    this case, we define
    \begin{quote}
      $\begin{array}{lll}
        {\it SB}'&= & \{B_1\}.
      \end{array}$
    \end{quote}
  \item Suppose, on the contrary, that ${\it acf}_1={\it nocut}$ 
                and $t_1={\it st}$. We define
    \begin{quote}
\hskip-1.0cm
     $\begin{array}{llll}
        {\it SB}'&= & \{B_2\} & \myif M_1 = 0\\

                 &= & \{\langle \beta_1,m_1,M_1,t_2\rangle\} 
                    & \myif M_1 \geq 1 \and M_2 =0\\

                 &= & \{\langle {\tt UNION}(\beta_1,\beta_2),
                                 m_1+m_2,M_1+M_2,t_2\rangle\} 
                    & \myif M_1 \geq 1 \and M_2 \geq 1\\
                 &&&\and \neg{\tt EXCLUSIVE}(\beta,\beta_1,\beta_2)\\

                 &= & \emptyset
                    & \myif M_1 \geq 1 \and M_2 \geq 1\\
                   &&&\and {\tt EXCLUSIVE}(\beta,\beta_1,\beta_2).
     \end{array}$
    \end{quote}
  \end{enumerate}
\item In the general case, we define
  \begin{quote}
    $
    \begin{array}{lll}
     {\it SB}'&= &
     \displaystyle\bigcup_{\renewcommand{\arraystretch}{0.1}
                                         \begin{array}{c}
                                        _{C\in\mbox
                                       {\footnotesize \tt SPLIT2}(C_1)}\\
                                        _{B\in{\it SB}_2}
                                        \end{array}
                           \renewcommand{\arraystretch}{1}}
                   {\tt CONC}(\beta,C,\{B\}).
    \end{array}$
  \end{quote}
\end{enumerate}

\subsection{Instantiation to {\tt Pattern}}
\label{sub:IP}

The domain of abstract substitutions {\tt Pattern} has been introduced
in \cite{MusumbuThesis} and it has been used in many of our
previous works, e.g., \cite{SPE,ACTA95}. The reader is referred
to \cite{TOPLAS} for a detailed description of the domain and of its
  abstract operations. 

\vskip0.3cm 
\noindent
{\bf The Abstract Domain  Pattern.}
\noindent
The version of {\tt Pattern} used in the experimental
evaluation of Section \ref{sub:EE} 
can be best viewed as an instantiation of the generic
pattern domain $\pag$  \cite{POPL94,SCP00} with  mode, sha\-ring, and
arithmetic  components.

The key intuition behind $\pag$ is to represent information on some
subterms occurring in a substitution instead of information on terms
bound to variables only.  More precisely, $\pag$ may associate the
following information with each 
conside\-red 
subterm: (1)
its {\em pattern}, which specifies the main functor of the
  subterm (if any) and the subterms which are its arguments;
its {\em properties}, which are left unspecified and are
given in the domain $\cal R$. 
 In addition
to the above information, each variable in the domain of the
substitution is associated with one of the subterms. It can be
expressed that two arguments have the same value (and hence
that two variables are bound together) by associating both arguments
with the same subterm.  It should be emphasized that the pattern
information may be void.  In theory, information on all subterms could
be kept but the requirement for a finite analysis makes this
impossible for almost all applications. As a consequence, the domain
shares some features with the depth-k abstraction \cite{Kanamori87},
although $\pag$ does not impose a fixed depth but adjusts it
dynamically through upper bound and widening operations. Note that the
identification of subterms (and hence the link between the structural
components and the $\cal R$-domain) is a somewhat arbitrary choice. In
\pag{}, subterms are identified by integer indices, say $1,\ldots,n$
if $n$ subterms are considered, and we denote sets of
indices by the symbol $I$.

More formally,
the pattern and same-value component
can be described as follows.
  The {\em pattern} component is a partial function 
${\it frm}: I \not\rightarrow {\it Pat}_I$, 
from the set of indices $I$ to the set
of patterns over $I$, i.e., elements of the form $f(i_1,\ldots,i_n)$,
where $f\in{\cal F}$ is a functor symbol of arity $n$
 and $i_1,\ldots,i_n \in I$. When the
pattern is undefined for an index $i$, we write {\it frm($i$) $=$ undef}.
The {\em same-value} component is a total function 
${\it sv}: D \rightarrow I$, 
where $D = \{x_1,\ldots,x_n\}$ is the domain of the abstract
substitution. 

A pattern component ${\it frm}: I \not\rightarrow {\it Pat}_I$
denotes a set of families $(t_i)_{i\in I}$
of terms as defined below.
%
\begin{quote}
  $\begin{array}{llll}
 {\it Cc}({\it frm})   & = &
 \{ (t_i)_{i\in I}\ |  &
 {\it frm}(i) = f(i_1, \dots, i_n) \,\Rightarrow\,
     t_i     = f(t_{i_1}, \dots, t_{i_n}),\\
& & &
 \forall i, i_1, \dots, i_n \in I, 
 \forall f \in {\cal F}\}.
  \end{array}$
\end{quote}

\noindent
In order to simulate
 unification with
occur-check,  we also assume that every pattern
component {\it frm}  satisfies the following condition:
 the relation $\subc\subseteq I\times I$ such that
$i\subc j$ if and only if ${\it frm}(i)$ is of the form 
$f(\dots, j, \dots)$ must be 
well-founded.

 A pair  $\langle {\it sv},{\it frm} \rangle$ with
${\it sv}: D \rightarrow I$ and ${\it frm}: I \not\rightarrow {\it Pat}_I$
is called
{\em structural abstract substitution};
it denotes a set of program substitutions as follows:
\begin{quote}
  $\begin{array}{lll}
 \Cc{\langle {\it sv},{\it frm} \rangle}   & = &
 \{ \theta \in {\it PS}_D \ | \ 
\exists
(t_i)_{i\in I}\in{\it Cc}({\it frm}):\;
x_j\theta = t_{{\it sv}(x_j)},\;
 \forall x_j \in D \}.
  \end{array}$
\end{quote}

The $\cal R$-domain is the generic part which specifies subterm
information by descri\-bing properties of a set of tuples $<t_1, \ldots,
t_n>$ where $t_1, \ldots, t_n$ are terms. As a consequence, defining
the $\cal R$-domain amounts essentially to defining a traditional
domain on substitutions and its operations. We now describe the
various components of the $\cal R$-domain which can be built as an
open product \cite{POPL94,SCP00}. 

The {\em mode} component is described in \cite{TOPLAS} and associates a
mode from the set $ {\it Modes} =
\{ {\tt var}, {\tt ground}, {\tt novar}, {\tt noground}, {\tt ngv},
{\tt gv}, {\tt any}\}$ with each subterm.
Formally, it is a total function ${\it mo}: I \rightarrow {\it Modes}$
whose concretization is defined as

\begin{quote}
  $\begin{array}{lll}
 {\it Cc}({\it mo}) & = &
   \{ (t_i)_{i\in I}\ |\; \ t_i \in{\it  Cc}({\it mo}(i)), \;\forall  i \in I \}.
  \end{array}$
\end{quote}

The {\em sharing} component maintains information about possible sharing
between pairs of subterms and is also described in \cite{TOPLAS}.
Formally, it is a symmetrical relation ${\it ps}\subseteq I\times I$
whose concretization is defined as 

\begin{quote}
  $\begin{array}{lll}
 {\it Cc}({\it ps}) & = &
   \{ (t_i)_{i\in I}\ |\; \ 
   {\it var}(t_i) \cap {\it var}(t_j) \Rightarrow {\it ps}(i,j), 
   \;\forall  i,j \in I \}.
  \end{array}$\\
\end{quote}

The {\em arithmetic} component is novel and aims at using arithmetic
predicates to detect mutual exclusion between clauses. It  approximates
information about arithmetic relationships by rational order
constraints, i.e., binary constraints of the form $i \ \delta \ j$ and
unary constraints of the form $i \ \delta \ c$, where $i,j$ are
indices, $\delta \in \{>,\geq,=,\neq,\leq,<\}$ and $c$ is an integer
constant. For instance, a built-in $X \geq Y + 2$ is approximated by a
constraint $X > Y$. 
Formally, an element {\it arithm}
is a set of rational order constraints over indices, whose
concretization  is defined as follows (a constraint being
satisfied only if the terms are numbers). 

\begin{quote}
  $\begin{array}{lll}
 {\it Cc}({\it arithm}) &= & \{
  (t_i)_{i\in I}\ |\; \ \forall \ i \ \delta \ j  \in  arithm:
\; t_i \ \delta \ t_j  \and  \forall \ i \ \delta \ c  \in 
arithm: \; t_i \ \delta \ c \}.  
  \end{array}$
\end{quote}

\vskip0.3cm
\noindent
{\bf The Operation EXCLUSIVE.}
We describe here the implementation of the ope\-ration {\tt EXCLUSIVE} on our 
domain of abstract substitutions. This operation was not present in our
previous  works.
It aims at detecting situations where
two output abstract sequences $B_1$ and $B_2$ are incompatible, given
that they both originate from the same abstract input
substitution $\beta$. Only the abstract substitution components
$\beta_1$ and $\beta_2$ of $B_1$ and $B_2$ are useful to detect such
situations. Thus the operation {\tt EXCLUSIVE} has three arguments
$\beta$, $\beta_1$, and $\beta_2$. (See its specification in
 Section~\ref{sub:GAO}.)

Let us first introduce the notion of
decomposition of a program substitution 
with respect to a structural abstract substitution.
It represents the family
of terms, occurring in the program substitution, that are given an
index by the structural abstract substitution.

\begin{definition}[Decomposition of a Program Substitution]
\label{def:DPS}

\noindent
Let $\langle {\it sv},{\it frm}\rangle $ be a structural abstract
substitution over domain
$D=\{x_1,\dots,x_n\}$ and set of indices $I$. 
Let also $\theta\in{\it Cc}\langle {\it sv},{\it frm}\rangle$. 
The {\em decomposition of $\theta$ 
          with respect to $\langle {\it sv},{\it frm}\rangle $} is the (unique)
family of terms $(t_i)_{i\in I}$ such that 
\begin{quote}
  $\begin{array}{lll}
   \theta=\{x_1/t_{{\it sv}(x_1)},\dots,x_n/t_{{\it sv}(x_n)}\} & \and
   &
   (t_i)_{i\in I}\in{\it Cc}({\it frm}).
  \end{array}$
\end{quote}

 Existence and unicity of the family $(t_i)_{i\in I}$ can be
proven by an induction argument that uses the fact that 
the relation $\subc$ over $I$ is well-founded. Unicity holds conditional to the fact that $I$ does not contain any "useless" element, i.e., for eve\-ry $i \in I$, there exists a variable $x_j \in D$ and a set of indices $i_1,\dots,i_k$ such that 
$i_1 = {\it sv}(x_j)$,
$i_1 \succ \dots \succ i_k$,
and $i_k = i$. From now on we assume that this condition always holds. 
\end{definition}

The next definition models a property of the structural abstract
substitutions obtained by performing any number of abstract
unification steps on another structural abstract substitution.


\begin{definition}[Instance of a Structural Abstract Substitution]
\label{def:ISAS}
\noindent
Let $\langle {\it sv},{\it frm}\rangle $ and 
$\langle {\it sv}',{\it frm}'\rangle $ be two structural abstract
substitutions over the same domain
$D=\{x_1,\dots,x_n\}$ and respective sets of indices $I$ and $I'$. 
Let also ${\it im}:I \rightarrow I'$ be a total function.
We say that 
{\em $\langle {\it sv}',{\it frm}'\rangle $ is an instance of
     $\langle {\it sv},{\it frm}\rangle $ with respect to {\it im}} if the
 following conditions hold:

\begin{enumerate}
\itemsep 2pt
\item ${\it sv}' = {\it im} \ \comp\  {\it sv}$;
\item for all $i,i_1,\dots,i_m\in I,\\
  \begin{array}{lll}
{\it frm}(i)= f(i_1,\dots,i_m) & \Rightarrow &
{\it frm}'({\it im}(i))= f({\it im}(i_1),\dots,{\it im}(i_m)).
  \end{array}$
\end{enumerate}
\noindent
Moreover, we say that 
{\em $\langle {\it sv}',{\it frm}'\rangle $ is an instance of
     $\langle {\it sv},{\it frm}\rangle $} if there exists a 
   function ${\it im}$ such that the conditions hold.
\end{definition}

The next property holds.

\begin{property}\label{prop:PI}
Let $\langle {\it sv},{\it frm}\rangle $ and
     $\langle {\it sv}',{\it frm}'\rangle $ 
be two structural abstract substitutions, and let
 ${\it im}: I \rightarrow I'$ be such that
$\langle {\it sv}',{\it frm}'\rangle $ is an instance of
$\langle {\it sv},{\it frm}\rangle $ with respect to {\it im}.
Let also $\theta\in{\it Cc}\langle {\it sv},{\it frm}\rangle $,
$\theta'\in{\it Cc}\langle {\it sv}',{\it frm}'\rangle $,  and
 $\sigma \in {\it SS}$. 
Finally, let $(t_i)_{i\in I}$ and $(t'_{i})_{i\in I'}$ be the
decompositions of $\theta$ and $\theta'$ with respect to 
$\langle {\it sv},{\it frm}\rangle $ and
     $\langle {\it sv}',{\it frm}'\rangle $, respectively.
Then we have
 \begin{quote}
   $\begin{array}{lll}
\theta' = \theta\sigma &\Rightarrow&
(t_i\sigma)_{i\in I} = (t'_{{\it im}(i)})_{i\in I}.
   \end{array}$
 \end{quote}
\end{property}
\noindent
The proof is a simple induction on the well-founded relation $\subc$,
induced on $I$ by~${\it frm}$.

The next definitions and properties are instrumental to the
implementation and correctness proof of the operation {\tt EXCLUSIVE}.

\begin{definition}[Exclusive Pair of Indices]
\label{def:EPI}
\noindent
Let $ {\it frm}_1$ and $ {\it frm}_2$ be two pattern components over sets
of indices $I$ and $J$, respectively. Let also $i\in I$ and  $j\in J$.
\begin{enumerate}
\itemsep 2pt
\item We say that $\langle i,j\rangle$ is {\em directly exclusive} with respect to 
$\langle{\it frm}_1,{\it frm}_2\rangle$ iff ${\it frm}_1(i)=
f(i_1,\ldots,i_p)$,  ${\it frm}_2(j)=g(j_1,\ldots,j_{q})$ 
and either $f \neq g$ or $p \neq q$.
\item We say that $\langle i,j\rangle$ is {\em exclusive} with respect to 
$\langle{\it frm}_1,{\it frm}_2\rangle$ iff $\langle i,j\rangle$ is
{\em directly exclusive} with respect to $\langle{\it frm}_1,{\it frm}_2\rangle$, 
or ${\it frm}_1(i)=f(i_1,\ldots,i_p)$, ${\it frm}_2(j)=f(j_1,\ldots,j_p)$ and there exists
$k: 1 \leq k \leq p$ such that $\langle i_k,j_k\rangle$ is exclusive with respect to 
$\langle{\it frm}_1,{\it frm}_2\rangle$.
\end{enumerate}
\end{definition}

\vskip-0.3cm
\begin{property}\label{prop:EPI}
Let $ {\it frm}_1$ and $ {\it frm}_2$ be two pattern components over sets
of indices $I$ and $J$, respectively.
Let $(t_i)_{i\in I}\in {Cc}({\it frm}_1)$ and
    $(t_j)_{j\in J}\in {Cc}({\it frm}_2)$. Let also  $i\in I$ and~$j\in J$.
    \begin{enumerate}
\itemsep 2pt
    \item If the pair $\langle i,j\rangle$ is {\em directly exclusive} with respect to 
          $\langle{\it frm}_1,{\it frm}_2\rangle$, then the terms $t_i$
          and $t_{j}$ are compound and they have distinct principal functors.
    \item If the pair $\langle i,j\rangle$ is {\em exclusive} with respect to 
          $\langle{\it frm}_1,{\it frm}_2\rangle$, then the terms $t_i$
          and $t_{j}$ are distinct ($t_i \neq t_{j}$).
    \end{enumerate}
\end{property}

We are now in position to provide the implementation of the operation
{\tt EXCLUSIVE} for the domain {\tt Pattern}.
We just show here a partial implementation which only uses the pattern,
same-value, and mode components but it gives the
idea behind the complete implementation. For additional details, the
 reader is referred to
\cite{Braem.Modard94}.

\vskip0.3cm
\noindent
{\em Operation} $\;{\tt
    EXCLUSIVE}:{\tt Pattern}\times {\tt Pattern}\times{\tt Pattern}\rightarrow {\it Bool}$\\
\noindent
Let $\beta,\beta_1,\beta_2$ be abstract substitutions over the same
domain $D$ and sets of indices $I$, $I_1$, and $I_2$, respectively.
Assume that $\langle {\it sv}_1, {\it frm}_1 \rangle$ and
$\langle {\it sv}_2, {\it frm}_2 \rangle$ 
are instances of $\langle {\it sv}, {\it frm} \rangle$ with respect to 
${\it im}_1$ and ${\it im}_2$, respectively.
The value of
{\tt EXCLUSIVE}$(\beta,\beta_1,\beta_2)$ is {\it true} if and only if there
exists
$i \in I $ such that
\begin{enumerate}
\itemsep 2pt
\item ${\it mo}(i) \in \{{\tt ngv},{\tt novar}\}$ and the pair
$\langle{\it im}_1(i),{\it im}_2(i)\rangle$ is directly exclusive 
with respect to $\langle {\it frm}_1,{\it frm}_2\rangle$, or
\item ${\it mo(i)} = {\tt ground}$ and the pair
$\langle{\it im}_1(i),{\it im}_2(i)\rangle$ is exclusive 
with respect to $\langle {\it frm}_1,{\it frm}_2\rangle$.
\end{enumerate}


\noindent
Correctness of the implementation 
follows from Properties \ref{prop:PI} and \ref{prop:EPI}; see 
\cite{sequence}.\\


\noindent
{\bf Prolog's Built-in Predicates.}
Prolog's built-in predicates such as test predicates 
({\tt var}, {\tt ground}, and the like) or arithmetic predicates
({\tt is}, {\tt <}, \dots) can be handled in essentially the same way
as abstract unification. Our implementation actually includes abstract
operations that deal with test and arithmetic predicates 
 \cite{Braem.Modard94}. Other built-in predicates can be
accommodated as well, including the predicates {\tt assert}
and {\tt retract}. However, the treatment of the latter predicates
assumes that dynamic predicates are disjoint from static predicates,
i.e., it assumes that the underlying program $P$ is not modified.
A more satisfactory treatment of dynamic predicates requires to
introduce a new abstract object representing the dynamic program;
this improvement is a topic for further work.

\subsection{Experimental Evaluation}\label{sub:EE}
The experimental results presented in this section provide evidence of
the fact that the approach presented in this paper allows one
to integrate predicate level analysis to existing variable
level analysis at a reasonable implementation cost.
Compari\-sons with other cardinality and determinacy analyses
can be found in Section \ref{sec:RDASBAI}.\\

\begin{table}
 \caption{Efficiency of the Cardinality Analysis}
\label{efficiency}
\centering
\begin{tabular}{||l||r|r||r|r|r|r||r|r|r|r||} 
\multicolumn{11}{c}{}  \\ 
\cline{1-11}
 \multicolumn{1}{||c||}{} & \multicolumn{2}{c||}{} & \multicolumn{4}{c||}{} &
\multicolumn{4}{c||}{}  \\
 \multicolumn{1}{||c||}{} &\multicolumn{2}{c||}{OR} & \multicolumn{4}{c||}{PC} &
\multicolumn{4}{c||}{PCA} \\ 
\cline{1-11}
& & & & & & & & & & \\
Programs &  I & T & I & T & IR & TR & I & T & IR & TR \\
\cline{1-11}
& & & & & & & & & & \\
Qsort &  
13 & 0.08 & 17 & 0.12 & 1.31 & 1.50 & 13 & 0.08 & 1.00 & 1.00\\
Qsort2 &
 15 & 0.08 & 19 & 0.12 & 1.27 &  1.50 & 15 & 0.09 & 1.00 & 1.13 \\
Queens &
15 & 0.07 & 18 & 0.08 & 1.20 &  1.14 & 18 & 0.10 &  1.20 & 1.43 \\
Press1 &
532 & 11.77 & 581 & 13.11 & 1.09 &  1.11 & 581 & 13.45 &  1.09 & 1.14 
\\
Press2 &
197 & 3.27 & 200 & 3.56 & 1.02 &  1.09 & 200 & 3.56 &  1.02 & 1.09 \\
Gabriel &
 78
 & 0.90 & 84 & 1.00 & 1.08 &  1.11 & 84 & 0.98 &  1.08 & 1.09 \\
Peep & 
132 & 3.21 & 131 & 18.85 & 0.99 &  5.87 & 131 & 19.08 &  0.99 & 5.94 \\
Read &
 432 & 23.91 & 458 & 25.32 & 1.06 &  1.06 & 458 & 25.37 &  1.06 & 1.06
\\
Kalah &
115 & 1.90 & 121 & 2.09 & 1.05 &  1.10 & 120 & 2.11 &  1.04 & 1.11 \\
Cs &  
79 & 2.19 & 91 & 3.05 & 1.15 &  1.39 & 90 & 3.02 &  1.14 & 1.38 \\
Plan &
36 & 0.21 & 38 & 0.30 & 1.06 &  1.43 & 38 & 0.27 &  1.06 & 1.29 \\
Disj &
 64 & 1.95 & 68 & 2.14 & 1.06 &  1.10 & 68 & 2.12 &  1.06 & 1.09 \\
Pg & 
38  & 0.32 & 40 & 0.36 & 1.05 &  1.13 & 39 & 0.35 &  1.03 & 1.09 \\
Boyer &
56  & 0.76 & 56 & 1.15 & 1.00 &  1.51 & 56 & 1.17 &  1.00 & 1.54 \\
Credit &
63 & 0.57 & 64 & 0.81 & 1.02 &  1.42 & 64 & 0.80 &  1.02 & 1.40\\ 
    &     & &    &      &      &        &    &      &      &     
\\
\cline{1-11}
& & & & & & & & & &\\
Mean & &   &   &   &  1.09  & 1.56  &   &   & 1.05  & 1.52  \\
\cline{1-11}
\end{tabular}
\end{table}

\noindent
{\bf Benchmarks.} Our experiments use our traditional benchmarks
except that cuts have been reinserted as in the original versions.
 In addition, some
new programs have been added.  {\tt Boyer} is a theorem-prover from the
DEC-10 benchmarks, {\tt Credit} is an expert system from \cite{Sterling86}.
  There are two versions of {\tt Qsort}
which differ in procedure {\tt Partition} which uses or does not use
auxiliary predicates for the arithmetic built-ins.
All the benchmarks are available by anonymous ftp from
ftp://ftp.info.fundp.ac.be/pub/users/ble/bench.p.
They have been run on a SUN SS-10/20.

\vskip0.3cm
\noindent

\noindent
{\bf Efficiency.} The efficiency results are reported in Table
\ref{efficiency}. Several algorithms are compared: {\tt OR} is the
original
{\tt GAIA} algorithm on {\tt Pattern} \cite{TOPLAS}, {\tt PC} is the
cardinality analysis with {\tt Pattern} and {\tt PCA} is {\tt PC} with
the abstraction for arithmetic predicates. 
I, T, IR and TR are the number of ite\-ra\-tions, the execution time
(in seconds),
the iteration's ratio and the time's ratio respectively.
The first interesting
point to notice is the slight increase (about 5\% on {\tt PCA}) in
iterations when moving from abstract substitutions to abstract
sequences, sho\-wing the effectiveness of our widening operator. Even
more important perhaps is the fact that the time overhead of the
cardinality analysis is small with respect to the traditional analysis:
{\tt PCA}
is 1.52 slower than {\tt OR}.  Note that in fact most programs enjoys
an even smaller overhead but {\tt Peep} is about 6 times slower than
{\tt
OR} in {\tt PCA}. This comes from many procedures with many clauses,
most of which being not surely cut; much time is spent in the
concatenation operation. Finally, note that adding
more functionality in the domain did not slow down the analysis by
much.

\begin{table}
\caption{Accuracy of the Cardinality Analysis}
\label{accuracy}
\centering
\begin{tabular}{||l|l|r||r|r||r|r||r|r||r|r||}
 \multicolumn{11}{c}{}  \\ 
\cline{1-11}
\multicolumn{1}{||c|}{}  &\multicolumn{1}{c|}{}   & \multicolumn{1}{c||}{} & \multicolumn{2}{c||}{} & \multicolumn{2}{c||}{} & 
\multicolumn{2}{c||}{} 
& \multicolumn{2}{c||}{} \\
\multicolumn{1}{||c|}{}  &\multicolumn{1}{c|}{}   & \multicolumn{1}{c||}{} & \multicolumn{2}{c||}{P} & \multicolumn{2}{c||}{C} & 
\multicolumn{2}{c||}{PC} 
& \multicolumn{2}{c||}{PCA} \\
\cline{1-11}
& & & & & & & & & & \\
Programs &  Query & NP & D & \%D & D & \%D & D & \%D & D & \%D \\
 \cline{1-11}
& & & & & & & & & & \\
Qsort & {\tt qsort(g,v)} & 3 & 0 & 0 & 0 & 0 & 0 & 0 & 3 & 100 \\
Qsort2 & {\tt qsort(g,v)} & 5 & 2 & 40 & 2 & 40 & 2 & 40 & 5 & 100 \\
Queens &{\tt queens(g,v)} &5 & 2 & 40 & 0 & 0 & 2 & 40 & 2 & 40 \\
Press1 & {\tt test\_press(v,v)}&
47 & 8 & 17 & 19 & 40 & 19 & 40 & 19 & 40 \\
Press2 & {\tt test\_press(v,v)}&
47 & 12 & 26 & 19 & 40 & 28 & 60 & 28 & 60 \\
Gabriel &  {\tt main(v,v)}&
17 & 0 & 0 & 4 & 24 & 4 & 24 & 4 & 24 \\
Peep &  {\tt comppeeppopt(g,v,g)}&
24 & 4 & 17 & 7 & 29 & 16 & 67 & 16 & 67 \\
Read & {\tt read(v,v)} &
46 & 11 & 24 & 27 & 59 & 31 & 67 & 31 & 67 \\
Kalah & {\tt play(v,v)}& 46 & 16 & 35 & 20 & 43 & 33 & 72 & 40 & 87 \\
Cs & {\tt pgenconfig(v)} & 32 & 11 & 34 & 7 & 22 & 11 & 34 & 13 & 41 \\
Plan & {\tt transform(g,g,v)} &
13 & 1 & 8 & 0 & 0 & 1 & 8 & 1 & 8 \\
Disj & {\tt top(v)} & 28 & 13 & 46 & 11 & 39 & 13 & 46 & 13 & 46 \\
Pg & {\tt pdsbm(g,v)} & 10 & 2 & 20 & 3 & 30 & 5 & 50 & 6 & 50 \\
Boyer & {\tt boyer(g)} & 24 & 0 & 0 & 20 & 83 & 20 & 83 & 20 & 83 \\
Credit & {\tt credit(a,a)} &
26 & 14 & 58 & 11 & 42 & 14 & 54 & 16 & 62 \\
       &   &   &      &    &    &  &  &  &       &         \\
\cline{1-11}
 & & & & & & & & & &\\
Mean &    &   & & 24 &   &  33 &  & 46 &   & 58 \\
\cline{1-11}
\end{tabular}
\end{table}

\vskip0.3cm
\noindent

\noindent
{\bf Accuracy.}
The accuracy results are reported in Table \ref{accuracy}. 
For each program we speci\-fy the initial query to which the abstract
interpretation algorithm is applied (we denote  by {\tt a}, {\tt g}
and {\tt v} the modes {\tt any}, {\tt ground} and {\tt var}, respectively).
Several
versions of the algorithm are compared with respect to their ability to
detect
determinacy of procedures, which was our primary motivation. {\tt P}
is using only the domain {\tt Pattern} (i.e., cuts are ignored), {\tt
C} is only using the cut (i.e.,  {\tt EXCLUSIVE} always returns false),
and {\tt PC, PCA} are defined as previously.
In the table, NP stands for the number of procedures and
 D and \%D denote
the number of  procedures and the percentage of
procedures, respectively,  that are detected to be
deterministic
 by the algorithms.
 There are several
interesting points to notice.  First, {\tt PCA} detects that 58\% of
the procedures are deterministic, although many of these programs in
fact use heavily the nondeterminism of Prolog. Most of the results are
optimal and a nice example is the program {\tt Kalah}. Second, the
cut and input/output patterns are really complementary to improve the
analysis. Input/output patterns alone give 41\% of the deterministic
procedures (i.e., those detected by {\tt PCA}), 
while the cut detects 57\%  of the deterministic
procedures. The abstraction of arithmetic predicates 
adds 21\% of deterministic procedures\footnote{Notice that
24/58=0.41, 33/58=0.57 and (58-46)/58=0.21.
 The inequality  41+57+21$\not=$100  can be understood by the fact
 that the analysis computed by P, C and A (the latter being the algorithm that
only considers the
 arithmetic predicates)
 are not completely exclusive.}. 
The main lesson here is that all
 components are of primary importance to obtain precise results.

\section{Retaled works on determinacy analysis}
\label{sec:RDASBAI}

Determinacy of logic programs in general and of Prolog programs in
particular
is an important research topic because determinate programs can be
implemented more efficiently than non-determinate programs 
(often, much more efficiently).
Several forms of determinacy have been identified, which lead to
different kinds of optimizations.
 In this section, we review a few
interesting papers on determinacy analysis at the light of our novel
framework for the abstract interpretation of Prolog.
The benefit of this study is twofold:
first, it sheds new light on these analyses in the
context of abstract interpretation;
second, it supports the claim that our proposal is appropriate to
integrate most existing analyses into a single framework.

\subsection{Sahlin's Determinacy Analysis for Full Prolog}\label{sub:SDAFP}

The analysis proposed by D. Sahlin  \shortcite{Sahlin91} aims at
detecting  procedures of a (full) Prolog program that are determinate
(i.e., they succeed at most once) or fully-determinate
(i.e., they succeed exactly once).
The analysis is developed in the context of the partial evaluator
Mixtus \cite{Sahlin.PhD} in order to detect situations where cuts
can be ``executed'' or removed.
 Sahlin's analysis  is not based on abstract interpretation;
 hence he provides a specific correctness proof for it. \\
In this section, we show that the determinacy analysis
proposed by Sahlin  \shortcite{Sahlin91}  is indeed an instance of our
framework over his abstract domain.

\vskip0.3cm
\noindent
{\bf Abstract Domains.}
\noindent
 Sahlin's analysis completely ignores information
on program variables. The
 abstract domains are
concerned with the sequence structure  only: substitutions are completely
ignored. Note that no abstract interpretation framework
available at the time of his writing
was adequate to his needs.

\vskip0.3cm
\noindent
{\bf Abstract Substitutions.}
Since program variables are ignored, we can assume a domain
${\it AS}$ consisting of an arbitrary single element.

\vskip0.3cm
\noindent
{\bf Abstract Sequences.}
Sahlin's analysis can be formalized in our framework by defi\-ning
 ${\it ASS}=\wp({\it AASS})$, where 
${\it AASS}=\{{\cal L}, 0, 1, 1', 2, 2'\}$\footnote{We 
choose to denote the elements of {\it AASS} by the same
  symbols as in \cite{Sahlin91}.}.
We call elements of {\it AASS}, {\em atomic abstract sequences}.
Their concretization is defined as follows:
\vskip0.2cm
\begin{quote}
  $\begin{array}{lllllll}
     {\it Cc}({\cal L}) & = & \{ <\bot> \} \\
     {\it Cc}(0)       & = & \{  \ <>  \   \} \\
   {\it Cc}(1)       & = & \{  S\in {\it PSS} \mid {\it Ns}(S)=1
                               \and S\mbox{\rm ~is~finite}\} \\
     {\it Cc}(1')      & = & \{  S\in {\it PSS} \mid {\it Ns}(S)=1
                               \and S\mbox{\rm ~is~incomplete}\} \\
      {\it Cc}(2)      & = &
          \{  S\in {\it PSS} \mid {\it Ns}(S)>1
                               \and S\mbox{\rm ~is~finite}\} \\
     {\it Cc}(2')      & = & 
          \{  S\in {\it PSS} \mid {\it Ns}(S)>1
                               \and S\mbox{\rm ~is~incomplete~or~infinite}\}
  \end{array}$
\end{quote}
\vskip0.2cm
\noindent
The concretization function ${\it Cc}:{\it ASS}\rightarrow \wp({\it
  PSS})$ is defined by:
\begin{quote}
  $\begin{array}{lll}
     {\it Cc}(B) &  = & 

      \bigcup_{b\in B} {\it Cc}(b).

  \end{array}$
\end{quote}
\noindent The relation $\leq$ on {\it ASS} is naturally defined as 
being set inclusion. The
concretization function is thus clearly monotonic.

\vskip0.3cm
\noindent
{\bf Abstract Sequences with Cut Information.}
We define the set ${\it ASSC}$ as being
 equal to $\wp({\it AASS}\times {\it CF})$.
 The elements of ${\it ASSC}$ are denoted by
${\cal L}_{n}$, $0_{n}$, $1_{n}$, $1'_{n}$, $2_{n}$, $2'_{n}$,
${\cal L}_{c}$, $0_{c}$, $1_{c}$, $1'_{c}$, $2_{c}$, $2'_{c}$,
in \cite{Sahlin91}, where the index $n$ stands for {\it nocut}, while the index
$c$ stands for {\it cut}.
The concretization function is defined in the obvious way.

\vskip0.3cm
\noindent
{\bf Extended Widening.}
\noindent
In order to instantiate our generic abstract interpretation algorithm
to the above domains, it remains to provide an implementation of the
various abstract operations. 
This can be done systematically from the specifications of the
operations
and the domain
definitions; we leave it as an exercise to the reader, except for 
the extended widening, whose implementation is not obvious.
The basic intuition behind the extended widening is that it should
``observe'' how the abstract sequences evolve between the consecutive iterations
 in order to ensure convergence when enough accuracy seems to be
attained.
In this abstract domain, the abstract sequence $B_i$ produced at
step $i$ may intuitively differ from $B_{i-1}$ by the fact that some
``incomplete'' elements (i.e., ${\cal L}$,  $1'$,  $2'$) can be removed
and replaced by more ``complete'' ones.
Of course the computation starts with $B_0=\{{\cal L}\}$.
Thus the algorithm waits until ``enough incomplete elements have been removed''
and then  accumulates the next iteration results to enforce termination.
This can be formalized by defining a pre-order 
$\cleq$ over {\it ASS}
such that $B_1\cleq B_2$ holds when $B_2$ only contains elements that
are ``more complete'' than some elements of $B_1$ and when, conversely,
$B_1$ only contains elements that
are ``less complete'' than some elements of $B_2$.
We first define the relation {\em is strictly less complete than} between
atomic abstract sequences by the table:

\begin{quote}
  $\begin{array}{llllllllll}

   {\cal L} \clt 0 &\; {\cal L} \clt 1 & \;{\cal L} \clt 1' & \;
{\cal L}   \clt 2 &\;
   {\cal L} \clt 2'&\;
   1' \clt 1      & \; 1'   \clt 2        & \; 1'  \clt  2' &\;
   2' \clt 2.    
  \end{array}$
\end{quote}

\noindent
Then, for all atomic abstract sequences $b_1$ and $b_2$, we say that
{\em $b_1$ is less complete than $b_2$}, denoted by
$b_1\cleq b_2$, if $b_1 = b_2$ or $b_1\clt b_2$.
This relation is lifted to general abstract sequences as follows:

\begin{definition}[Computational Pre-Ordering]
\label{def:CO}
\noindent Let $B_1,B_2\in {\it ASS}$. By definition,
\begin{quote}
  $\begin{array}{lll}
   B_1\cleq B_2 & \myiff & (\forall  b_1\in B_1,\,
                         \exists b_2\in B_2 \mbox{ such that }
 b_1 \cleq b_2) \and\\
                      & &   (\forall  b_2\in B_2,\,
                         \exists b_1\in B_1
\mbox{ such that } b_1 \cleq b_2).
\end{array}$
\end{quote}

 We  write $B_1\clt B_2$ to denote the condition
($B_1 \cleq B_2 \and B_2 \not\cleq B_1$).
\end{definition}
\vskip-0.3cm

We are now in position to define the extended widening.

\begin{definition}
[Extended Widening for Sahlin's Domain: $B'=B_{\it new}\nabla B_{\it old}$]
\label{imp:EWSD}
\begin{quote}
  $\begin{array}{llll}
   B' & = & B_{\it new}         & \myif   B_{\it old} \clt  B_{\it new},\\
       & = & B_{\it new}\cup B_{\it old} & \otherwise.
  \end{array}$
\end{quote}
\end{definition}

In fact, the above operation does not
fulfill, strictly speaking, the requirements for being an extended
widening. It works however if we have $B_{\it old}\cleq B_{\it new}$ each time it is
applied. This is normally the case if the other abstract
operations are carefully implemented, since each iteration of the
abstract interpretation algorithm should
 replace every element in
$B_{\it old}$ by one or several more complete elements.
Before stating 
what  is it actually achieved by the 
operation $\nabla$, we need two definitions.

\begin{definition}[Equivalent Abstract Sequences]
\label{def:EAS}
Let $B_1, B_2\in {\it ASS}$. By definition,
\begin{quote}
  $\begin{array}{lll}
   B_1 \eqass B_2 & \myiff & B_1 \cleq B_2 \and     B_2 \cleq B_1  .
  \end{array}$
\end{quote}
\end{definition}

 The relation $\eqass$ is an equivalence because $\cleq$ is a
pre-order. It can be shown that $\eqass$ determines $42$ equivalence
classes, of which 28 are a singleton (e.g., $\{\{{\cal L}, 0, 1'\}\}$),
10 have 2 elements (e.g., $\{\{{\cal L}, 0, 2'\},\{{\cal L}, 0, 1',
2'\}\}$),
and 4 have 4 elements (e.g., $\{\{{\cal L}, 0, 2\},
                                \{{\cal L}, 0, 2, 2'\},
                                \{{\cal L}, 0, 1', 2\},
                               \{{\cal L}, 0, 1', 2, 2'\}\}$).
It is also important to note that distinct equivalent abstract
sequences always have different concretizations.

\begin{definition}[Strengthened Computational Ordering]
\label{prop:SCO}
\noindent
Let $B_1, B_2\in {\it ASS}$. By definition,
\begin{quote}
  $\begin{array}{lll}
     B_1 \scleq B_2 & \myiff & B_1\clt B_2 \myor 
                          (B_1 \eqass B_2 \and B_1 \subseteq B_2).
   \end{array}$
\end{quote}
\end{definition}

The relation $\scleq$ is an order;  every ascending
sequence
$B_1 \scleq B_2 \scleq \dots \scleq B_i \dots\,$ is stationary 
 since {\it ASS} is finite.

\begin{property}[Conditional Convergence of the
  Extended Widening]
\label{prop:CCEW}
Let $\{B_i\}_{i\in{\bf N}}$ and $\{B'_i\}_{i\in{\bf N}}$ be two
sequences of elements of {\it ASS} such that
\begin{enumerate}
\itemsep 3pt
\item $B'_i \cleq B_{i+1},$ for all $i\in{\bf N}$;
\item $B'_{i+1}=B_{i+1}\nabla B'_i,$ for all $i\in{\bf N}$.
\end{enumerate}

\noindent Then we have $B_i\leq B'_i$, for all $i\in{\bf N}^\ast$, 
and the sequence
$\{B'_i\}_{i\in{\bf N}}$ is stationary.
\end{property}

\begin{proof}
The fact that $B_i\leq B'_i$, for all $i\in{\bf N}^\ast$, is a direct
consequence of the definition of the operation $\nabla$.
Moreover, the hypotheses on the sequences ensure that 
$B'_1\scleq B'_2 \scleq \dots \scleq B'_i \dots\,$; thus the sequence
$\{B'_i\}_{i\in{\bf N}}$ is stationary.
\end{proof}

\vskip0.3cm

If all abstract operations are congruent with respect to $\cleq$
\footnote{We would have written {\em monotonic} if the relation $\cleq$ was an
  order, not a pre-order only.}, 
each iteration of the 
abstract interpretation algorithm ensures that $B_{\it old}\cleq B_{\it new}$,
where $B_{\it old}$ is the current value in {\it sat} and $B_{\it new}$ is the newly
computed
abstract sequence. Thus, Proper\-ty~\ref{prop:CCEW} guarantees
termination of the abstract interpretation algorithm.
Congruence of the abstract operations with respect to $\cleq$
is ensured if they are
``as accurate as possible'' (which is achieved in \cite{Sahlin91});
however,
proving this property entails a lot of work.
 A
simpler solution  consists of testing whether $B_{\it old}\cleq B_{\it new}$
actually holds before each application of the extending widening. If
the condition does not hold, we switch to a cruder form of widening, which
simply merges all successive results.

\vskip0.3cm
\noindent
{\bf
Comparison with our Cardinality Analysis.}
The determinacy information inferred by means of Sahlin's domain is in general
less accurate than our cardinality analysis (except maybe in some
partial evaluation contexts). For instance, with the former domain, it
is not possible to detect mutually exclusive clauses except when cuts
occur in the clauses. As illustrated in Section \ref{sub:IP}, the
 information provided by the abstract substitution
component of our domain is instrumental to detect sure failure, sure
success, and mutual exclusion, which all contribute to improve the
accuracy of the determinacy (or cardinality) analysis.
Nevertheless, the specific information about the sequence structure is
finer grained in Sahlin's domain than in ours. Consider the abstract
sequence $\{ {\cal L}, 1\}$; it is approximated, in our domain, by 
$\langle 0, 1, {\it pt} \rangle$, which is actually equivalent to
$\{ {\cal L}, 0, 1, 1'\}$. Thus, it could be interesting to design a
domain for abstract sequences similar to our cardinality domain, 
where  
the sequence component coincides with Sahlin's domain.

\vspace{-0.2cm}
\subsection{Giacobazzi and Ricci's Analysis of Determinate Computations}
\label{sub:GRADC}

The work of  Giacobazzi and  Ricci  \shortcite{Giacobazzi92},
is also worth being reviewed in our context.
They propose an analysis of functional dependencies \cite{Mendelzon85}
between procedure arguments
of the success set of pure logic programs. Their
ana\-ly\-sis is a bottom-up abstract interpretation, based on
\cite{Barbuti90,FLMP89}. The analysis also infers groundness
information and is intended to be used for parallel logic program
optimization.
In our comparison, we focus on the functional dependencies and we
 simplify the presentation in order to concentrate on the
salient points. 
First, we provide a definition of functional dependency tailored to
our framework. The definitions use some notions from Section \ref{sub:IP}.

\begin{definition}[Functional Dependency]
\label{def:FD}
\noindent
Let $\langle {\it sv}, {\it frm} \rangle$ be a structural abstract
substitution over domain $D$ and set of indices $I$. A {\em functional
dependency for} $\langle {\it sv}, {\it frm} \rangle$,
denoted by $J\rightarrow j$, is a pair consisting of a subset $J$ of
$I$ and an index $j\in I$. 

\noindent Let $S\in{\it PSS}_D$ be a program substitution sequence
such that ${\it Subst}(S)\subseteq {\it Cc}\langle {\it sv}, {\it frm}
\rangle$.
We say that the functional dependency $J\rightarrow j$ {\em holds  in
$S$ for $\langle {\it sv}, {\it frm} \rangle$}, if for all fami\-lies of
terms $(t_i)_{i\in I}$, $(t'_i)_{i\in I}$ that are decompositions of
some program substitutions of ${\it Subst}(S)$, the
following implication is true:
\begin{quote}
  $\begin{array}{lll}
(t_i)_{i\in J}=(t'_i)_{i\in J} & \Rightarrow & t_j=t'_j.
  \end{array}$
\end{quote}
\end{definition}

Then we  define an abstract domain to express functional
dependencies.

\begin{definition}[Abstract Sequences with Functional
  Dependencies]
\label{def:ASFD}

\noindent  {\em An abstract sequence with functional
  dependencies} is a triple $\langle {\it sv}, {\it frm}, {\it fd}
\rangle$ where $\langle {\it sv}, {\it frm} \rangle$ is a structural abstract
substitution over  domain $D$ and set of indices $I$, and {\it fd}
is a set of functional dependencies for $\langle {\it sv}, {\it frm} \rangle$. 
The concretization function for abstract sequences with functional 
 dependencies is defined by
\vskip0.2cm
 \begin{quote}

   $\begin{array}{lll}
   {\it Cc}\langle {\it sv}, {\it frm}, {\it fd} \rangle & = &

\left\{\begin{array}{l}

      S\in {\it PSS}_D \ \ 

     \begin{array}{|l}

         {\it Subst}(S)\subseteq \Cc{\langle {\it sv}, {\it frm} \rangle} \and\\

         J\rightarrow j \mbox{\rm ~holds~in~}
         S \mbox{\rm~for~} \langle {\it sv}, {\it frm} \rangle,\\

         \mbox{\rm for~every~}J\rightarrow j\in {\it fd}.
                                  
     \end{array}

\end{array}\right\}.

   \end{array}$
 \end{quote}

\end{definition}
\vskip0.3cm

In fact, the functional dependency component {\it fd} is best
viewed as an additional component to the cardinality domain defined in Section
\ref{sec:FASAS}, since its usefulness for determinacy analysis depends
on the availability of mode information. Let $S\in{\it CPSS}_D$ be a
canonical program substitution sequence. We say that $S$ is {\em functional}
if the set ${\it Subst}(S)$ is empty or is a singleton.
Such sequences model the behavior of procedures that cannot produce
two or more distinct solutions. Assume that $S$ is the output sequence
corresponding to the input substitution $\theta$, for some procedure $p$.
Assume that $\theta\in{\it Cc}\langle {\it sv}, {\it frm} \rangle$ and
$S\in{\it Cc}\langle {\it sv}', {\it frm}', {\it fd}' \rangle$ where 
$\langle {\it sv}', {\it frm}' \rangle$ is more instantiated than
$\langle {\it sv}, {\it frm} \rangle$
. We can
infer that $S$ is functional if there exists $J\subseteq I'$ such
that ${\it fd}'$ contains a functional dependency of the form
$J\rightarrow i$, for every $i\in {\it sv}'(D)$, and if every term $t_j$
corresponding to an index $j\in J$ in a program substitution of $S$ is
not more instantiated than the corresponding term in $\theta$. The
latter information is easily deduced if we know, for instance, that
$t_j$ is ground or is a variable. Thus adding a functional dependency
component to our cardinality domain allows us to infer that output
program substitution sequences are functional.

It is important to point out that the new component {\it fd} expresses a
property of program substitution sequences, not a property of (single) program
substitutions.
 It is meaningless  to use functional dependencies
in a domain of abstract substitutions, because a set of functional
dependencies determines a (two valued) condition
 on a set of program substitution.
Either the set verifies the condition, then no constraint is added, or
it does not and the set is rejected as a whole.
Thus, a component {\it fd} defines a {\em set of sets of} program
substitutions.
As a consequence, functional
dependencies cannot be handled by previous top-down abstract
interpretation frameworks such as \cite{Bruynooghe91,TOPLAS,Marriott89,%
Mellish87,Muthukumar92,Warren92,Winsborough92}.
However the abstract interpretation framework used by \cite{Giacobazzi92}
is bottom-up and abstracts the success set of the program. The result
of an analysis represents a set of possible success sets, i.e.,
{\em a set of sets of output patterns}, which is similar to a set of
sets of program substitutions.
As far as we know, it is the first time that this difference of
expressivity between bottom-up and (previous) top-down abstract
interpretation frameworks is pointed out in the literature. 
The comparison usually concentrates on the fact that bottom-up
frameworks are {\em goal independent}, i.e., they provide information on
the program as a whole, while top-down frameworks are {\em goal dependent},
i.e., they provide information about the program {\em and} a given initial
goal. We believe that a more fundamental difference lies in the fact that
 top-down frameworks are {\em functional}, i.e., they abstract
the behavior of a program by a function between sets of sets, while bottom-up
frameworks are {\em relational}, i.e., they abstract  the behavior of
a program by a set of relations. The difference between the two
approaches has been previously put forward 
by Cousot and Cousot \shortcite{CousotJLC92}, but not in the context of logic programs.
The functional approach can easily focus on small parts of the program
behavior but looses the dependencies between inputs and outputs;
the converse holds for the relational approach.
Our novel framework is basically functional, but the domain of
abstract sequences is in some sense relational; thus the framework allows us to combine the
advantages of both approaches.

\subsection{Debray and Warren's Analysis of Functional Computations}

In the previous section, we have shown that functional dependencies
are useful to infer that an output program substitution sequence is functional,
 i.e., does not contain two or more distinct program substitutions.
Such a sequence may contain several occurrences of the same program
substitution, however.
The importance of functional computations for logic program
optimization was advocated early by Debray and Warren 
\shortcite{Debray89a}. 
In this paper, these authors propose a sophisticated algorithm to infer
functional computations of a logic program. The analysis exploits
functional dependencies and mode information, as well as a set of
sufficient conditions to detect mutually exclusive clauses.
Their algorithm is not based on abstract interpretation and assumes
that functional dependencies and mode information are given from
outside.
Thus the algorithm considers an annotated program; it uses a set
$\{ \bot, {\bf true}, {\bf false} \}$ where $\bot$ is an initializing value, 
{\bf true} means that a procedure is functional and {\bf false} means
that it is not known whether the procedure is functional. Hence, the
set can be viewed as a domain of abstract sequences, with
concretization function 
${\it Cc} : \{ \bot, {\bf true}, {\bf false}\} \rightarrow \wp({\it CPSS})$ 
defined by
\begin{quote}
  $\begin{array}{lll}
   {\it Cc}(\bot)       & = & \{<\bot>\};\\
   {\it Cc}({\bf true})  & = & \{S\in{\it CPSS}\mid \ 
                               {\it Subst}(S) 
                                \mbox{\rm~is~empty~or~is~a~singleton.}
                             \};\\
   {\it Cc}({\bf false}) & = & {\it CPSS}.
        \end{array}$
\end{quote}

\noindent
All aspects of their analysis can be accommodated in our approach by providing
suitable abstract domains. An abstract domain consisting of our cardinality
domain augmented with a functional dependency component  would
probably be fairly accurate. 
Moreover, in our approach,  all analyses can be performed at the same
time and interact with each other, making it possible to get
a better accuracy.

\section{Conclusion}
\label{conclusion}

This paper has introduced a novel abstract interpretation framework,
capturing the depth-first search strategy and the cut operation of
Prolog. The framework is based on the notion of substitution sequences
and the abstract semantics is defined as a pre-consistent post-fixpoint
of the abstract transformation. Abstract interpretation algorithms
need chain-closed domains and a special widening operator to compute
the semantics. This approach overcomes some of the limitations of
previous frameworks. In particular, it broadens the applicability of
the abstract interpretation approach to new analyses and 
can potentially improve 
the precision of existing analyses.
On the practical side, in this paper, we have only shown
that our approach allows one to integrate -
efficiently and at a low conceptual cost - a predicate
level analysis (i.e., determinacy analysis)
to variable level analyses classically handled by abstract
interpretation. However, the improvement on classical analyses
is marginal because, due to our design choices for the
abstract sequence domain 
(i.e., a simple extension of {\tt Pattern}), the new system
behaves almost as the previous
version of GAIA for variable level analyses.
Nevertheless, the new framework opens a door for defining
and exploiting more sophisticated domains for abstract sequences.

\bibliographystyle{tlp}
\bibliography{Biblio}

\begin{thebibliography}{}

\bibitem[\protect\citename{Apt, }1997]{Apt97}
Apt,~K.~R. (1997)
\emph{From Logic Programming to Prolog}.
International Series in Computer Science, Prentice Hall.


\bibitem[\protect\citename{Barbuti \emph{et al.}, }1993]{BarbutiControl}
Barbuti,~R.,  Codish,~M., Giacobazzi,~R. and Levi,~G. (1993)
Modelling Prolog control.
\emph{Journal of Logic and Computation}, 3~(6): 579--603.


\bibitem[\protect\citename{Barbuti and Giacobazzi, }1992]{BG89b}
Barbuti,~R. and Giacobazzi,~R. (1992)
A bottom-up polymorphic type inference in logic programming.
\emph{Science of Computer Programming}, 19~(3): 281--313.

\bibitem[\protect\citename{Barbuti \emph{et al.}, }1993]{Barbuti90}
Barbuti, R., Giacobazzi,~R.  and Levi, G. (1993)
A general framework for semantics-based bottom-up abstract
  interpretation of logic programs.
\emph{ACM Transactions on Programming Languages and Systems (TOPLAS)},
  15~(1): 133-181.


\bibitem[\protect\citename{Baudinet, }1992]{Baudinet}
Baudinet,~M. (1992)
Proving  termination properties of Prolog programs: a semantic approach.
\emph{Journal of Logic Programming}, 14~(1\&2): 1--29.


\bibitem[\protect\citename{Bossi and Cocco, }1999]{BC99}
Bossi,~A. and Cocco,~N. (1999)
Successes in logic Programs.
In P.~Flener (editor), \emph{Proc. of the 8th  International Workshop on 
 Logic-Based Program Synthesis and Transformation
  ({LOPSTR}'98), Lecture Notes in Computer Science, 1559}, 
pp.~219--239. Springer-Verlag.





\bibitem[\protect\citename{Braem \emph{et al.}, }1994]{cardinality}
Braem,~C., Le~Charlier,~B.,  Modard,~S. and  Van~Hentenryck, P. (1994)
Cardinality analysis of Prolog.
In M.~Bruynooghe (editor), \emph{Proc. of the International Logic
  Programming Symposium ({ILPS}'94)},
pp.~457--471. MIT Press.


\bibitem[\protect\citename{Braem and Modard, }1994]{Braem.Modard94}
Braem,~C. and Modard,~S. (1994)
Abstract interpretation for Prolog with cut: cardinality analysis.
Master's thesis, Institut d'Informatique, University of Namur,
  Belgium.

\bibitem[\protect\citename{Bruynooghe, }1991]{Bruynooghe91}
Bruynooghe,~M. (1991)
A practical framework for the abstract interpretation of logic
  programs.
\emph{Journal of Logic Programming}, 10~(2): 91--124.

\bibitem[\protect\citename{Bueno and Hermenegildo, }1991]{Bueno91b}
Bueno,~F. and Hermenegildo,~M. (1991)
Results on automatic translation from prolog
to the Andorra kernel language.
Technical Report, Facultad Informatica UPM,
 Universidad Politecnica de Madrid, Spain.


\bibitem[\protect\citename{Cabeza~Gras and Hermenegildo, }1994]{SAS94.Gras}
Cabeza~Gras,~D. and   Hermenegildo,~M. (1994)
Extracting non-strict independent And-parallelism
using sharing and freeness information.
In \cite{Book.SAS94}, pp.~297--313.



\bibitem[\protect\citename{Chang \emph{et al.}, }1985]{Chang85}
Chang,~J.~H.,  Despain,~A.~M. and  DeGroot,~D. (1985)
And-parallelism of logic programs based on a static
data dependency analysis. 
In {\em Proc. of the 30th IEEE Compcon Spring
({COMPCON}'85)}. IEEE  Press.


\bibitem[\protect\citename{Codish \emph{et al.}, }1991]{Codish91}
Codish,~M., Dams,~D. and Yardeni,~E. (1991)
Derivation and safety of an abstract unification algorithm for
  groundness and aliasing analysis.
In K.~Furukawa (editor), \emph{Proc.  of the 8th International
  Conference on Logic Programming ({ICLP}'91)},
pp.~79--93.  MIT Press.

\bibitem[\protect\citename{Codognet and Fil\`e, }1992]{Codognet92a}
Codognet,~P. and Fil\`{e},~G. (1992)
Computations, abstractions and constraints in logic programs.
 In \emph{Proc. of IEEE  International Conference on
  Computer Languages (ICCL'92)}. IEEE Press.

\bibitem[\protect\citename{Corsini, }1991]{Corsini91}
Corsini,~M.-M. (1991)
(Yet) an abstract domain and unification for accurate groundness and
  sharing analysis based on graphs traversing. 
In \emph{{ICLP}'91 Pre-Conference Workshop on Semantics-Based Analysis
  of Logic Programs}, {INRIA} Rocquencourt.

\bibitem[\protect\citename{Cortesi and Fil\`e, }1991]{Cortesi}
Cortesi,~A. and Fil\`{e},~G. (1991)
Abstract interpretation of logic programs: an abstract domain for
  groundness, sharing, freeness and compoundness analysis. 
In \emph{Proc. of
 the Symposium on Partial Evaluation and Semantics-Based
  Program Manipulation ({PEPM}'91)}, SIGPLAN Notices
26~(9): 52--61. 


\bibitem[\protect\citename{Cortesi \emph{et al.}, }1991]{Cortesi91}
Cortesi,~A., Fil\`e,~G. and Winsborough,~W. (1991)
Prop revisited: propositional formula as abstract domain for
  groundness analysis.
In \emph{Proc. of the 6th Annual IEEE Symposium on Logic in
  Computer Science (LICS'91)},  pp.~322--327. 
IEEE Computer Society Press.


\bibitem[\protect\citename{Cortesi \emph{et al.}, }1994]{POPL94}
Cortesi,~A., Le~Charlier, B., and Van~Hentenryck, P. (1994)
Combinations of abstract domains for logic programming.
 In \emph{Proc. of the  21th {ACM SIGPLAN-SIGACT} Symposium on
  Principles of Programming Languages ({POPL}'94)}, pp.~227--239.
ACM Press.



\bibitem[\protect\citename{Cortesi \emph{et al.}, }1995]{Graph.JLP}
Cortesi,~A., Le~Charlier,~B. and Van~Hentenryck,~P. (1995)
Type analysis of Prolog using type graphs.
\emph{Journal of Logic Programming}, 22~(3): 179--209.

\bibitem[\protect\citename{Cortesi \emph{et al.}, }2000]{SCP00}
Cortesi,~A., Le~Charlier, B., and Van~Hentenryck, P. (2000)
Combinations of abstract domains for logic programming:
 open product and generic pattern construction.
\emph{Science of Computer Programming}, to appear.



\bibitem[\protect\citename{Cousot and Cousot, }1977]{Cousot77}
Cousot,~P. and Cousot,~R. (1977)
Abstract Interpretation: a unified lattice model for static analysis
  of programs by construction or approximation of fixpoints.
 In {\em Proc.  of the 4th {ACM SIGPLAN-SIGACT} Symposium on
  Principles of Programming Languages ({POPL}'77)}, pp.~238--252.
ACM Press.



\bibitem[\protect\citename{Cousot and Cousot, }1979]{Cousot79a}
Cousot,~P. and Cousot,~R. (1979)
Systematic design of program analysis frameworks.
 In {\em Proc.  of the 6th {ACM SIGPLAN-SIGACT} Symposium on
  Principles of Programming Languages ({POPL}'79)}, pp.~269--282.
ACM Press.

 

\bibitem[\protect\citename{Cousot and Cousot, }1992a]{Cousot92}
Cousot,~P. and Cousot,~R. (1992)
Abstract interpretation and application to logic programs.
\emph{Journal of Logic Programming}, 13~(2\&3): 103--179.

\bibitem[\protect\citename{Cousot and Cousot, }1992b]{CousotJLC92}
Cousot,~P. and Cousot,~R. (1992)
Abstract interpretation frameworks.
\emph{Journal of Logic and Computation}, 2~(4): 511--547.


\bibitem[\protect\citename{Cousot and Cousot, }1992c]{Cousot92c}
Cousot,~P. and Cousot,~R. (1992)
Comparing of the Galois connection and widening/narrowing
  approaches to abstract interpretation (invited paper).
In M.~Bruynooghe and M.~Wirsing (editors), \emph{Proc. of the
  4th International Workshop on Programming Language Implementation and
  Logic Programming ({PLILP}'92), Lecture Notes in Computer Science, 631}, 
pp.~269--295. Springer-Verlag.


\bibitem[\protect\citename{Cousot and Cousot, }1994]{CousotICCL94}
Cousot,~P. and Cousot,~R. (1994)
Higher-order abstract interpretation (and application to comportment
  analysis generalizing strictness, termination, projection and PER analysis of
  functional languages). (Invited paper).
In \emph{Proc. of IEEE  International Conference on
  Computer Languages ({ICCL}'94)}, pp.~95--112.
 IEEE Press.



\bibitem[\protect\citename{Dawson \emph{et al.}, }1993]{Ramakrishnan93}
Dawson,~S.,  Ramakrishnan,~C.~R.,  Ramakrishnan,~I.~V. and  Sekar,~R.~C.
(1993)
Extracting determinacy in logic programs.
In D.~S.~Warren (editor), \emph{Proc.  of the 10th International
  Conference on Logic Programming ({ICLP}'93)},
pp.~424--438.  MIT Press.


\bibitem[\protect\citename{Debray, }1989]{D89}
Debray,~S.~K. (1989)
Static inference of modes and data dependencies in logic programs.
\emph{ACM Transactions on Programming Languages and Systems (TOPLAS)},
  11~(3): 418--450.



\bibitem[\protect\citename{Debray \emph{et al.}, }1997]{DLH97}
Debray,~S.~K.,  L\'opez-Garc\'ia,~P., Hermenegildo,~M. (1997)
Non-failure analysis for logic programs.
In L.~Naish (editor), \emph{Proc.  of the 14th International
  Conference on Logic Programming ({ICLP}'97)},
pp.~48--62.  MIT Press.




\bibitem[\protect\citename{Debray and Mishra, }1988]{Debray88b}
Debray,~S.~K. and Mishra,~P. (1988)
Denotational and operational semantics for Prolog.
\emph{Journal of Logic Programming}, 5~(1): 61--91.

\bibitem[\protect\citename{Debray and Warren, }1988]{Debray88c}
Debray,~S.~K. and  Warren,~D.~S. (1988)
Automatic mode inference for logic programs.
\emph{Journal of Logic Programming}, 5~(3): 207--229.

\bibitem[\protect\citename{Debray and Warren, }1989]{Debray89a}
Debray,~S.~K. and Warren,~D.~S. (1989)
Functional computations in logic programs.
\emph{ACM Transactions on Programming Languages and Systems (TOPLAS)},
  11~(3): 451--481.




\bibitem[\protect\citename{De~Bruin and De~Vink, }1989]{debruin}
De~Bruin,~A. and De~Vink,~E.
Continuation semantics for Prolog with cut.
In J. D\'iaz and F. Orejas (editors),
\emph{Proc. of the International Joint Conference on Theory and
Practice of Software Develpment ({TAPSOFT}'89),
Lecture Notes in Computer Science, 351}, pp.~178--192. Springer-Verlag.


\bibitem[\protect\citename{Englebert \emph{et al.}, }1993]{SPE}
Englebert,~V., Le~Charlier,~B., Roland,~D. and Van~Hentenryck,~P.
(1993)
Generic abstract interpretation algorithms for Prolog: two
  optimization techniques and their experimental evaluation.
\emph{Software Practice and Experience}, 23~(4): 419--459.

\bibitem[\protect\citename{Falaschi \emph{et al.}, }1989]{FLMP89}
Falaschi,~M., Levi,~G., Martelli,~M. and Palamidessi,~C. (1989)
Declarative modeling of the operational behaviour of logic
  languages.
\emph{Theoretical Computer Science}, 69~(3): 289--318.

\bibitem[\protect\citename{Fil\`e and Ranzato, }1994]{FileILPS94}
Fil\`e,~G. and Ranzato,~F. (1994)
Improving abstract interpretations by systematic lifting to the
  powerset.
In M.~Bruynooghe (editor), 
 \emph{Proc. of the International Logic
  Programming Symposium ({ILPS}'94)},
pp.~655--669. MIT Press.



\bibitem[\protect\citename{Fil\`e and Rossi, }1993]{FileRossi}
Fil\`e,~G. and Rossi,~S. (1993)
Static analysis of Prolog with cut.
In A.~Voronkov (editor),
\emph{Proc. of  4th International
Conference on Logic Programming and Automated Reasoning ({LPAR}'93),
Lecture Notes in Computer Science
 698}, pp.~134--145. Springer--Verlag.

\bibitem[\protect\citename{Gang and Zhiliang, }1986]{GZ86}
Gang,~Y. and Zhiliang,~X. (1986)
An efficient type system for Prolog.
In H.~J.~Kugler (editor). {\em Proc. of the 10th IFIP World Computer Congress,
Information Processing 86}, 
pp.~355--359. North-Holland/IFIP.

\bibitem[\protect\citename{Getzinger, }1994]{Getzinger.SAS94}
Getzinger,~T.W. (1994)
The costs and benefits of abstract interpretation-driven Prolog
  optimization.
In \cite{Book.SAS94}, pp.~1--25.



\bibitem[\protect\citename{Giacobazzi and Ricci, }1990]{Giacobazzi90}
Giacobazzi,~R. and Ricci,~L. (1990)
Pipeline optimizations in AND-parallelism by
abstract interpretation.
In  D. S. Warren and P.~Szeridi (editors), 
\emph{Proc.  of the 7th International
  Conference on Logic Programming ({ICLP}'90)},
pp.~291--305.  MIT Press.

\bibitem[\protect\citename{Giacobazzi and Ricci, }1992]{Giacobazzi92}
Giacobazzi,~R. and Ricci,~L. (1992)
Detecting determinate computations by bottom-up abstract
  interpretation.
In B.~Krieg-Br\"uckner (editor),
{\em Proc. of the 4th European Symposium
on Programming, {ESOP'92}, Lecture Notes in Computer Science
 582}, pp.~167--181. Springer--Verlag.


\bibitem[\protect\citename{Hermenegildo, }1986]{Hermenegildo86}
Hermenegildo,~M.~V. (1986)
An abstract machine for restricted AND-parallel execution of
logic programs.
In E.~Y.~Shapiro (editor), \emph{Proc.  of the 3rd International
  Conference on Logic Programming ({ICLP}'86),
 Lecture Notes in Computer Science 225},
pp.~25--40.   Springer--Verlag.




\bibitem[\protect\citename{Hermenegildo \emph{et al.}, }1992]{Hermenegildo92}
Hermenegildo,~M.~V., Warren,~R. and Debray,~S.~K. (1992)
Global flow analysis as a practical compilation tool.
\emph{Journal of Logic Programming}, 13~(4): 349--367.

 

\bibitem[\protect\citename{Jacob and Langen, }1989]{Jacobs89}
Jacobs,~D. and Langen,~A. (1989)
Accurate and efficient approximation of variable aliasing in logic
  programs.
 In E.~L.~Lusk and R.~A.~Overbeek (editors), \emph{Proc. of the
  North American Conference on Logic Programming ({NACLP}'89)}, 
pp.~154--165. MIT Press.


\bibitem[\protect\citename{Jacob and Langen, }1992]{Jacobs92}
Jacobs,~D. and Langen,~A. (1992)
Static analysis of logic programs for 
independent AND parallelism.
\emph{Journal of Logic Programming}, 13~(2\&3): 291--314.



\bibitem[\protect\citename{Janssens and Bruynooghe, }1992]{Janssens92}
Janssens,~G. and Bruynooghe,~M. (1992)
Deriving descriptions of possible values of program variables by
  means of abstract interpretation.
\emph{Journal of Logic Programming}, 13~(2\&3): 205--258.

\bibitem[\protect\citename{Jensen and Mogensen, }1990]{Jensen90}
Jensen,~T.~P. and  Mogensen.~T.~\AE. (1990)
A backwards analysis for compile-time garbage collection.
In  N.~Jones (editor),
{\em Proc. of the 3th European Symposium
on Programming, ({ESOP}'90), Lecture Notes in Computer Science
 432}, pp.~227--239. Springer--Verlag.


\bibitem[\protect\citename{Jones and Mycroft, }1984]{JonesMy}
Jones,~N.~D. and Mycroft,~A. (1984)
Stepwise development of operational and denotational semantics for
  Prolog. In \emph{Proc.  of the International
  Symposium  on Logic Programming ({SLP}'84)},
pp.~281--288.  IEEE-CS.


\bibitem[\protect\citename{Jones and S{\o}ndergaard, }1987]{Jones87}
Jones,~N.~D. and S{\o}ndergaard,~H. (1987)
A semantic-based framework for the abstract interpretation of
Prolog.
In S.~Abramsky and C.~Hankin (editors) \emph{Abstract Interpretation
  of Declarative Languages}, pp.~123--142. 
Ellis Horwood.



\bibitem[\protect\citename{Kanamori and Kawamura, }1987]{Kanamori87}
Kanamori,~T. and Kawamura,~T. (1987)
Analysing success patterns of logic programs by abstract hybrid
  interpretation.
Technical Report 279, ICOT, Tokyo, Japan.


\bibitem[\protect\citename{Kanamori and Horiuchi, }1985]{KH85}
Kanamori,~T. and Horiuchi,~K. (1985)
Type inference in Prolog and its application.
In A.~K.~Joshi (editor),
\emph{Proc. of  9th International Joint Conference
on Artificial Intelligence,  ({IJCAI}'85)  IJCAI}, pp.~704--709.
Morgan Kaufmann.



\bibitem[\protect\citename{Kieburtz, }1983]{K83}
Kieburtz,~R.~B. (1983)
Precise typing of abstract data type specification.
In {\em Proc.  of the 10th {ACM} Symposium on
  Principles of Programming Languages ({POPL}'83)}, pp.~109--116.
ACM Press.


\bibitem[\protect\citename{Klu\'zniak, }1987]{Kluzniak87}
Klu\'zniak,~F. (1987)
Type synthesis for ground Prolog.
In J.-L.~Lassez (editor), \emph{Proc.  of the 4th International
  Conference on Logic Programming ({ICLP}'87)},
pp.~788--816.  MIT Press.



\bibitem[\protect\citename{Klu\'zniak, }1988]{Kluzniak88}
Klu\'zniak,~F. (1988)
Compile-time garbage collection for ground Prolog.
In  R.~A.~Kowalski and K.~A.~Bowen (editor), \emph{Proc.  of the 5th
International Conference on Logic Programming ({ICLP}'88)},
pp.~1490--1505.  MIT Press.



\bibitem[\protect\citename{Le Charlier, }1994]{Book.SAS94}
Le~Charlier,~B. (Ed.). (1994)
\emph{Proceedings of the 1st 
International Static 
Analysis Symposium ({SAS'94}), 
Lecture Notes in Computer Science
864}. Springer-Verlag.


\bibitem[\protect\citename{Le Charlier \emph{et al.}, }1993]{WSA93}
Le~Charlier,~B., Degimbe,~O., Michel,~L. and Van~Hentenryck,~P. (1993)
Optimization techniques for general purpose fixpoint algorithms:
  practical efficiency for the abstract interpretation of Prolog.
In P.~Cousot, M.~Falaschi, G.~Fil\`e and A.~Rauzy
 (editors) {\em Proc. of the 3rd International
  Workshop on Static Analysis ({WSA}'93),
Lecture Notes in Computer Science
 724}, pp.~15--26. Springer--Verlag.


\bibitem[\protect\citename{Le Charlier \emph{et al.}, }1999]{RP-98-002}
Le~Charlier,~B.,  Lecl\`ere,~C.,  Rossi,~S. and  Cortesi,~A. (1999)
Automated verification of Prolog programs.
\emph{Journal of Logic Programming}, 39~(1--3): 3--42.



\bibitem[\protect\citename{Le Charlier \emph{et al.}, }1991]{ICLP91AI}
Le~Charlier,~B., Musumbu,~K. and Van~Hentenryck,~P. (1991)
A generic abstract interpretation algorithm and its complexity
  analysis.
In K.~Furukawa (editor), \emph{Proc.  of the 8th International
  Conference on Logic Programming ({ICLP}'91)},
pp.~64--78.  MIT Press.





\bibitem[\protect\citename{Le Charlier \emph{et al.}, }1996]{Cut_rep95}
Le~Charlier,~B. and Rossi,~S. (1996)
Sequence-based abstract semantics of Prolog.
Technical Report RR-96-001, Facult\'es Universitaires Notre-Dame de
  la Paix, Institut d'Informatique, Namur, Belgium.


\bibitem[\protect\citename{Le Charlier \emph{et al.}, }1994]{CUT94}
Le~Charlier,~B., Rossi,~S. and Van~Hentenryck,~P. (1994)
An abstract interpretation framework which accurately handles
Prolog search-rule and the cut.
In M.~Bruynooghe (editor), \emph{Proc. of the International Logic
  Programming Symposium ({ILPS}'94)},
pp.~157--171. MIT Press.


\bibitem[\protect\citename{Le Charlier \emph{et al.}, }1997]{sequence}
Le~Charlier,~B., Rossi,~S. and Van~Hentenryck,~P. (1997)
Sequence-based abstract interpretation of Prolog.
Technical Report RR-97-001, Facult\'es Universitaires Notre-Dame de
  la Paix, Institut d'Informatique, Namur, Belgium.


\bibitem[\protect\citename{Le Charlier and Van Hentenryck, }1993]{universal}
Le~Charlier,~B. and Van~Hentenryck,~P. (1993)
A general top-down fixpoint algorithm (revised version).
Technical Report RR-93-022, Facult\'es Universitaires Notre-Dame de
  la Paix, Institut d'Informatique, Namur, Belgium.


\bibitem[\protect\citename{Le Charlier and Van Hentenryck, }1994]{TOPLAS}
Le~Charlier,~B. and Van~Hentenryck,~P. (1994)
Experimental evaluation of a generic abstract interpretation
  algorithm for Prolog.
\emph{ACM Transactions on Programming Languages and Systems (TOPLAS)},
  16~(1): 35--101.


\bibitem[\protect\citename{Le Charlier and Van Hentenryck, }1995]{ACTA95}
Le~Charlier,~B. and Van~Hentenryck,~P. (1995)
Reexecution in abstract interpretation of {P}rolog.
\emph{Acta Informatica}, 32~(3): 209--270.



\bibitem[\protect\citename{Leivant, }1983]{L83}
Leivant,~D. (1983)
Polymorphic type inference.
 In \emph{Proc. of the  10th {ACM} Symposium on
  Principles of Programming Languages ({POPL}'83)}, pp.~88--98.
ACM Press.


\bibitem[\protect\citename{Lloyd, }1987]{Lloyd}
 Lloyd,~J. W. (1987)
\emph{Foundations of Logic Programming}.
Springer Series: Symbolic Computation--Artificial Intelligence.
  Springer-Verlag, second edition.


\bibitem[\protect\citename{Marien and Demoen, }1989]{Marien89}
Marien, A. and Demoen,~B. (1989)
On the management of choicepoint and environment frames in the WAM.
In E.~L. Lusk and R. A. Overbeek (editors) {\em Proc. of the
  North American Conference on Logic Programming ({NACLP}'89)}, 
pp.~1030--1047. MIT Press.

\bibitem[\protect\citename{Marien \emph{et al.}, }1989]{Marien89b}
Marien,~A.,  Janssens,~G.,  Mulkers,~A. and  Bruynooghe.~M.
(1989)
The impact of abstract interpretation: an experiment in code
generation.
In G.~Levi and M.~Martelli (editors), \emph{Proc.  of the 6th International
  Conference on Logic Programming ({ICLP}'89)},
pp.~33--47.   MIT Press.



\bibitem[\protect\citename{Marriott, }1993]{Marriott93}
Marriott,~K. (1993)
Frameworks for abstract interpretation.
\emph{Acta Informatica}, 30~(2): 103--129.


\bibitem[\protect\citename{Marriott and S{\o}ndergaard, }1989a]{Marriott89}
Marriott,~K. and S{\o}ndergaard,~H. (1989)
Notes for a tutorial on abstract interpretation of logic programs.
 North American Conference on Logic Programming ({NACLP}'89).


\bibitem[\protect\citename{Marriott and S{\o}ndergaard, }1989b]{Marriott89b}
Marriott,~K. and ~S{\o}ndergaard,~H. (1989)
Semantics-based dataflow analysis of logic programs.
In G.~Ritter (editor) {\em Proc. of the IFIP 11th World Computer 
Congress, Information Processing 89}, 
pp.~601--606. North-Holland/IFIP.


\bibitem[\protect\citename{Maier, }1991]{Meier91}
Meier,~M. (1991)
Recursion versus iteration in Prolog.
In K.~Furukawa (editor), \emph{Proc.  of the 8th International
  Conference on Logic Programming ({ICLP}'91)},
pp.~157--169.  MIT Press.


\bibitem[\protect\citename{Mellish, }1987]{Mellish87}
Mellish,~C. (1987)
Abstract interpretation of Prolog programs.
In S.~Abramsky and C.~Hankin (editors) {\em Abstract Interpretation
  of Declarative Languages}, chapter~8, 
pp.~181--198.
Ellis Horwood Limited.
  

\bibitem[\protect\citename{Mendelzon, }1991]{Mendelzon85}
Mendelzon,~A.~O. and Wood,~P.~T. (1991)
Functional dependencies in Horn Clause Queries.
\emph{ACM Transactions on Database Systems (TODS)},
  16~(1): 31--55.




\bibitem[\protect\citename{Mulkers, }1991]{Mulkers91a}
Mulkers,~A. (1991)
Deriving live data structures in logic programs by means of
  abstract interpretation.
PhD thesis, Department of Computer Science, Katholieke Universiteit
  Leuven, Belgium.


\bibitem[\protect\citename{Mulkers \emph{et al.}, }1990]{Mulkers90}
Mulkers,~A., Winsborough,~W. and Bruynooghe,~M. (1990)
Analysis of shared data structures for compile-time garbage
  collection in logic programs.
In  D.S. Warren and P.~Szeridi (editors), 
\emph{Proc.  of the 7th International
  Conference on Logic Programming ({ICLP}'90)},
pp.~747--762.  MIT Press.




\bibitem[\protect\citename{Musumbu, }1990]{MusumbuThesis}
Musumbu,~K. (1990)
Interpr\'etation abstraite de programmes Prolog.
PhD thesis, Institute of Computer Science, University of Namur,
  Belgium.




\bibitem[\protect\citename{Muthukumar and Hermenegildo, }1991]{Muthukumar91}
Muthukumar,~K. and Hermenegildo,~M. (1991)
Combined determination of sharing and freeness of program variables
  through abstract interpretation.
In K.~Furukawa (editor), \emph{Proc.  of the 8th International
  Conference on Logic Programming ({ICLP}'91)},
pp.~49--63.  MIT Press.


\bibitem[\protect\citename{Muthukumar and Hermenegildo, }1992]{Muthukumar92}
Muthukumar,~K. and Hermenegildo,~M. (1992)
Compile-time derivation of variable dependency using abstract
  interpretation.
\emph{Journal of Logic Programming}, 13~(2\&3): 315--347.


\bibitem[\protect\citename{Mycroft and  O'Keefe, }1984]{MK84}
Mycroft,~A. and O'Keefe,~R.~A. (1984)
A polymorphic type system for Prolog.
\emph{Artificial Intelligence}, 23~(3): 295--307.


\bibitem[\protect\citename{Nilsson, }1990]{Nilsson90a}
Nilsson,~U. (1990)
Systematic semantic approximations of logic programs.
In P.~Deran\-sart and J.~Ma{\l}uszy\'nski (editors),
 \emph{Proc. of the
  International Workshop on Programming Language Implementation and
  Logic Programming ({PLILP}'90), Lecture Notes in Computer Science, 456}, 
pp.~293--306. Springer-Verlag. 



\bibitem[\protect\citename{Plotkin, }1981]{Plotkin81}
Plotkin,~G.~D. (1981)
A structural approach to operational semantics.
Technical Report DAIMI FN-19, CS Department, University of Aarhus.


\bibitem[\protect\citename{Sahlin, }1991]{Sahlin91}
Sahlin,~D. (1991)
Determinacy analysis for full Prolog.
In \emph{Proc. of
 the Symposium on Partial Evaluation and Semantics-Based
  Program Manipulation ({PEPM}'91)}, SIGPLAN Notices
26~(9): 23--30. 

\bibitem[\protect\citename{Sahlin, }1993]{Sahlin.PhD}
Sahlin,~D. (1993)
Mixtus: an automatic partial evaluator for full Prolog.
\emph{New Generation Computing}, 12~(1): 7--51.



\bibitem[\protect\citename{Schmidt, }1988]{Schmidt88}
Schmidt,~D.~A. (1988)
\emph{Denotational Semantics}.
Allyn and Bacon, Inc.

\bibitem[\protect\citename{Somogyi, }1987]{So86}
Somogyi,~Z. (1987)
A system of precise models for logic programs.
In  E.~Shapiro (editor), \emph{Proc.  of the Fourth International
  Conference on Logic Programming ({ICLP}'87)},
pp.~769--787. MIT Press.


\bibitem[\protect\citename{Spoto, }2000]{Spoto}
Spoto,~F.  (2000)
Operational and goal-independent denotational
semantics for Prolog with cut.
\emph{Journal of Logic Programming}, 42~(1): 1--46.



\bibitem[\protect\citename{Sterling and Shapiro, }1986]{Sterling86}
Sterling,~L. and Shapiro,~E. (1986)
\emph{The Art of Prolog: Advanced Programming Techniques}.
 MIT Press, Cambridge Mass.


\bibitem[\protect\citename{Stoy, }1977]{Stoy77}
Stoy,~J. (1977)
Denotational semantics: the Scott-Strachey approach to
  programming language theory.
MIT Press, Cambridge Mass.


\bibitem[\protect\citename{Tamaki, }1986]{Tamaki86}
Tamaki,~H. and Sato,~T. (1986)
{OLD}-resolution with tabulation.
In E.~Y.~Shapiro (editor), \emph{Proc.  of the 3rd International
  Conference on Logic Programming ({ICLP}'86),
 Lecture Notes in Computer Science 225},
pp.~84--98.   Springer--Verlag.


\bibitem[\protect\citename{Taylor, }1989]{Taylor89}
Taylor,~A. (1989)
Removal of dereferencing and trailing in Prolog compilation.
In G.~Levi and M.~Martelli (editors), \emph{Proc.  of the 6th International
  Conference on Logic Programming ({ICLP}'89)},
pp.~48--60.   MIT Press.


\bibitem[\protect\citename{Ueda, }1987]{Ueda}
Ueda,~K. (1987)
Making exhaustive search programs deterministic, part II.
In J.-L.~Lassez (editor), \emph{Proc.  of the 4th International
  Conference on Logic Programming ({ICLP}'87)},
pp.~356--375.  MIT Press.



\bibitem[\protect\citename{Van~Hentenryck \emph{et al.}, }1993]{WSA93_Granularity}
Van~Hentenryck,~P., Degimbe,~O., Le~Charlier,~B. and Michel,~L. (1993)
The impact of granularity in abstract interpretation of
  Prolog.
In P.~Cousot, M.~Falaschi, G.~Fil\`e and A.~Rauzy
 (editors) {\em Proc. of the 3rd International
  Workshop on Static Analysis ({WSA}'93),
Lecture Notes in Computer Science
 724}, pp.~1--14. Springer--Verlag.


\bibitem[\protect\citename{Van Roy \emph{et al.}, }1987]{Vanroy87}
Van~Roy,~P.,  Demoen,~B., and  Willems,~Y.~D. (1987)
Improving the execution  speed of compiled Prolog
with modes, clause selection, and determinism.
In H.~Ehrig, R.~A.~Kowalski, G.~Levi and U.~Montanari (editors),
\emph{Proc. of the International Joint Conference on Theory and
Practice of Software Develpment ({TAPSOFT}'87),
Lecture Notes in Computer Science, 250}, pp.~111--125. Springer-Verlag.


\bibitem[\protect\citename{Van Roy and Despain, }1992]{Vanroy92}
Van~Roy,~P. and Despain,~A.~M. (1992)
High-performance computing with the Aquarius compiler.
\emph{IEEE Computer}, 25~(1): 54-68.



\bibitem[\protect\citename{Warren, }1992]{Warren92}
Warren,~D.~S. (1992)
Memoing for logic programs.
\emph{Communications of the {ACM}}, 35~(3): 93--111.

\bibitem[\protect\citename{Warren \emph{et al.}, }1988]{Warren88}
Warren,~R., Hermenegildo,~M.~V. and Debray,~S.~K. (1988)
On the practicality of global flow Analysis of Logic Programs.
In  R.~A.~Kowalski and K.~A.~Bowen (editor), \emph{Proc.  of the 5th
International Conference on Logic Programming ({ICLP}'88)},
pp.~349--366.  MIT Press.


\bibitem[\protect\citename{Winsborough, }1992]{Winsborough92}
Winsborough,~W. (1992)
Multiple specialization using minimal-function graph semantics.
\emph{Journal of Logic Programming}, 13~(2\&3) :259--290.


\bibitem[\protect\citename{Xu and Warren, }1988]{XW88}
Xu,~J. and  Warren,~D.~S. (1988)
A type inference system for {P}rolog.
In  R.~A.~Kowalski and K.~A.~Bowen (editor), \emph{Proc.  of the 5th
International Conference on Logic Programming ({ICLP}'88)},
pp.~604--619.  MIT Press.


\bibitem[\protect\citename{Yardeni and Shapiro, }1991]{YS91}
Yardeni,~E. and Shapiro,~E. (1991)
A type System for logic programs.
\emph{Journal of Logic Programming}, 10~(1/2/3\&4): 125--153.

\end{thebibliography}

\newpage
\appendix
\section*{Appendix}
We complete here the description of the abstract operations
started in Section \ref{sub:GAO}. 
The correctness proofs of all the abstract operations can be found in
\cite{sequence}. The definitions below have been added in order to allow
the reader to check the details of the examples in Section \ref{sub:GAO}.
\\

\noindent
{\bf Extension at  Clause Entry}:\ \  
     {\tt EXTC}$(c,\cdot):{\it AS}_D
\rightarrow{\it ASSC}_{D'} $ \ \\
The implementation reuses the homonymous operation
from the previous framework,
which is specified as follows.
\\

\noindent
{\it Operation}$\;$
{\tt EXTC}$(c,\cdot):{\it AS}_D
                       \rightarrow{\it AS}_{D'} $\\
Let $\beta\!\in\!{\it AS}_{D}$,
$\theta\!\in\!{\it CPS}_{D}$, and $\theta'\!\in\!{\it PS}_{D'}$
such that $x_i\theta'=x_i\theta$ 
                   $(\forall i:1\leq i \leq n)$ and
                    $x_{n+1}\theta'$, \dots , $x_{m}\theta'$ are
               distinct standard variables
                   not belonging to ${\it codom}(\theta).$
Then
$$
\begin{array}{lll}
\theta\in{\it Cc}(\beta) & \Rightarrow & [\![\theta']\!]\in
{\it Cc}({\tt EXTC}(c,\beta)).
\end{array}
$$
Hence, the {\tt EXTC} operation on sequences is defined by
$$
\begin{array}{lll}
{\tt EXTC}(c,\beta) & = & 
  \langle \langle {\tt EXTC}(c,\beta), 1, 1, {\it st} \rangle
         , {\it nocut} \rangle.
\end{array}
$$

\vskip0.3cm
\noindent
{\bf Restriction at Clause Exit}:\ \ 
 {\tt RESTRC}$(c,\cdot):{\it ASSC}_{D'}
                         \rightarrow{\it ASSC}_D $ \ \\
The treatment of this operation is similar to the previous one.
We first specify the abstract substitution version of the operation.
\\

\noindent
{\it Operation}$\;$
{\tt RESTRC}$(c,\cdot):{\it AS}_{D'}
                       \rightarrow{\it AS}_{D} $
\\
Let $\beta\!\in\!{\it AS}_{D'}$ and
$\theta\!\in\!{\it CPS}_{D'}$. We have
$$
\begin{array}{lll}
\theta\in{\it Cc}(\beta) & \Rightarrow & [\![\theta_{|D}]\!]\in
{\it Cc}({\tt RESTRC}(c,\beta)).
\end{array}
$$
Hence, the {\tt RESTRC} operation on sequences is defined by
$$
\begin{array}{lll}
{\tt RESTRC}(c,C) & = & 
  \langle {\tt RESTRC}(c,\beta), m, M, {\it acf} \rangle.
\end{array}
$$

\vskip0.3cm
\noindent
{\bf Restriction before a Call}:\ \ 
 {\tt RESTRG}$(l,\cdot):{\it AS}_{D'}
                         \rightarrow{\it AS}_{D'''} $\ \\
This  operation is simply inherited from the previous framework.
\\

\noindent
{\bf Unification of a Variable and a Functor}:\ \ 
   {\tt UNIF-FUNC}$(f,\cdot): {\it AS}_{D} 
                     \rightarrow {\it ASS}_{{D}}$\ \\ 
The treatment of this operation is identical to the treatment of the
{\tt UNIF-VAR} ope\-ration and  is thus omitted.
\\

\noindent
{\bf Extension of the Result of a Call}:\ \  
{\tt EXTGS}$(l,\cdot,\cdot):{\it ASSC}_{D'}\times {\it ASS}_{D'''}
                         \rightarrow{\it ASSC}_{D'} $\ 
\\
This operation reuses the operation {\tt EXTG} from the previous framework.
The reused operation has to fulfill the specification just below.
\\

\noindent
{\it Operation}$\;$
{\tt EXTG}$(l,\cdot,\cdot):{\it AS}_{D'}\times {\it AS}_{D'''}
                         \rightarrow{\it AS}_{D'} $
\\
Let $\beta_1\in{\it AS}_{D'}$ and $\beta_2\in{\it AS}_{D'''}$.
Let $\theta_1\in {\it CPS}_{D'}$ and $\theta_2\in {\it PS}_{D'''}$ be
such that $x_{i_j}\theta_1=x_j\theta_2$
$(\forall j : 1\leq j\leq {n'})$. 
Let $\sigma\in {\it SS}$ such that 
${\it dom}(\sigma)\subseteq {\it codom}(\theta_2)$.
Let $\{z_1,\ldots,z_r\}={\it codom}(\theta_1)
                             \setminus{\it codom}(\theta_2)$.
Let $y_{1},\ldots,y_{r}$ be distinct standard variables
      not belonging to ${\it codom}(\theta_1)\cup {\it
        codom}(\sigma)$.
Let $\rho=\{z_1/y_{1},\ldots,
                    z_r/y_{r},
                    y_{1}/z_1,\ldots,
                    y_{r}/z_r\}$.
Under these assumptions,
\begin{quote}$
\begin{array}{rcl}
       \left.
       \begin{array}{r}
       \theta_1\in {\it Cc}(\beta_1),\\
       \theta_2\sigma\in{\it Cc}(\beta_2)
       \end{array}
       \right\}
               &\Rightarrow&
        [\![\theta_1\rho\sigma]\!] \in 
                       {\it  Cc}({\tt EXTG}(l,\beta_1,\beta_2)).
\end{array}$
\end{quote}

\noindent
The implementation of {\tt EXTGS} is as follows.

\begin{quote}
  $\begin{array}{llll}
   \beta' & = & {\tt EXTG}(l,\beta_1,\beta_2);                     \\

   m'     & = & m_1 m_2         & \myif t_2={\it st},              \\
          & = & \min(1,m_1) m_2 & \otherwise;                      \\

   M'     & = & \min(1,M_1) M_2 & \myif t_2={\it snt},             \\
          & = & M_1 M_2         & \otherwise;                      \\

   t'     & = & {\it snt}       & \myif t_1={\it snt}
                                  \myor (t_2={\it snt}
                                         \and m_1\geq 1),         \\
          & = & {\it st}        & \myif t_1={\it st}
                                  \and (t_2={\it st}
                                         \myor M_1=0),            \\
          & = & {\it pt}        & \otherwise;                     \\

   {\it acf}'
          & = & {\it acf}.      
   \end{array}$
\end{quote}

\vskip0.3cm
\noindent
{\bf Operation}$\;$ ${\tt SEQ}: {\it ASSC}_{D}
                         \rightarrow{\it ASS}_{D}$\ 
\\
We define 
$${\tt SEQ}(\langle B, {\it acf}\rangle) \  = \  B.$$

\vskip0.3cm
\noindent
{\bf Operation} $\;$ ${\tt SUBST}: {\it ASSC}_{D'}
                         \rightarrow{\it AS}_{D'}$\\ 
We define
$$
{\tt SUBST}(\langle \langle \beta, m, M, t \rangle, {\it acf}\rangle)\
=\  \beta.$$

\end{document}